\documentclass[prb,twocolumn,showpacs]{revtex4}%
\usepackage{graphicx}
\usepackage{url}
\usepackage{amsfonts}
\usepackage{amsmath}
\usepackage{amssymb}%
\input{bold_greek}
\begin{document}
\title[Fluctuations on the market]{Theory of market fluctuations}
\author{S.V. Panyukov}
\affiliation{\emph{P.N. Lebedev Physics Institute, Russian academy of Science, Leninskiy
pr., 53, Moscow, 117924, Russia}}

\begin{abstract}
We propose coalescent mechanism of economic grow because of redistribution of
external resources. It leads to Zipf distribution of firms over their sizes,
turning to stretched exponent because of size-dependent effects, and predicts
exponential distribution of income between individuals.

We present new approach to describe fluctuations on the market, based on
separation of hot (short-time) and cold (long-time) degrees of freedoms, which
predicts tent-like distribution of fluctuations with stable tail exponent
$\mu=3$ ($\mu=2$ for news). The theory predicts observable asymmetry of the
distribution, and its size dependence. For financial markets the theory
explains first time ``market mill''\ patterns, conditional distribution,
``D-smile'', z-shaped response, ``conditional double dynamics'', the skewness
and so on.

We propose a set of Langeven equations for the market, and derive equations
for multifractal random walk model. We find logarithmic dependence of price
shift on the volume, and volatility patterns after jumps. We calculate
correlation functions and Hurst exponents at different time scales. We show,
that price experiences fractional Brownian motion with chaotically switching
of sub- and super-diffusion, and calculate corresponding probabilities,
response functions, and risks.

\end{abstract}
\maketitle
\tableofcontents

\section{Introduction}

First question behind any research is why do we need it? There are no unique
approach in econophysics, and the number of different approaches grows
exponentially with time. How can we decide, which of them is ``correct'', if
by construction, any one well describes empirical facts?

The answer is simple: in no way. All of them are equivalent at regions of
their applicability. But these regions are very different, and only several
theories able to describe a large variety of empirical facts. Extrapolating,
one can assume that only one theory can predict all important phenomena:
Market mill patterns, multifractality of fluctuations, volatility patterns,
different Hurst exponents above and below a time $\tau_{x}$, and many other
facts. Some of them can be described in different ways, but not the hookup of
all facts.

What criteria should satisfy such theory? At first sight, it is mathematical
rigor. The most striking example is the Flory approach in polymer physics,
which is absolutely ``wrong''\ mathematically, but extremely well describing
all known situations. All multiple attempts to (im)prove it were failed. We
conclude, that the rigor of the theory is usually ``inversely
proportional''\ to intuition.

Well, what kind of the theory should not be? If any new fact or their series
need to introduce additional terms or ideas into the theory, the later can be
considered as a collection of facts, arbitrary ordered according to the test
of the author. We think, the real theory must predict in future yet unknown
facts (at present this criterium is equivalent to extremely wide region of its
applicability), to be as rigor as possible, and use minimum of initial assumptions.

We do not know other criteria of the ``validity''\ of the theory, and this is
the reason why the theory \emph{must} describe all known trustable facts.
Present paper can be considered as an attempt to follow this criteria. Only
one main idea lays in the basis of our theory of market fluctuations: we
assume, that they can be described as random walk motion at all time scales.
In the case of financial market, it is random trading at all time horizons
from seconds to tenths years.

Our theory can be considered as an attempt to make a step from numerous
descriptive approaches toward a physical Langeven formulation of the
``econophysical''\ problem. This is why we emphasize analogies with other
branches of physics, which may confuse econo-physicists otherwise. Although we
show, that multi-time random trading allows to explain most of market
dynamics, it may be extended later in many directions.

As a strategy line, for each problem we try to construct a simplified model of
such multi-time random motion, capturing the most of physics. As the result,
we left with several parts of the whole puzzle, strongly inter-correlated with
each other. It is the reason of unusual length of this paper, which can not be
cut into several independent small parts.

\section{Firms, cities and income distributions\label{FIRMS}}

\subsection{Is there thermodynamics of the market?}

Econophysics studies physical problems in economics, and most of its results
were obtained from analogy with thermodynamics. One of classical problems of
econophysics, the firm grow, is usually described by the model of stochastic
firm growing\cite{Gibrat}. In order to explain empirically observed Zipf
distribution of firm sizes\cite{Pe-PRE-96} it is proposed to introduce the
lower reflecting boundary in the space of firm sizes, which stabilizes the
distribution to a power law\cite{LS-96}. Unfortunately, this explanation is
inconsistent for firms of one or several employers, well described by the same
empirical Zipf distribution.

Different models of internal structure of firms were proposed for the
stochastic mechanism of firm growing. Hierarchical tree-like model of firm was
studied in Refs. \cite{Am-PRL-98,Le-PRL-98}. A model of equiprobable
distribution of all partitions of a firm was introduced in Ref.\cite{Su-Ph-02}%
. Both models neglect the effect of competition between different firms. The
random exchange of resources between firms was taken into consideration in
``saving''\ models\cite{Sa-80}. In Refs. \cite{Cha-95,Dra-00,Cha-04,Sl-04} the
process of stochastic firm grow and loss was considered by analogy with
scattering processes in liquids and gases. The distribution of firms over
their sizes in different countries was studied in Ref.\cite{Ra-Ph-00}.

The theory of firms is usually called microeconomics, and from economical
point of view it is hard to consider the stochasticity as the moving force of
economic grow. While in thermodynamics the stochasticity originates from
interaction with a huge ``thermostat'', there are no such thermostat for the
market, which subsists only because of activity of its direct participants.

This puzzle forces us to develop a ``mean field''\ theory of firm growing,
neglecting any fluctuation processes. We show, that the moving force of
evolution on the market are not thermal-like excitations, but the supply of
external resources, which are (re-)distributed between different firms.
Exhaustion of the resource kills this (part of the) market, while appearance
of a new resource gives rise to a new market. The process of firm growing and
mergence is similar to coalescence of droplets of a new phase, when
stochasticity plays only minor role.

In section~\ref{MEAN} we show that the coalescence theory predicts Pareto
power low for the distribution of firm sizes. We propose self-similar
tree-like model of firms in section~\ref{MODEL}. This model is solved in
Appendix~\ref{SLEZOV}, and we show, that it explains empirically observable
time dependence of the Pareto exponent for the world income.

The formal resemblance of observable exponential distribution of the income
between individuals to Boltzmann statistics was used in Ref.\cite{DY-01} to
justify the applicability of methods of equilibrium thermodynamics. But how
can all sectors of country economics and services always be in thermal
equilibrium? In section~\ref{INCOME} we propose an alternative explanation,
based on unified tax policy in the whole country: the coalescent approach
predicts, as a by-product, the exponential income distribution, even without
invention of thermal equilibrium. This distribution is valid for the majority
of the population, and statistical fluctuations are only responsible for power
tails of its upper part (1--3\%).

Countries with different financial policy have different ``effective
temperature''\ of the distribution, which can be equilibrated\ only after
unification of their financial policies, even without establishment of a
``heat death''\ -- global thermal equilibrium. Although one may consider the
perpetual trade deficit of US as consequence of the fundamental second law of
thermodynamics\cite{DY-01}, it would be more natural to explain it by
financial policy, directed on attraction of resources to the country.

Econophysics is not only one field, deceptively resembling thermodynamics, we
have to mention also a sand, turbulence and other macroscopic systems, which
form complex dissipative structures in the response on some external forces.
Although such ``open\ systems''\ can not be characterized by thermodynamic
potentials, the process of dissipation is accompanied by the rise of
information entropy. We calculate the entropy of the market and show, that it
can only increase with time, since the market irreversibly absorbs external
information (there is deep analogy with physics of decoherence, discussed in Conclusion).

``Thermodynamic-type''\ models predict asymptotically Gaussian distribution of
firm grow rates, while the real distribution has tent-like shape. In order to
reproduce it, in Ref. \cite{St-JPh-97} an artificial potential was introduced
in diffusion equation, restoring the firm size to a certain reference value,
at which the grow rate abruptly changes its sign. In this paper we elaborate a
new approach to study dynamics of temporal dissipative structures on the
market, which do not use these artificial assumptions.

In section~\ref{FLUCT} we introduce new general approach to study market
fluctuations. Main ideas of this approach will be first formulated for the
problem of firm grow. The market is the system with multiple (quasi-)
equilibrium states, characterized by extremely wide spectrum of relaxation
times. By analogy with glasses, for given observation (coarse graining) time
interval $\tau$ we can divide all degrees of freedom of the market into
``hot''\ and ``cold''\ ones, depending on their relaxation times. Hot degrees
of freedom are in equilibrium, and they generate high frequency fluctuations
because of uncertainty on the market, while cold degrees of freedom are not
equilibrated, and evolve on times large with respect to $\tau$. As in the case
of spin-glasses, high degeneracy of quasi-equilibriums in the market is
reflected in the presence of a gauge invariance. Any averages should be
defined in two stages: first, the annealed averaging over hot\ degrees of
freedom, and then quenched averaging over cold\ degrees of freedom.

We demonstrate, that our theory reproduces empirically observable (in general,
asymmetric) tent-like distribution of firms over their grow rates. In
section~\ref{FAT} we show, that this distribution has fat tail with stable
exponent $\mu$, equals to the number of essential degrees of freedom of the
noise ($\mu=3$ for Markovian statistics of hot degrees of freedom, and $\mu=2$
for uncorrelated noise).

\subsection{Mean field theory\label{MEAN}}

Dynamics of firm growing is similar to kinetics of growing of droplets of a
new phase. Large firms can absorb smaller ones, and they can grow or leave the
business, by analogy with resorption and growing of droplets in the
supersaturated solution. Below we use this analogy to construct a new theory,
not relying on stochastic mechanisms of firm growth. Entropic and
microeconomic interpretations of our theory are discussed in
Appendixes~\ref{THERMO} and~\ref{MACRO}.

\subsubsection{Zipf distribution\label{PARETO}}

For definiteness sake we define the firm size as the number $G$ of its
employees. In general, it could be any resource, shared between different
firms on the market. According to economic approach (analog of the mean field
approach in physics) firms can hire or loose the staff only through the
``reservoir''\ of unemployments of value $U\left(  t\right)  $ at time $t$.
Diffusion processes lead to finite value $U_{\ast}>0$ of the ``natural
unemployment''. ``Actual unemployment''\ $U$ is the sum of\ $U_{\ast}%
$\cite{Fr--68} and the ``market unemployment'', $\Delta\left(  t\right)  $:
\[
U\left(  t\right)  =U_{\ast}+\Delta\left(  t\right)  .
\]

The equation of the resource balance can be written in the form%
\begin{equation}
Q\left(  t\right)  =U\left(  t\right)  +\int Gf\left(  G,t\right)  dG,
\label{Qf}%
\end{equation}
where $Q\left(  t\right)  $ is the supply of external resources. The
probability distribution function (PDF) $f\left(  G,t\right)  $ of firm sizes
is determined by the continuity equation,
\begin{equation}
\dfrac{\partial f\left(  G,t\right)  }{\partial t}=-\dfrac{\partial}{\partial
G}\left[  \dfrac{dG}{dt}f\left(  G,t\right)  \right]  , \label{kinet}%
\end{equation}
where $dG/dt$ is the rate of ordered motion in the space of firm sizes.
Diffusion contribution in Eq.~(\ref{kinet}) is negligible in coalescent
regime. According to the famous Gibrat's observation\cite{Gibrat} the relative
grow rate of the firm,%
\begin{equation}
\dfrac{1}{G}\dfrac{dG}{dt}=r_{G} \label{r}%
\end{equation}
do not depend on its size, $G$. In the case of full employment, $\Delta=0$,
the average numbers of people getting a job and leaving it are the same, and
there are no source for firm grow, $r_{G}=0$. At small $\Delta$ we can hold
only linear term in the series expansion of the grow rate $r_{G}=q\Delta$ in
powers of $\Delta$ with constant $q$.

To solve the set of equations~(\ref{Qf}) --~(\ref{r}) we substitute
Eq.~(\ref{r}) with $r_{G}=q\Delta$ into Eq.~(\ref{kinet}), and find its
general solution%
\[
f\left(  G,t\right)  =\dfrac{1}{G}\chi\left[  \ln\dfrac{G}{G_{0}}-q\int
_{0}^{t}\Delta\left(  t^{\prime}\right)  dt^{\prime}\right]  ,
\]
where $G_{0}$ is the firm size at initial time $t=t_{0}$ and $\chi$ is
arbitrary function. Substituting this solution into the balance
equation~(\ref{Qf}) and introducing new variable of integration $u=\ln\left(
G/G_{0}\right)  $, we find%
\begin{equation}
Q\left(  t\right)  =U\left(  t\right)  +G_{0}\int e^{u}\chi\left[  u-q\int
_{0}^{t}\Delta\left(  t^{\prime}\right)  dt^{\prime}\right]  du. \label{QU}%
\end{equation}

Consider the case of power growing of external resources,%
\begin{equation}
Q\left(  t\right)  =Q_{0}t^{m}. \label{Qt}%
\end{equation}
For general time dependence $Q\left(  t\right)  $ its logarithmic rate $m$ is
determined by expression
\begin{equation}
m=\dfrac{d\ln Q\left(  t\right)  }{d\ln t}. \label{mQ}%
\end{equation}
In the case of small unemployment value, $U\ll Q$, general solution of
Eq.~(\ref{QU}) takes exponential form, $\chi\left(  u\right)  =\chi
_{0}e^{-\kappa u}$. Substituting this expression into Eq.~(\ref{QU}) and
taking into account that the distribution $f\left(  G,t\right)  $ can not
depend on initial firm size, $G_{0}$, we find $\kappa=1$ and%
\[
Q_{0}t^{m}=\chi_{0}\ln\dfrac{G_{\max}}{G_{\min}}\exp\left[  q\int_{0}%
^{t}\Delta\left(  t^{\prime}\right)  dt^{\prime}\right]  ,
\]
where $G_{\min}\ $and $G_{\max}$ are maximal and minimal firm sizes on the
market. The solution of this equation has the form%
\begin{equation}
\Delta\left(  t\right)  =m/\left(  qt\right)  ,\qquad\chi_{0}=Q_{0}/\ln\left(
G_{\max}/G_{\min}\right)  . \label{Dt}%
\end{equation}
First of Eqs.~(\ref{Dt}) predicts, that the economic grow, see Eq.~(\ref{Qt}),
leads to less actual unemployment, $\Delta$, in qualitative agreement with the
famous macroeconomic ``Fillips curve''. Close quantitative relation between
the coalescent theory and the Fillips low is established in
Appendix~\ref{MACRO}.

\begin{figure}[tb]
\begin{center}
\includegraphics[
height=1.923in,
width=3.105in
]{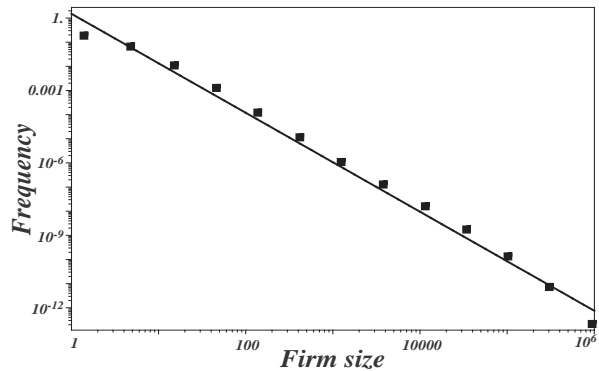}
\end{center}
\caption{Size distribution of U.S. business firms in 1997 (Census
data)\cite{Ax-01}. Straight line corresponds to power law distribution
$F\left(  G\right)  \sim G^{-\gamma}$ with exponent $\gamma=1.059$.}%
\label{Rate}%
\end{figure}

We conclude, that for any monotonically increasing function $Q\left(
t\right)  $ the distribution of firms over their sizes $G$ has Zipf form:%
\begin{equation}
f\left(  G,t\right)  =\dfrac{Q\left(  t\right)  }{\ln\left(  G_{\max}/G_{\min
}\right)  }\dfrac{1}{G^{2}}. \label{Zipf}%
\end{equation}
This dependence was really observed for extremely wide range of firm sizes,
see Fig.~\ref{Rate}, where empirically observable distribution%
\begin{equation}
F\left(  G,t\right)  \equiv\dfrac{\int_{G}^{G_{\max}}f\left(  G,t\right)
dG}{\int_{G_{\min}}^{G_{\max}}f\left(  G,t\right)  dG}\sim\dfrac{1}{G}
\label{FG}%
\end{equation}
is plotted. The Zipf distribution\cite{Pe-PRE-96}~(\ref{Zipf}) is valid for
the entire range of US firms\cite{Ax-01} (from $G_{\min}=1$ to $G_{\max
}=10^{6}$) with Pareto exponent very close to unity. 

The same mechanism may be responsible for power distribution function of
cities over their population, the amount of assets under management of mutual
funds\cite{Ga-03}, banks\cite{PuHa-04} and so on. In the analysis of city
population in different countries, the exact form of Zipf's law~(\ref{FG}) was
confirmed in 20 out of 73 countries\cite{Soo-02}. Deviations from this low
will be studied in next section.

\subsubsection{Stretched exponent\label{MODEL}}

The Pareto exponent~(\ref{FG}) can deviate from $1$ because of ineffective
management, strong influence of industry effects on small firms and so on.
With increasing size, these effects gradually trail off, while remaining
international, national and regional shocks equally affect all firms. Assuming
self-similiarity of firm structure, the variation of the firm size can be
described by Master equation
\begin{equation}
r\equiv G^{-1}dG/dt=r_{G}-pG^{-\beta}, \label{dxdt}%
\end{equation}
with constant $p$ and $\beta$.

To derive Eq.~(\ref{dxdt}), consider the firm as the self-similar
tree\cite{St-JPh-97} of $n$ generations, each of $G_{0}\gg1$ branches. The
size $G_{0} $ of each subdivision is described by the same type of
equation~(\ref{dxdt}),%
\begin{equation}
r_{0}=G_{0}^{-1}dG_{0}/dt=r_{0}-p_{0}G_{0}^{-\beta_{0}}. \label{ddd}%
\end{equation}
Substituting the estimation $G\simeq G_{0}^{n}$ for the size of the whole tree
in Eq.~(\ref{dxdt}) and comparing with Eq.~(\ref{ddd}), we find the relation
between coefficients of Master equations~(\ref{dxdt}) and~(\ref{ddd}):
\[
\beta=\beta_{0}/n,\quad r_{G}=r_{0}n,\quad p=p_{0}n.
\]

In Appendix~\ref{MACRO} we show that while the Gibrat grow rate $r_{G}$ is
fixed by economic factors, the coefficient $p$ of job destruction can
experience strong random fluctuations $\Delta p$. Neglecting fluctuations of
$r_{G}$ in Eq.~(\ref{dxdt}) we find that fluctuations in size are inversely
correlated to the size with an exponent $\beta$:%
\begin{equation}
\Delta r=-\Delta pG^{-\beta}. \label{rG}%
\end{equation}
In order to estimate the exponent $\beta_{0}$, consider a hypothetical
structureless firm with $n=1$ of the size $G=G_{0}\gg1$. Fluctuations of its
size are characterized by Gaussian exponent $\beta=\beta_{0}=1/2$. The
exponent $\beta$ of real firms takes small values $\beta=0.15-0.21$%
\cite{Le-PRL-98}, corresponding to the number of tree generations $n=1/\left(
2\beta\right)  =3-4$. Using Eq.~(\ref{rG}) we find the dependence of the
standard deviation of grow rate $\Delta r$ on the firm size,
\begin{equation}
\left\langle \Delta r^{2}\right\rangle ^{1/2}=\sigma G^{-\beta}, \label{dr}%
\end{equation}
where $\sigma\equiv\left\langle \Delta p^{2}\right\rangle ^{1/2}$ does not
depend on firm size $G$. This relation is in excellent agreement with
empirical data\cite{St-JPh-97,St-Na-96}.

The condition $r=0$~(\ref{dxdt}) determines the critical firm size
\begin{equation}
G_{c}=\left(  p/r_{G}\right)  ^{1/\beta}. \label{Gc}%
\end{equation}
Small firms with $G<G_{c}$ collapse with time and may leave from the business
(or reach a certain fluctuation size), while large firms with $G>G_{c}$ grow.
In Appendix~\ref{THERMO} we find the entropy $S\left(  G\right)  $ of the firm
of size $G$, and show that $G=G_{c}$ corresponds to its minimum, and also to
the minimum point of a ``U-shaped''\ average cost curve in the conventional
economic theory (Appendix~\ref{MACRO}). We also derive maximum entropy
principle for the market (Appendix~\ref{THERMO}), which is known as the most
foundational concepts of Gibbs systems.

In Appendix~\ref{SLEZOV} we show that the solution of Eqs.~(\ref{Qf}%
),~(\ref{kinet}) with the rate~(\ref{dxdt}) has stretched exponent form:
\begin{equation}
F\left(  G\right)  =\exp\left[  -\left(  1/\beta-m\right)  \allowbreak\left(
G/G_{c}\right)  ^{\beta}\right]  . \label{Str}%
\end{equation}
Taking the limit $\beta\rightarrow0$ we reproduce Eq.~(\ref{FG}). It is shown
that stretched exponent is the best fitting approximation for many observable
distributions (size of cities, population of different countries, popularity
of executors, lifetime of different species, strength of earthquakes, indices
of quoting, number of coauthors, relative rates of protein synthesis and many
others\cite{La-98,Da-02,Ne-01}), which are determined by the competition of
units for common resources. At small but finite $\beta\ll1$ expanding $\left(
G/G_{c}\right)  ^{\beta}\simeq1+\beta\ln\left(  G/G_{c}\right)  $ in
Eq.~(\ref{Str}) we find
\begin{equation}
F\left(  G\right)  \sim G^{-\gamma},\qquad\gamma=1-\beta m. \label{Par}%
\end{equation}

We conclude, that the exponent $\gamma$ of Pareto distribution is, in general,
not universal and depends on current rate $m\left(  t\right)  $ of external
supply, Eq.~(\ref{mQ}). This conclusion can be verified by empirical
observations: typically, the value of this exponent is in the interval
$0.7<\gamma<1$. For example, the size distribution of Danish production
companies with ten or more employees follows a rank-size distribution with
exponent $\gamma=0.741$\cite{KN-01}.

\begin{figure}[tb]
\begin{center}
\includegraphics[
height=1.8049in,
width=3.3079in
]{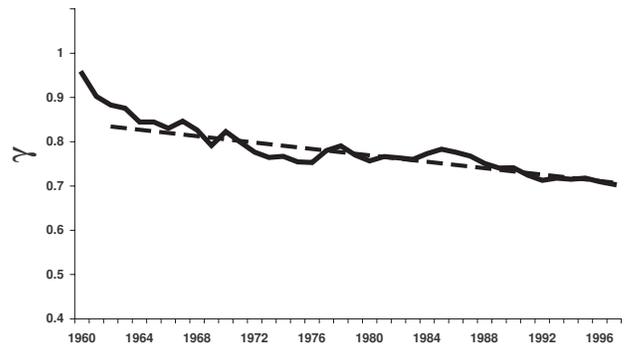}
\end{center}
\caption{Temporal path of the exponent $\gamma$ (continuous line), and its
approximation by linear dependence (dotted line)\cite{GU-EB-03}.}%
\label{World}%
\end{figure}

To confirm the dependence of the exponent $\gamma$ on the supply rate
$m\left(  t\right)  $, consider the distribution of world income across
different countries. We assume, that countries could be described by the same
Master equation~(\ref{dxdt}) as large firms. Exponential growing of consumable
resources leads to linear time dependence of $m\sim t$, see Eq.~(\ref{mQ}). As
the result, the exponent $\gamma$ linearly decreases with time, in good
agreement with empirical observations, see Fig.~\ref{World}. Assuming, that
$Q\left(  t\right)  $ doubles every 12 years, we estimate $\beta\simeq0.1$,
corresponding to a reasonable number $n\simeq5$ of hierarchical management
ranks in the ``typical''\ country. 

\subsubsection{Income distribution\label{INCOME}}

In order to find the distribution of income between individuals we first
introduce the most important economic terms. The total income per state, $Q$,
is shared between all individuals $\left\{  G\right\}  $ and the state
expenses, $U\left(  t\right)  $, according to the balance equation~(\ref{Qf}).
There are some minimal expenses of the state, $U_{\ast}$, and the inequality
$\Delta=U-U_{\ast}>0$ is usually regulated indirectly, through taxes, which
determine the relative income rate, $r_{G}=q\Delta$, of individuals.
Therefore, the income $G$ can be described by a generalization of the Master
equation~(\ref{r}),
\begin{equation}
\dfrac{dG}{dt}=r_{G}G-p. \label{G1}%
\end{equation}
The last term describes the rate of losses (living-wage), the same for all
individuals (linear in $G$ losses renormalize $r_{G}$). Since Eq.~(\ref{G1})
has the form of Eq.~(\ref{dxdt}) with $\beta=1$, from Eq.~(\ref{Str}) we get
exponential distribution of the income:%
\begin{equation}
f\left(  G,t\right)  \sim e^{-G/T},\quad T=p/\left[  r_{G}\left(  1-m\right)
\right]  . \label{expon}%
\end{equation}
According to Eq.~(\ref{dt}) of Appendix~\ref{MACRO} the average income (the
``temperature\cite{SY-05}'')\ $T$ linearly grows with time, in good agreement
with empirical observations\cite{SY-05}, and also rises with the supply rate
$m$~(\ref{mQ}). It is small for countries with low living wage $p $, producing
high inequality in incomes.

Analysis of empirical data shows\cite{DY-01}, that for approximately $95\%$ of
the total population, the distribution is exponential, while the income of the
top $5\%$ individuals is described by a power-law~(\ref{Par}) with time
dependent Pareto index $\gamma$. This tail is because of speculation in
stocks, when the income is proportional to the volume of sale/buy $G\sim V$.
The distribution of large volumes is power tailed, $P\left(  V\right)  \sim
V^{-\gamma}$. The exponent $\gamma$ is not universal, it depends on individual
stocks with typical value $\gamma\simeq3/2$, in good agreement with observable
values\cite{SY-05} $\gamma=1.4-1.8$ (changing of the most profitable stocks
leads to variations in $\gamma$).

In general, the income may come from different sources. In the case of $n$
independent sources convolution of $n$ exponential distributions gives the
Gamma distribution $P_{n}\left(  G\right)  \sim G^{n}e^{-G/T}$, which better
describes Russian Rosstat data of salary distribution.

\subsection{Fluctuation theory\label{FLUCT}}

\subsubsection{Cold and hot degrees of freedom}

Our approach to the description of fluctuations on the market is related to
the main idea of microeconomic theory, based on independent study of
``short-time''\ and ``long-time''\ periods of firm growth. The separation of
time scales also has deep analogy with methods of study of complex physical
systems with a wide spectrum of relaxation times, as glasses. For given
observation time $\tau$ degrees of freedoms of such systems can be divided
into ``hot''\ and ``cold''\ ones. Hot degrees of freedoms fluctuate in the
short-time period ($t<\tau$) given that cold degrees of freedoms are fixed and
can only vary in the long-time period ($t>\tau$). Instead of consideration of
slow dynamics of one system in the long-time period one usually study
statistical properties of an ensemble of such systems at the given time $t$.

We apply this approach to find PDF of grow rates of firms, which have
different dynamics in ``short-time''\ and ``long-time''\ periods. In order to
establish general expression for oscillations of the parameter $\Delta
p\left(  t\right)  $~(\ref{dxdt}) it is instructive to consider first
single-harmonic case. General expression $\Delta p\left(  t\right)  =\sqrt
{2}a\cos\left(  \omega t+\phi\right)  $ can be expanded over two basis
functions $\xi^{\prime}\left(  t\right)  =\cos\left(  \omega t\right)  $ and
$\xi^{\prime\prime}\left(  t\right)  =\sin\left(  \omega t\right)  $:
\begin{equation}
\Delta p\left(  t\right)  =\sqrt{2}a^{\prime}\xi^{\prime}\left(  t\right)
+\sqrt{2}a^{\prime\prime}\xi^{\prime\prime}\left(  t\right)  \equiv\sqrt
{2}\left(  \mathbf{a},\mathbf{\xi}\left(  t\right)  \right)  . \label{dqt1}%
\end{equation}
which are orthogonal:%
\begin{equation}
\left\langle \xi^{2}\right\rangle =\left\langle \left(  \xi^{\prime}\right)
^{2}\right\rangle +\left\langle \left(  \xi^{\prime\prime}\right)
^{2}\right\rangle =1,\qquad\left\langle \xi^{\prime}\xi^{\prime\prime
}\right\rangle =0. \label{xx12}%
\end{equation}
Here $\left\langle \cdots\right\rangle $ means time average. Instead of two
real basis functions it is convenient to introduce one complex function
$\mathbf{\xi}\left(  t\right)  =\xi^{\prime}\left(  t\right)  +i\xi
^{\prime\prime}\left(  t\right)  $ and complex amplitude $\mathbf{a}%
=a^{\prime}+ia^{\prime\prime}=ae^{i\phi}$, in terms of which the scalar
product in Eq.~(\ref{dqt1}) is given by expression $\left(  \mathbf{a}%
,\mathbf{\xi}\right)  =\operatorname{Re}\left(  \mathbf{a}^{\ast}\mathbf{\xi
}\right)  $. In the following we use bold notations both for vectors and
complex numbers.

In general case, the frequency of quick oscillations $\omega\gtrsim\tau^{-1}$
of $\mathbf{\xi}\left(  t\right)  $ (as well as its amplitude) randomly varies
with time. Real and imaginary parts of $\mathbf{\xi}$ can be considered as
random values normalized by condition~(\ref{xx12}), where $\left\langle
\cdots\right\rangle $ has the meaning annealed averaging over the noise
$\mathbf{\xi}\left(  t\right)  $. Complex amplitude $\mathbf{a}$\ is fixed in
the short-time period, and can be considered as random variable in the
long-time period (or for the ensemble of different firms for given time $t$).
The random function $\mathbf{\xi}\left(  t\right)  $ and the amplitude
$\mathbf{a}$ describe hot and cold degrees of the freedom of the market, respectively.

Notice, that $\Delta p\left(  t\right)  $~(\ref{dqt1}) is invariant with
respect to ``gauge''\ transformation
\begin{equation}
\mathbf{\xi}\rightarrow\mathbf{\xi}e^{i\varphi},\quad\mathbf{a}\rightarrow
\mathbf{a}e^{i\varphi}, \label{calibr_1}%
\end{equation}
with constant $\varphi$, reflecting high degeneracy of market
quasi-equilibrium states.

\subsubsection{Double Gaussian model}

We first calculate PDF of fluctuations $\Delta p$,
\begin{equation}
\mathcal{P}\left(  x\right)  \equiv\overline{\left\langle \delta\left[
x-\Delta p\left(  t\right)  \right]  \right\rangle }. \label{Pdq}%
\end{equation}
The bar means ensemble (quenched for the time $\tau$) averaging over
amplitudes $\mathbf{a}$ of fluctuations of different firms. The main
assumption of ``Double Gaussian model''\ is extremely simple: since tactics of
firms at the short-time period is determined by large number of essentially
independent factors, we assume Gaussian statistics of random variable
$\mathbf{\xi}$ at time horizon $\tau$ (due to centeral limit theorem). But two
different firms (or the same firm at two different time intervals $\tau$) will
have, in general, different amplitude of fluctuations $\mathbf{a}$ at the
long-time strategy horizon. Since the strategy of firms is also determined by
large number of independent random factors, we assume Gaussian statistics of
the random amplitude $\mathbf{a}$ with dispersion $\sigma^{2}=\overline{a^{2}%
}$.

Due to the gauge invariance~(\ref{calibr_1}) the noise and the amplitude PDFs
could depend only on moduli $\xi=\left\vert \mathbf{\xi}\right\vert $ and
$a=\left\vert \mathbf{a}\right\vert $. In this section we assume, that hot
($\mathbf{\xi}$) and cold ($\mathbf{a}$) random variables are independent with
zero average and Gaussian weights%
\begin{equation}
\mathcal{Q}_{G}\left(  \xi\right)  =\dfrac{1}{\pi}e^{-(\xi^{\prime})^{2}%
-(\xi^{\prime\prime})^{2}},\quad\dfrac{1}{\pi\sigma^{2}}e^{-[\left(
a^{\prime}\right)  ^{2}+\left(  a^{\prime\prime}\right)  ^{2}]/\sigma^{2}}
\label{Gauss}%
\end{equation}
respectively.

Fourier transform can be used to calculate the averages:%
\[
\mathcal{P}\left(  x\right)  =\int G\left(  k\right)  e^{-ikx}\dfrac{dk}{2\pi
},\quad G\left(  k\right)  =\overline{\left\langle e^{i\sqrt{2}k\left(
a^{\prime}\xi^{\prime}+a^{\prime\prime}\xi^{\prime\prime}\right)
}\right\rangle }%
\]
We first calculate the average over Gaussian normalized $\xi^{\prime}$ and
$\xi^{\prime\prime}$ and get $G\left(  k\right)  =\overline{\exp
\{-k^{2}[(a^{\prime})^{2}+(a^{\prime\prime})^{2}]/2\}}$. Calculating the
average over $a^{\prime}$ and $a^{\prime\prime}$, we get $G\left(  k\right)
=\left(  1+\sigma^{2}k^{2}/2\right)  ^{-1}$. The last step -- is to take the
inverse Fourier transform of this $G\left(  k\right)  $:%
\begin{equation}
\mathcal{P}\left(  x\right)  =\int_{-\infty}^{\infty}\dfrac{\cos\left(
kx\right)  }{1+\sigma^{2}k^{2}/2}\dfrac{dk}{2\pi}=\dfrac{1}{2\sigma}%
e^{-\sqrt{2}\left\vert x\right\vert /\sigma}. \label{Pexp}%
\end{equation}
Exponential distribution of firm grow rates~(\ref{Pexp}) was really observed
for typical fluctuations $x=\Delta p=-\Delta rG^{\beta}$, see Eq.~(\ref{rG}),
with the exponent $\beta=0.15$. We conclude, that tent-like exponential
distribution of firm grow rates is the consequence of Gaussian statistics of
all degrees of freedom (hot and cold) of the market.

\subsubsection{Asymmetry of PDF\label{ASSYM}}

The assumption of Double Gaussian model about independence of cold and hot
variables is, in general, too strong, and the noise $\mathbf{\xi}\left(
t\right)  $ is (anti)correlated with the amplitude $\mathbf{a}$. Taking such
anticorrelations into account, we can write general expression for the noise,
satisfying gauge transformation~(\ref{calibr_1}):
\begin{equation}
\mathbf{\xi}\left(  t\right)  =\mathbf{\tilde{\xi}}\left(  t\right)
-\zeta\mathbf{a}/\alpha,\qquad\alpha^{2}=\overline{a^{2}}, \label{xi_av}%
\end{equation}
where $\zeta>0$ is the dimensionless correlation factor and random variable
$\mathbf{\tilde{\xi}}\left(  t\right)  $ is not correlated with $\mathbf{a}$,
and has zero average, $\left\langle \mathbf{\tilde{\xi}}\left(  t\right)
\right\rangle =0$. In the case $\zeta=0$ positive and negative fluctuations of
firm grow rate, $\Delta r$, have equal probability, while in the case of
positive $\zeta>0$ firms will in average grow (because of grow of external
resources, see section~\ref{MEAN}).

At economic level anticorrelations between firm tactics and
strategy~(\ref{xi_av}) reflect the fact that firms prefer to have tactical
losses with the hope to get a profit at strategy horizons (say, by pressing
out business rivals). And firms (and countries), aimed at the maximum instant
profit without significant investments in the short time period will
eventually get losses in the long time period.

Repeating our calculations for the model~(\ref{xi_av}), we again find
exponential distribution~(\ref{Pexp})%
\begin{equation}
\mathcal{P}_{0}\left(  x|\sigma\right)  =\frac{1}{\alpha\sqrt{2\left(
1+\zeta^{2}\right)  }}\left\{
\begin{array}
[c]{cc}%
e^{-\sqrt{2}x/\sigma_{+}} & \text{for}\ x>0\\
e^{\sqrt{2}x/\sigma_{-}} & \text{for}\ x<0
\end{array}
\right.  , \label{Pxpm}%
\end{equation}
but with different widths $\sigma_{\pm}$ ($\sigma_{+}<\sigma_{-}$) of positive
and negative PDFs, and the dispersion $\sigma$:%
\begin{equation}
\sigma_{\pm}=\alpha\left(  \sqrt{1+\zeta^{2}}\mp\zeta\right)  ,\ \sigma
^{2}=\left(  1+2\zeta^{2}\right)  \alpha^{2}. \label{spm}%
\end{equation}
The average of this distribution is shifted to negative $\allowbreak\Delta p$,
corresponding to systematic tendency to grow:
\begin{equation}
\allowbreak\overline{\left\langle \Delta p\right\rangle }=-\sqrt{2}\alpha
\zeta,\quad\overline{\left\langle \Delta r\right\rangle }=-\overline
{\left\langle \Delta p\right\rangle }G^{-\beta}>0. \label{Shift}%
\end{equation}

\begin{figure}[tb]
\begin{center}
\includegraphics[
height=2.5391in,
width=3.275in
]{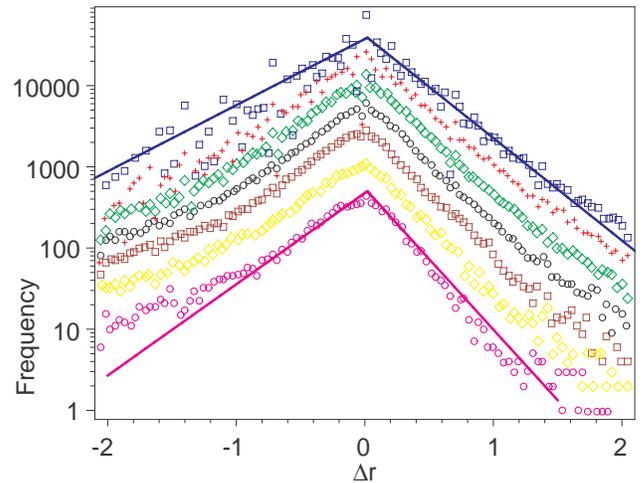}
\end{center}
\caption{The distribution of grow rates of US firms in 1998-2003 for seven
size groups from $G_{up}=8-15$ through $G_{down}=512-1023$\cite{PeAxTe-06}.
Comparing with the theory we use the same correlation factor $\zeta=0.23$ and
varied only one parameter $\sigma$ ($\left\langle \Delta r^{2}\right\rangle
^{1/2}=0.62,0.45$ and $0.4,0.3$ respectively for upper and lower curves.
Deviations from exponential dependence ar large $\left\vert \Delta
r\right\vert $ will be explained in section~\ref{FAT}.}%
\label{Gr1}%
\end{figure}

Such asymmetrical exponential distribution was really observed in the analysis
of empirical data in Ref.\cite{PeAxTe-06} for large averaging intervals ($5$
years, see Fig.~\ref{Gr1}).\ In Fig.~\ref{Gr1} the $x$-axis is in units of
$\Delta r$~(\ref{rG}), and not $\Delta p$. Empirical value $\zeta=0.23$, and
for typical $\left\langle \Delta r^{2}\right\rangle ^{1/2}=0.5$ we reproduce
the observed mean $\allowbreak\overline{\left\langle \Delta r\right\rangle
}=0.16$. 

\subsubsection{Fat tails\label{FAT}}

One of the most prominent features of PDF, the fat tail, is usually attributed
to large volatility fluctuations (in different stochastic volatility and
multifractal models). In this section we show, that the tail originates from
large jumps of the noise, and not of the volatility. This new mechanism
predicts universal tail exponent $\mu=3$ for stock jumps, independent on the
coarse graining time interval $\tau$.

Fluctuations $\Delta p\left(  t\right)  $ of the Double Gaussian model are
characterized by random variable $\mathbf{\xi}$, which is Gaussian at the time
interval $\tau$ and normalized by the condition $\left\langle \mathbf{\xi}%
^{2}\right\rangle =1$~(\ref{xx12}). The problem is that even if we normalize
Gaussian variable for given time interval $\tau$, this normalization will be
broken at next time intervals because of the intermittency effect: relatively
rare, but large picks of fluctuations. The only way to normalize $\mathbf{\xi
}\left(  t\right)  $ for all times is to divide it
\begin{equation}
\mathbf{\xi}\left(  t\right)  =\mathbf{\xi}_{0}\left(  t\right)  /\sigma
_{0}\left(  t\right)  . \label{xi_1}%
\end{equation}
by the mean squared average
\begin{equation}
\sigma_{0}^{2}\left(  t\right)  =\sum\nolimits_{k}w_{k}\xi_{0}^{2}\left(
t-k\tau\right)  . \label{so2}%
\end{equation}
$\sigma_{0}\left(  t\right)  $ slowly varies at time interval $\tau$, and
therefore, random variable $\mathbf{\xi}\left(  t\right)  $ leaves Gaussian at
time scale $\tau$. The division of $\mathbf{\xi}_{0}\left(  t\right)  $ by
$\sigma_{0}\left(  t\right)  $ removes from general Gaussian process
$\mathbf{\xi}_{0}\left(  t\right)  $ the long-time (at the time scale $\tau$)
trend (long-time variations of the amplitude), leaving only high frequency components.

Standard definition of the mean square $\sigma_{0}^{2}$ assumes that weights
$w_{k}$ in Eq.~(\ref{so2}) do not depend on $k$, and we reproduce our previous
result~(\ref{Pexp}) for PDF. But this definition must be corrected, since
there are no any fundamental value of dispersion $\sigma_{0}^{2}$, which can
only be estimated from the knowledge of past values of $\xi_{0}^{2}$. As the
first step, we have to put $w_{k}=0$ at $k\leqslant0$ and get%
\begin{equation}
\sigma_{0}^{2}\left(  t\right)  =w_{1}\xi_{0}^{2}\left(  t-\tau\right)
+\sum\nolimits_{k>1}w_{k}\xi_{0}^{2}\left(  t-k\tau\right)  . \label{s2}%
\end{equation}
In the case of totally uncorrelated events $\sigma_{0}^{2}$ is determined only
by the ``reference''\ value of $\xi_{0}^{2}\left(  t-\tau\right)  $ at
previous time interval, and all $w_{k}\longrightarrow0$ at $k>1$. Terms with
$k>1$ describe the effect of correlations of events, leading to variations
$\Delta p\left(  t\right)  $.

Second, hot variable $\mathbf{\xi}\left(  t\right)  $ can vary only on the
time scale small with respect to $\tau$. Therefore, all $w_{k}\rightarrow0$ at
$k>2$, and random variable $\mathbf{\xi}\left(  t\right)  $ has Markovian
statistics with correlations only between neighbour time intervals $\tau$.
Otherwise it will depend on many time intervals time $k\tau$ in the past,
which is prohibited by definition of hot variable $\mathbf{\xi}\left(
t\right)  $.

And the last: the only information known in future about past fluctuations, is
the very increment $\Delta p$, which depends only on one component $\xi
_{0}^{\prime}=\left(  \mathbf{\xi}_{0},\mathbf{a}\right)  /a$ of $\mathbf{\xi
}_{0}$ along the vector $\mathbf{a}$. The information about corresponding
``perpendicular''\ component $\xi^{\prime\prime}$ do not enter to the
increment, and is lost. Therefore, we should drop the contribution of $\left(
\xi_{0}^{\prime\prime}\right)  ^{2}$ from correlation terms with $k>1$ in
Eq.~(\ref{s2}): $\xi_{0}^{2}=\left(  \xi_{0}^{\prime}\right)  ^{2}+\left(
\xi_{0}^{\prime\prime}\right)  ^{2}\longrightarrow\left(  \xi_{0}^{\prime
}\right)  ^{2}$. After all these corrections we left with expression for the
mean square in Eq.~(\ref{xi_1}):%
\begin{equation}
\sigma_{0}^{2}\left(  t\right)  =w_{1}\xi_{0}^{2}\left(  t-\tau\right)
+w_{2}\left[  \xi_{0}^{\prime}\left(  t-2\tau\right)  \right]  ^{2} \label{st}%
\end{equation}

Although we get similar results for any quickly decaying weights $w_{k}$,
calculations are much simplified in the case of equal weights $w_{1}%
=w_{2}=1/2$ and all $w_{k}=0$ at $k>1$. In order to calculate PDF of the
noise~$\mathbf{\xi}\left(  t\right)  $~(\ref{xi_1}), we rewrite it in the form%
\begin{align}
\mathcal{Q}\left(  \xi\right)   &  =\int_{0}^{\infty}d\sigma_{0}\pi\left(
\sigma_{0}\right)  \left\langle \delta\left[  \mathbf{\xi}-\mathbf{\xi}%
_{0}/\sigma_{0}\right]  \right\rangle \nonumber\\
&  =\int_{0}^{\infty}d\sigma_{0}\pi\left(  \sigma_{0}\right)  \dfrac
{\sigma_{0}^{2}}{\pi}e^{-\xi^{2}\sigma_{0}^{2}}, \label{Qq}%
\end{align}
where we take the average over Gaussian variable $\mathbf{\xi}_{0}$. The
probability distribution of the random variable $\sigma_{0}$~(\ref{st}) is
$\pi\left(  \sigma\right)  =2\sigma\left\langle \delta\left(  \sigma
^{2}-\sigma_{0}^{2}\right)  \right\rangle $. Using exponential representation
of this $\delta$-function, we get%
\[
\pi\left(  \sigma\right)  =\dfrac{\sigma}{\pi}\int\dfrac{dse^{is\sigma^{2}}%
}{\left(  1+is/2\right)  ^{3/2}}=\sqrt{\dfrac{2}{\pi}}\sigma^{2}%
e^{-\tfrac{\sigma^{2}}{2}}.
\]
Substituting this expression into Eq.~(\ref{Qq}), we come to Student noise
distribution:%
\begin{equation}
\mathcal{Q}\left(  \xi\right)  =\frac{3}{\pi}\left(  1+2\xi^{2}\right)
^{-5/2}. \label{Qstir}%
\end{equation}

Using this distribution function, we finally get%
\begin{equation}
\mathcal{P}\left(  x|\sigma\right)  =\dfrac{6}{\sqrt{\pi}\sigma}%
e^{\frac{x^{2}}{2\sigma^{2}}}D_{-4}\left(  \sqrt{2}\frac{x}{\sigma}\right)  ,
\label{PStir_1}%
\end{equation}
where $D$ is the parabolic cylinder function. The central part of this
distribution has exponential shape~(\ref{Pexp}), while its tale has power
dependence:%
\begin{equation}
\mathcal{P}\left(  x\right)  \sim\left\vert x\right\vert ^{-1-\mu}%
,\qquad\left\vert x\right\vert \gg\sigma. \label{pow}%
\end{equation}
with the tail exponent $\mu=3$, well outside the stable L\'{e}vy range
($\mu<2$). One can show, that this exponent does not depend on relation
between weights $w_{1}$ and $w_{2}\sim1$ in Eq.~(\ref{st}) for Markovian
noise. But in the absence of noise correlations, $w_{2}\rightarrow0$, we get
the effective exponent $\mu\rightarrow2$.

\begin{figure}[tb]
\begin{center}
\includegraphics[
height=2.4327in,
width=3.3321in
]{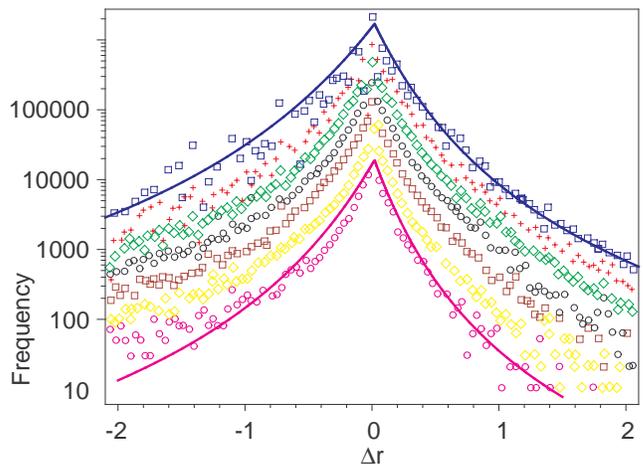}
\end{center}
\caption{The distribution of grow rates of US firms in
1998-1999\cite{PeAxTe-06}, the same parameters as in Fig.~\ref{Gr1}. Tail
exponent $\mu=3$. The varied parameter $\left\langle \Delta r^{2}\right\rangle
^{1/2}=0.45$ and $0.3$ respectively for upper and lower curves.}%
\label{Gr2}%
\end{figure}

If we take into account correlations between the noise and the amplitude (see
Eq.~(\ref{xi_av}) and discussion therein), $\left\langle \mathbf{\xi}%
_{0}\right\rangle =-\zeta\mathbf{a}/\alpha$, and after some calculations we
get simple expression for PDF:%
\begin{align}
\mathcal{P}\left(  x\right)   &  =\int_{0}^{\infty}d\sigma_{0}\pi\left(
\sigma_{0}\right)  \mathcal{P}_{0}\left(  x|\sigma/\sigma_{0}\right)
=\nonumber\\
&  \dfrac{1}{\alpha\sqrt{1+\zeta^{2}}}\left\{
\begin{array}
[c]{cc}%
\sigma_{+}\mathcal{P}\left(  x|\sigma_{+}\right)  & \text{for}\ x>0\\
\sigma_{-}\mathcal{P}\left(  x|\sigma_{-}\right)  & \text{for}\ x<0
\end{array}
\right.  , \label{Ppm}%
\end{align}
where functions $\mathcal{P}_{0}\left(  x|\sigma\right)  $ and $\mathcal{P}%
\left(  x|\sigma\right)  $ are defined in Eqs.~(\ref{Pxpm}) and~(\ref{PStir_1}%
), and $\sigma_{\pm}$ are given in Eq.~(\ref{spm}). We show in Fig.~\ref{Gr2}
that Eq.~(\ref{Ppm}) with $\zeta=0.23$ allows to explain both the asymmetry
and the shape of empirical PDF for different size groups. The size dependence
of both Fig.~\ref{Gr1} and Fig.~\ref{Gr2} follows Eq.~(\ref{rG}) with exponent
$\beta\simeq0.1$ and universal $\Delta p$. 

Now we study the stability of the exponent $\mu$ for different time periods
$\tau$. The total increment $\Delta p=\sqrt{2}\left(  \mathbf{a},\mathbf{\xi
}\right)  $ for two joint intervals $\tau$ is the sum of corresponding
increments $\Delta p_{i}=\sqrt{2}\left(  \mathbf{a}_{i},\mathbf{\xi}%
_{i}\right)  $ for each of these intervals. Since the amplitude $\mathbf{a}$
in Eq.~(\ref{dqt1}) slowly varies on the time scale $\tau$, we take it the
same for both intervals, $\mathbf{a}_{1}=\mathbf{a}_{2}$, and so the noise
$\mathbf{\xi}$ is proportional to the sum of noises $\mathbf{\xi}_{i}$ for
these intervals. Each of $\mathbf{\xi}_{i}$ can be represented in the form of
Eq.~(\ref{xi_1}), and corresponding dispersions $\sigma_{1}$ and $\sigma_{2}$
depend on the same (but shifted over time) series of Gaussian variables
$\mathbf{\xi}_{0}$, Eq.~(\ref{st}). Calculating the distribution function of
the sum $\mathbf{\xi=\xi}_{1}+\mathbf{\xi}_{2}$, we find
\begin{align}
\mathcal{Q}\left(  \xi\right)   &  =\int_{0}^{\infty}d\sigma_{1}d\sigma
_{2}\iint ds_{1}ds_{2}\dfrac{\sigma_{0}^{2}}{\pi}e^{-\xi^{2}\sigma_{0}^{2}%
}\times\nonumber\\
&  \dfrac{\sigma_{1}\sigma_{2}}{\pi^{2}}\dfrac{1}{\left(  1+is_{1}/2\right)
\left(  1+is_{2}/2\right)  }\times\label{Qxi}\\
&  \dfrac{e^{is_{1}\sigma_{1}^{2}+is_{2}\sigma_{2}^{2}}}{\sqrt{1+i\left(
s_{1}+s_{2}\right)  /2}},\nonumber
\end{align}
where%
\begin{equation}
\dfrac{1}{\sigma_{0}^{2}}=\dfrac{1}{\sigma_{1}^{2}}+\dfrac{1}{\sigma_{2}%
^{2}\left(  1+is_{1}/2\right)  }. \label{s01}%
\end{equation}
The tail of the distribution~(\ref{Qxi}) is determined by small $\tilde
{\sigma}$, corresponding to large $\left\vert s_{1}\right\vert \gg\left\vert
s_{2}\right\vert \sim1$. As the result we find, that the distribution
$\mathcal{Q}\left(  \xi\right)  \sim\xi^{-5}$ for the time interval $2\tau$ is
characterized by the same exponent $\mu=3$, as each of $\mathbf{\xi}_{i} $ for
the time interval $\tau$. The only difference is that this asymptotic behavior
can be reached at larger $\xi$, with respect to the distribution function of
$\mathbf{\xi}_{i}$.

This observation explains why the fat tail in Fig.~\ref{Gr1} for five year
period is shifted to higher $\left\vert \Delta r\right\vert $, with respect to
Fig.~\ref{Gr2} for one year period data. Experimental observation of the
stability\ of the exponent $\mu=3$ for widely different economies, as well as
for different time periods\cite{SGPS-02} $\tau$, gives strong experimental
support of our theory. The stability originates in nonlinear correlations of
the noise, see Eq.~(\ref{xi_1}), while linear correlations vanish,
$\left\langle \left(  \mathbf{\xi}_{1},\mathbf{\xi}_{2}\right)  \right\rangle
=0$. To demonstrate the importance of such correlations, assume, that the
noise $\mathbf{\xi}_{i}$ has tail exponent $\mu$, and is uncorrelated at
neighboring intervals $\tau$. Than the exponent of $\mathbf{\xi\sim\xi}%
_{1}+\mathbf{\xi}_{2}$ for the interval $2\tau$ is equal $2\mu$, and not $\mu
$, as follows from our model.

The systematic study of the distribution of annual growth rates by industry
was performed in Ref. \cite{TeAx-SB-05} using Census U.S. data. It is shown,
that all sectors but finance can be fitted by exponential
distribution~(\ref{Pexp}). We checked the data for finance sector, and show
that they can be well fitted by Eq.~(\ref{PStir_1}) with exponent $\mu=3$.

\subsection{ Main results\label{NERG}}

In this section we considered evolution of the market as the result of
competition of different firms for external resources, by analogy with
coalescent regime in physics of supersaturated solutions. This analogy allows
to find informational entropy of the market, and prove the principle of
maximum entropy.

We demonstrate that in coalescent regime for Gibrat mechanism of firm growing
the distribution of firms over their sizes follows the Pareto power low with
the exponent $\gamma=1$ (Zipf distribution). Taking into account size effects,
it turns to stretched exponent distribution, which also describes different
processes, related to competition of units for common resources. Coalescent
mechanism is also responsible for observable exponential distribution of the
income between individuals. The production of real firms can be taken into
account by vector models, by analogy with multicomponent solutions.

We propose the theory of market fluctuations, based on separation of all
degrees of freedom of the market into cold and hot ones. For Gaussian
statistics of all degrees of freedom such separation leads to experimentally
observable exponential PDF of firm grow rates. We also prove, that this
distribution has power tail with universal stable exponent $\mu=3$.

We find analytical expression for PDF, and show, that it reproduces observable
shape and asymmetry of the distribution of firm grow rates, which is related
to existing anticorrelations between tactics of firms at short-time horizon
and their strategy at long-time horizon. In next section we apply this
approach to study price fluctuations on financial markets.

\section{Financial market\label{FINANCE}}

Dynamics of fluctuations is determined by the spectrum of relaxation times of
the system. When all times are small with respect to the observation time
interval $\tau$, the state of the market at time $t+\tau$ depends only on its
state at previous time $t$, and dynamics is Markovian random process.
Short-range correlations of price fluctuations on the market can be studied
using stochastic volatility models\cite{ARCH}, but in order to describe real
markets with multi-time dynamics, the model should take infinite-range
correlations into account\cite{BoBo-05}, and has ``infinite''\ number of
correction terms. In addition, to take empirically observable excess of
volatility into account, one has to go at the boundary of stability of such models.

The real market has enormous number of (quasi-) equilibrium states and
extremely wide spectrum of relaxation times, by analogy with
turbulence\cite{Turbul} and glasses. Multifractal properties of time series
can be described by phenomenological Multifractal Random Walk model\cite{MRW}.
Although this model well characterizes scaling behavior of price fluctuations,
it can not capture correlations at neighboring time intervals, which determine
``conditional dynamics of the market''\ and can be described by the bivariate
probability distribution of price increments\cite{MaHu-04}.

In previous section \ref{FLUCT} we show, that the increment of the random
value $P\left(  t\right)  $ of the time series
\begin{equation}
\Delta_{\tau}P\left(  t\right)  \equiv P\left(  t+\tau\right)  -P\left(
t\right)  \label{xi}%
\end{equation}
has the form of scalar product of two-component random vectors -- the noise
$\mathbf{\xi}\left(  t\right)  $ and its amplitude $\mathbf{a}\left(
t\right)  $:
\begin{equation}
\Delta_{\tau}P\left(  t\right)  =\sqrt{2}\left(  \mathbf{a}\left(  t\right)
,\mathbf{\xi}\left(  t\right)  \right)  . \label{dP}%
\end{equation}
Hot variables $\mathbf{\xi}\left(  t\right)  $ vary at the scale small with
respect to $\tau$, while characteristic times of cold variables $\mathbf{a}%
\left(  t\right)  $ are large with respect to $\tau$. The time $\tau$ plays
the role of the effective temperature: at minimal trade-by-trade time,
$\tau\simeq\tau_{k}$, the price is almost frozen, while in the opposite limit
$\tau>\tau_{0}$ it has random walk statistics. In the intermediate time
interval $\tau_{k}<\tau<\tau_{0}$ (of many decades) the market has
``restricted''\ ergodicity: only hot degrees of freedom are exited, while cold
degrees of freedom are frozen and determine the amplitude $\mathbf{a}$ of
price fluctuations.

Here we apply this approach to calculate PDF of price increments, as well as
various conditional distributions and their moments. The dependence of
parameters of these distributions on observation time $\tau$ will be studied
later, in section~\ref{RG}. In section~\ref{HOT_COLD} we introduce hot and
cold degrees of freedom of the market. Two simplified models are formulated
and solved in sections~\ref{LONG} and~\ref{SHORT}. ``Markovian''\ model takes
short-time correlations into account and neglects the effect of long-time
challenges. ``Effective market''\ model captures such effects, but neglects
any short-time correlations because of trader activity. Although both these
models capture essential part of observable phenomenons of price fluctuations
(extremely small linear correlations -- the Bachelier's first law,
``dependence-induced volatility smile'', ``compass rose''\ pattern\cite{CL-96}
and so on), they can not describe all the variety of such ``stylized
facts''\cite{Styl}.

In section~\ref{ASYMM} we introduce Double Gaussian model, that takes all
correlation effects into account, and show that it allows to explain the
behavior of different types of stocks\cite{Le3-06}. Analytical solution of
this model is derived in Appendix~\ref{PDF}. We demonstrate, that this
solution reproduces all observable types of ``market mill''\ patterns and
gives the mysterious $z$-shaped response of the market for all kinds of
asymmetry of bivariate PDF, as well as other fine characteristics of this
distribution. We also show that our theory allows to explain empirically
observable Markovian ``double dynamics''\ of signs of returns on the
market\cite{BoMa-06}.

\subsection{Cold and hot degrees of freedom\label{HOT_COLD}}

The idea of hot and cold degrees of freedom of the market is qualitatively
supported by empirical observations: It is shown in Ref. \cite{LM-01}, that
the amplitude of fluctuations for ensemble (quenched) averaging significantly
exceeds the amplitude of fluctuations for time (annealed) averaging. This
observation can be interpreted as the result of the presence of cold degrees
of freedom, which remain ``frozen''\ when considering time fluctuations of hot
degrees of freedom. In the case of ensemble averaging such cold degrees of
freedom become ``unfrozen'', increasing the amplitude of price fluctuations
with respect to its time average value.

Following Ref. \cite{Le2-06} consider two consecutive price increments, $x$
(push) and $y$ (response) for the time intervals $\tau$:%
\[
x=\Delta_{\tau}P\left(  t\right)  ,\qquad y=\Delta_{\tau}P\left(
t+\tau\right)  .
\]
According to Eq.~\ref{dP} price increments can be written in the form of the
scalar products:%
\begin{equation}
x=\sqrt{2}\left(  \mathbf{a}_{1},\mathbf{\xi}_{1}\right)  ,\qquad y=\sqrt
{2}\left(  \mathbf{a}_{2},\mathbf{\xi}_{2}\right)  , \label{dp}%
\end{equation}
of complex noises $\mathbf{\xi}_{1}=\mathbf{\xi}\left(  t\right)
,\mathbf{\xi}_{2}=\mathbf{\xi}\left(  t+\tau\right)  $ and complex amplitudes
$\mathbf{a}_{1}=\mathbf{a}\left(  t\right)  ,\mathbf{a}_{2}=\mathbf{a}\left(
t+\tau\right)  $. Complex random walk $\mathbf{\xi}\left(  t\right)  $ in the
``tactic''\ space describes ``impatient''\ agents. Complex random walk
$\mathbf{a}\left(  t\right)  $ in the ``strategy''\ space can be thought of as
a result of slow variation of composition of the population of such agents on
the market, as well as the activity of ``patient''\ agents.

Moduli of complex variables $\mathbf{\xi}_{i}$ and $\mathbf{a}_{i}$ are
normalized as:
\begin{equation}
\left\langle \xi_{i}^{2}\right\rangle =1,\qquad\overline{a_{i}^{2}}=\sigma
^{2}, \label{n_xi0}%
\end{equation}
$\sigma$ is the dispersion of price fluctuations%
\begin{equation}
\overline{\left\langle \Delta_{\tau}P^{2}\left(  t\right)  \right\rangle
}=\overline{\left\langle \Delta_{\tau}P^{2}\left(  t+\tau\right)
\right\rangle }=\sigma^{2}. \label{ortho}%
\end{equation}
Eqs.~(\ref{dp}) are invariant with respect to ``gauge''\ transformation of
noise and amplitude variables, Eq.~(\ref{calibr_1}).

We will characterize correlations of price increments by uni- and bivariate
PDFs:%
\begin{align}
\mathcal{P}\left(  x\right)   &  \equiv\overline{\left\langle \delta\left[
x-\Delta_{\tau}P\left(  t\right)  \right]  \right\rangle }=\int dy\mathcal{P}%
\left(  x,y\right)  ,\label{Px}\\
\mathcal{P}\left(  x,y\right)   &  \equiv\overline{\left\langle \delta\left[
x-\Delta_{\tau}P\left(  t\right)  \right]  \delta\left[  y-\Delta_{\tau
}P\left(  t+\tau\right)  \right]  \right\rangle }. \label{Pxy}%
\end{align}
Using exponential representation of $\delta$-function, these expressions can
be rewritten in the form%
\begin{align}
\mathcal{P}\left(  x\right)   &  =\int_{-\infty}^{\infty}\frac{dk}{2\pi
}e^{-ikx}G\left(  k,0\right)  ,\label{Pint0}\\
\mathcal{P}\left(  x,y\right)   &  =\int_{-\infty}^{\infty}\frac{dk}{2\pi}%
\int_{-\infty}^{\infty}\frac{dp}{2\pi}e^{-ikx-ipy}G\left(  k,p\right)  ,
\label{Pint}%
\end{align}
where $G\left(  k,p\right)  $ is the Fourier component of PDF%
\begin{equation}
G\left(  k,p\right)  \equiv\overline{\left\langle e^{ik\Delta_{\tau}P\left(
t\right)  +ip\Delta_{\tau}P\left(  t+\tau\right)  }\right\rangle }.
\label{Gkqd}%
\end{equation}

The variable $y$ may be interpreted as the response on initial push $x$, which
is characterized by conditional PDF
\begin{equation}
\mathcal{P}\left(  y|x\right)  =\dfrac{\mathcal{P}\left(  x,y\right)
}{\mathcal{P}\left(  x\right)  },\ \mathcal{P}\left(  x\right)  \equiv
\int\frac{dk}{2\pi}e^{-ikx}G\left(  k,0\right)  , \label{Pyx}%
\end{equation}
The average conditional response is%
\begin{align}
\left\langle y\right\rangle _{x}  &  =\int_{-\infty}^{\infty}dyy\mathcal{P}%
\left(  y|x\right) \label{yav}\\
&  =\frac{i}{\mathcal{P}\left(  x\right)  }\int\frac{dk}{2\pi}e^{-ikx}\left.
\frac{\partial G\left(  k,p\right)  }{\partial p}\right\vert _{p=0}.\nonumber
\end{align}
The width of the conditional PDF $\mathcal{P}\left(  y|x\right)  $ is
characterized by the conditional mean-square deviation%
\begin{align}
\sigma_{x}^{2}  &  \equiv\int dy\left(  y-\left\langle y\right\rangle
_{x}\right)  ^{2}\mathcal{P}\left(  y|x\right) \label{scond}\\
&  =-\frac{1}{\mathcal{P}\left(  x\right)  }\int\frac{dk}{2\pi}e^{-ikx}\left.
\frac{\partial^{2}G\left(  k,p\right)  }{\partial p^{2}}\right\vert
_{p=0}.\nonumber
\end{align}
Large $\sigma_{x}$ correspond to a large variety of the behaviors, the
``volatility''. The dependence of $\sigma_{x}$ on $x$ reflects the volatility
clustering: $\sigma_{x}$ should not depend of $x$ if there is no volatility clustering.

The conditional response~(\ref{yav}) and PDF~(\ref{Pxy}) depend of
correlations between noises $\mathbf{\xi}_{i}$ and their amplitudes
$\mathbf{a}_{i}$ in two time intervals. Before formulating general model (see
section~\ref{ASYMM}), that takes all such correlations into account, it would
be instructive to study some simple limits.

\subsection{Markovian model\label{LONG}}

\begin{figure}[tb]
\begin{center}
\includegraphics[
height=2.7294in,
width=2.7294in
]{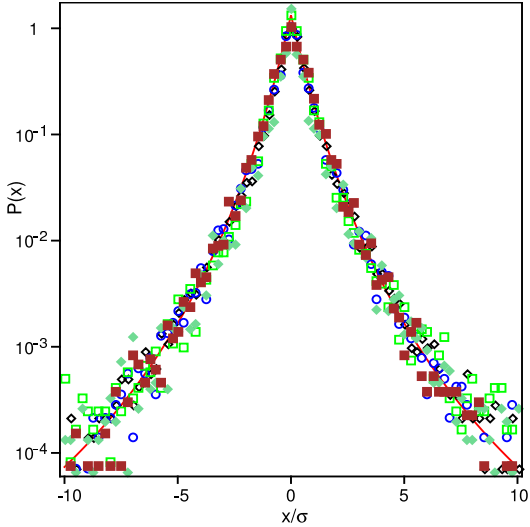}
\end{center}
\caption{PDF of Russian financial market (finam.ru, 2006) for $\tau=5\min$
($\diamondsuit$- EESR, $\bigcirc$ - LKOH, $\square$ - RTKM, $\blacklozenge$ -
SBER, $\blacksquare$ - SNGS), solid line shows theoretical
prediction~(\ref{PStir_1}) with $\mu=3$.}%
\label{Distrib}%
\end{figure}

We first consider the case when the amplitude $\mathbf{a}\left(  t\right)  $
is not correlated with external challenges at strategy horizons, and
$\mathbf{a}_{1}=\mathbf{a}_{2}$ for two neighboring time intervals. We also
assume that the noise is not correlated with the amplitude, but take into
account short range correlations of the noise, $\left\langle \left(
\mathbf{\xi}_{1},\mathbf{\xi}_{2}\right)  \right\rangle =\varepsilon$. For
this Markovian model we find Eq.~\ref{PStir_1} for the probability
distribution, which describes very well Russian financial market for
$\tau=5\min$, see Fig.~\ref{Distrib}. For Gaussian noise we find exponential PDF~(\ref{Pexp}) of price
fluctuations, which is really observed for high frequency
fluctuations\cite{SPY-04}.

Averaging the Fourier component of PDF~(\ref{Gkqd}) over fluctuations of
Gaussian amplitude $\mathbf{a}_{1}=\mathbf{a}_{2}$ and noise $\mathbf{\xi}%
_{i}$ we find $G\left(  k,p\right)  =\left[  1+\sigma^{2}\left(  k^{2}%
/2+p^{2}/2+\varepsilon kp\right)  \right]  ^{-1}$. Calculating the Fourier
transformation of this function~(\ref{Pint}), we get the distribution function%
\begin{equation}
\mathcal{P}_{t}\left(  x,y\right)  =\dfrac{1}{\pi\sigma^{2}\sqrt
{1-\varepsilon^{2}}}K_{0}\left[  \sqrt{\tfrac{2\left(  x^{2}\allowbreak
+\allowbreak y^{2}-2\varepsilon\allowbreak xy\right)  }{\sigma^{2}\left(
1-\varepsilon^{2}\right)  }}\right]  , \label{P0xy}%
\end{equation}
where $K_{0}$ is the Bessel function. Calculating the integral~(\ref{yav})
with function~(\ref{P0xy}), we find the conditional response
\begin{equation}
\left\langle y\right\rangle _{x}=\varepsilon x. \label{ACOR}%
\end{equation}
Linear dependence~(\ref{ACOR}) with $\varepsilon<0$ well agrees with data for
Russian market, what can be interpreted as indication that Russian investors
are oriented only on current benefits, mostly ignoring opening possibilities
at strategy horizons. Although linear response~(\ref{ACOR}) is typical for
ACOR group of stocks with $\varepsilon<0$ (according to classification of Ref.
\cite{Le3-06}), this model can not describe essentially nonlinear response of
other groups of stocks.

\subsection{Effective market model\label{SHORT}}

In general, the amplitude $\mathbf{a}$ is varied in response to unpredictable
external challenges. We first study this effect in the model of
``Effective\ market'', neglecting correlations between noise $\mathbf{\xi}%
_{i}^{0}$ in two consecutive time intervals $\tau$, but taking into account
random variations of its amplitude $\mathbf{a}_{i}^{0}$:$\qquad$%
\begin{equation}
\left\langle \left(  \mathbf{\xi}_{i}^{0},\mathbf{\xi}_{j}^{0}\right)
\right\rangle =\delta_{ij},\quad\overline{\left(  \mathbf{a}_{1}%
^{0},\mathbf{a}_{2}^{0}\right)  }=\nu\overline{(\mathbf{a}_{1}^{0})^{2}}%
=\nu\overline{(\mathbf{a}_{2}^{0})^{2}}, \label{12}%
\end{equation}
where $\nu$ is dimensionless correlation parameter, $0<\nu<1$. As in Markovian
model we ignore (anti)correlations between noise and amplitude. Correlations
of the noise $\mathbf{\xi}_{i}^{0}$ are induced by trader activity, while the
change of a stock price in the model of Effective market is determined only by
an external information, which may be considered as uncorrelated random process.

PDF $\mathcal{P}\left(  x\right)  $ of this model is proportional to the
parabolic cylinder function with power tail exponent $\mu=2$ (see
Eq.~(\ref{pow})). Non Gaussian character of the noise PDF can be ignored when
considering the central parts of price distributions, when $\mathcal{P}\left(
x\right)  $ takes exponential form~(\ref{Pexp}). In order to calculate
bivariate PDF, we substitute equation~(\ref{dp}) in~(\ref{Gkqd}) and perform
the averaging over fluctuations of Gaussian variables $\mathbf{a}_{i}^{0}$:%
\begin{equation}%
\begin{array}
[c]{c}%
G\left(  k,p\right)  =\left\langle \exp\left[  -\sigma^{2}k^{2}(\xi_{1}%
^{0})^{2}/2-\right.  \right. \\
\left.  \left.  \sigma^{2}p^{2}(\xi_{2}^{0})^{2}/2-\nu\sigma^{2}kp\left(
\mathbf{\xi}_{1}^{0},\mathbf{\xi}_{2}^{0}\right)  \right]  \right\rangle .
\end{array}
\label{Gkq1}%
\end{equation}
The averaging over noise $\mathbf{\xi}_{i}^{0}$ is performed with Gaussian
PDF~(\ref{Gauss}), and the integral over $k$ and $p$ in expression~(\ref{Pint}%
) is calculated expanding the function $G\left(  k,p\right)  $ in powers of
$\nu$:%
\begin{equation}
\mathcal{P}_{0}\left(  x,y\right)  =\sum\nolimits_{l=0}^{\infty}\nu
^{2l}\mathcal{P}_{l}\left(  x\right)  \mathcal{P}_{l}\left(  y\right)  .
\label{Ppp}%
\end{equation}
Here $\mathcal{P}_{0}\left(  x\right)  =\mathcal{P}\left(  x\right)  $ is
given by Eq.~(\ref{Pexp}), and functions $\mathcal{P}_{l}\left(  x\right)  $
are defined by:%
\begin{equation}
\mathcal{P}_{l}\left(  x\right)  =\frac{1}{l!}\left.  \frac{d^{l}}{dz^{l}%
}\left[  \frac{1}{\sqrt{z}}\mathcal{P}\left(  \frac{x}{\sqrt{z}}\right)
\right]  \right\vert _{z=1}. \label{Pl}%
\end{equation}

PDF $\mathcal{P}_{0}\left(  x,y\right)  $ is symmetrical with respect to
independent transformations of its variables, $x\rightarrow-x,y\rightarrow-y$,
and also with respect to time reversal transformation, which corresponds to
push-response interchange, $x\longleftrightarrow y$. This function is not
analytical in origin, and the geometry of equiprobability levels can be
approximated by $\left\vert x\right\vert ^{\lambda}+\left\vert y\right\vert
^{\lambda}=const$, where $\lambda\simeq1$ near origin and $\lambda\simeq2$ far
away from it. \begin{figure}[tb]
\begin{center}
\includegraphics[
height=2.1369in,
width=2.9741in
]{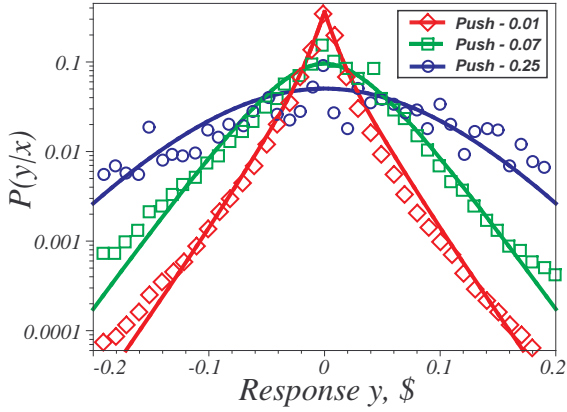}
\end{center}
\caption{Profiles of conditional PDF $\mathcal{P}\left(  y|x\right)  $ at
different $x$ and $\nu=0.95$ ($\sigma=\$0.04$) in comparison with empirically
observable profiles\cite{Le1-06}}%
\label{Dist}%
\end{figure}

\begin{figure}[tb]
\begin{center}
\includegraphics[
height=2.019in,
width=2.643in
]{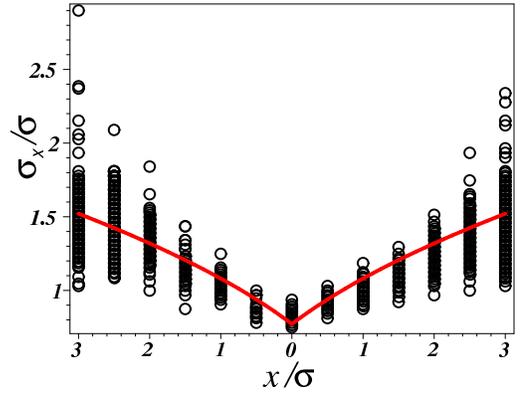}
\end{center}
\caption{Conditional mean-squared deviation as function of $x=\Delta p$; the
result of Gaussian model with $\nu=0.95$ and empirically observable
D-smile\cite{LeTr-06}.}%
\label{Condition}%
\end{figure}

Profiles of conditional distribution~(\ref{Pyx}) are shown for different $x$
in Fig.~\ref{Dist}. With the rise of the push $x$ the response becomes more
flat in origin, in good agreement with empirical data. Slight deviations
between the theory and data at large $\left\vert y\right\vert \gg\sigma$ are
related to non-Gaussian character of the noise (leading to power tail), see
Eq.~\ref{PStir_1} for more details. Calculating integral~(\ref{scond}) in the
case of Gaussian noise, we get the conditional mean-square deviation,%
\begin{equation}
\sigma_{x}^{2}=\sigma^{2}\left[  1+\tfrac{1}{2}\nu^{2}\left(  \sqrt
{2}\left\vert x\right\vert /\sigma-1\right)  \right]  . \label{Scond}%
\end{equation}
This function is plotted in Fig.~\ref{Condition}. It demonstrates the so
called ``dependence-induced volatility smile''\ (``D''-- smile), well known
from empirical data\cite{LeTr-06}. At small $\left\vert x\right\vert
\lesssim\sigma$ the standard deviation of the response~(\ref{Scond}) is
smaller than the unconditional standard deviation $\sigma$, while at large
$\left\vert x\right\vert \gtrsim\sigma$ it is larger. 

The shape of conditional PDF can also be characterizes by the kurtosis,
proportional to fourth momentum. One can show that, in agreement with
empirical data, the kurtosis of theoretical conditional PDF $\mathcal{P}%
_{0}\left(  y|x\right)  $ decreases with the rise of $\left\vert x\right\vert
$. We conclude, that Effective market model captures main features of the
market behavior, but it is enable to describe finite response of real stocks.

\subsection{Double Gaussian model\label{ASYMM}}

In this section we generalize Effective market model to take into account
short-range correlations at strategy horizon because of activity of traders.
Such correlations lead to the exponent $\mu=3$ of the power tail of PDF
(section~\ref{FAT}), and relatively weakly affects the central part of PDF:
for time interval $\tau$ about several minutes it is estimated as about
$5\%$\cite{Le2-06}. We assume normal distribution of noise fluctuations
$\mathbf{\xi}_{i}$ for the central part of PDF, and neglect the effect of
noise-amplitude anticorrelations, which is small at short $\tau$, and leads to
gain/loss asymmetry (see section~\ref{DOUBLE}).

For given noise variables $\mathbf{\xi}_{i}$ we introduce random variables
$\mathbf{\xi}_{i}^{0}$ of Effective market model, which form orthogonal basis
in the space of random functions $\mathbf{\xi}_{i}$, see Eq.~(\ref{12}).
Expanding price fluctuations $\Delta P\left(  t\right)  $ and $\Delta P\left(
t+\tau\right)  $ over this basis, we get:
\begin{align}
\Delta_{\tau}P\left(  t\right)   &  =\sqrt{2}\left(  \mathbf{a}_{1}%
^{0},\mathbf{\xi}_{1}^{0}\right)  +\sqrt{2}\left(  \mathbf{\varepsilon}%
_{1},\mathbf{\xi}_{2}^{0}\right)  ,\label{d1}\\
\Delta_{\tau}P\left(  t+\tau\right)   &  =\sqrt{2}\left(  \mathbf{a}_{2}%
^{0},\mathbf{\xi}_{2}^{0}\right)  +\sqrt{2}\left(  \mathbf{\varepsilon}%
_{2},\mathbf{\xi}_{1}^{0}\right)  . \label{d2}%
\end{align}
We consider amplitudes $\mathbf{a}_{i}^{0}$ of Effective market as Gaussian
random variables, Eq.~(\ref{12}). Non-diagonal amplitudes $\mathbf{\varepsilon
}_{i}$ describe the shift of equilibrium on the market because of trader
activity. Since there are only two independent amplitudes, $\mathbf{a}_{1}$
and $\mathbf{a}_{2}$, for two time intervals, the amplitudes
$\mathbf{\varepsilon}_{1}$ and $\mathbf{\varepsilon}_{2}$, can be expanded
over two diagonal amplitudes $\mathbf{a}_{1}^{0}$ and $\mathbf{a}_{2}^{0}$:%
\begin{equation}
\mathbf{\varepsilon}_{i}\mathbf{=}\sum\nolimits_{j}\tilde{c}_{ij}%
\mathbf{a}_{j}^{0}. \label{b}%
\end{equation}
In the case $\mathbf{\varepsilon}_{1}=0$ or $\mathbf{\varepsilon}_{2}=0$ this
Double Gaussian model is reduced to Markovian model (section~\ref{LONG}), and
in the case $\mathbf{\varepsilon}_{1}=\mathbf{\varepsilon}_{2}=0$ -- to the
model of Effective market (section~\ref{SHORT}).

PDF of this model is calculated in Appendix~\ref{PDF}:%
\begin{equation}
\mathcal{P}\left(  x,y\right)  =\mathcal{P}_{0}\left(  x\cos\phi_{+}-y\sin
\phi_{-},y\cos\phi_{-}+x\sin\phi_{+}\right)  , \label{Pppp}%
\end{equation}
where $\mathcal{P}_{0}$ is PDF of Effective market model, Eq.~(\ref{Ppp}). The
distribution~(\ref{Pppp}) depends on only four independent parameters: the
dispersion $\sigma$, the correlator of the amplitude $\nu$ ($0<\nu<1$), and
two angles $\phi_{-}$ and $\phi_{+}$, depending on starting parameters
$\left\{  \tilde{c}_{ij}\right\}  $ of our model. The correlator $\nu$
describes the ``elasticity''\ of the market to external challenges at the
strategy horizon. The angles $\phi_{-}$ and $\phi_{+}$ control the feedback
between trader expectations and real price changes at the tactic horizon.
Their difference, $\varepsilon=\phi_{+}-\phi_{-}$, is taken as small parameter
of our theory, which controls the correlator of neighboring price increments%
\begin{equation}
\overline{\left\langle \Delta_{\tau}P\left(  t\right)  \Delta_{\tau}P\left(
t+\tau\right)  \right\rangle }=\sigma^{2}\varepsilon. \label{corr}%
\end{equation}
Eq.~(\ref{Pppp}) turns to corresponding expression~(\ref{P0xy}) for Markovian
model in the limit $\nu\rightarrow1$, and reproduces Eq.~(\ref{Ppp}) of
Effective market model for $\phi_{-}=\phi_{+}=0$.

The sum~(\ref{Ppp}) goes only over even $2l$ because of neglect of
noise-amplitude correlations, $\left\langle \mathbf{\eta}_{i}\right\rangle =0
$. In general, there are correlations between noise and amplitude, described
by a factor $\zeta$ (see section~\ref{ASSYM}). Such correlations (studied in
section~\ref{DOUBLE}) break the symmetry of the conditional average
$\left\langle y\right\rangle _{x}$ with respect to positive and negative $x$,
and are responsible for the so called Leverage effect\cite{Leverage}.

In our theory we have an hierarchy of small parameters, $\zeta\ll\left\vert
\varepsilon\right\vert \ll\phi\ll1$. PDF of Double Gaussian model with all
nonzero $\zeta,\phi,\varepsilon\neq0$ has no symmetries at all. In the case
$\zeta=0$ but\textbf{\ }$\phi,\varepsilon\neq0$ there is a symmetry
$\mathcal{P}\left(  x,y\right)  =\mathcal{P}\left(  -x,-y\right)  $,
corresponding to rotations on the angle $\pi$ in the plane $\left(
x,y\right)  $. In the case $\zeta=\varepsilon=0$ but $\phi\neq0$, when there
are no linear correlations of price~(\ref{corr}) at adjacent time intervals,
PDF~(\ref{Ppp}) remains symmetrical only with respect to mixed transformation,
$\mathcal{P}\left(  x,y\right)  =\mathcal{P}\left(  -y,x\right)  $,
corresponding to rotations by the angle $\pi/2$ in the plane $\left(
x,y\right)  $. The change of sign of $y $ in the above equation is related to
reversion of the time: on reversed time scale one can think about losses in
future, $y<0$, as about gains in the ``past''. This approximate push-response
invariance was established first time from the analysis of empirical
data\cite{Le1-06}. And finally, in the case $\zeta=\phi=\varepsilon=0$ the
function $\mathcal{P}\left(  x,y\right)  $ acquires the total symmetry
$x\rightarrow-x,y\rightarrow-y$ and~$x\longleftrightarrow y$ of Effective
Market model.

\subsubsection{Market MILL, ACOR and COR stocks}

\begin{figure}[tb]
\begin{center}
\includegraphics[
height=3.033in,
width=3.163in
]{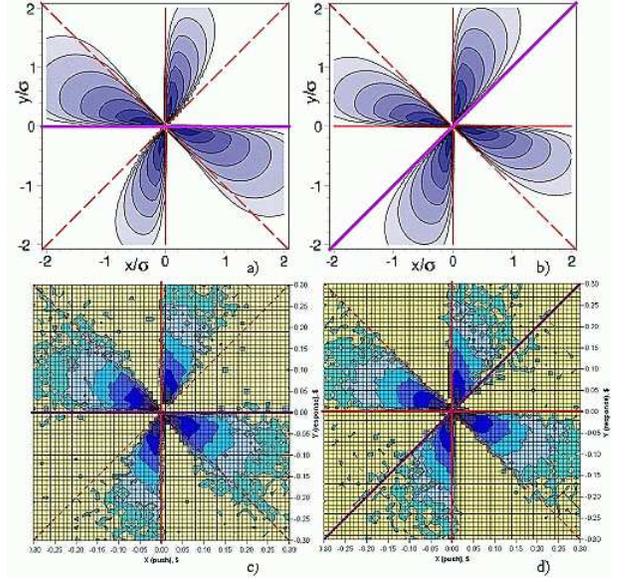}
\end{center}
\caption{Two-dimensional projection of $\log_{4}\mathcal{P}^{a}\left(
x,y\right)  $ with respect to $y=0$ axes a) and $y=x$ axes b) for $\nu=0.95,$
$\phi_{-}=8^{\circ}$ and $\phi_{+}=8.7^{\circ}$. For comparison sake we show
corresponding empirically observed pictures c) and d)\cite{Le3-06}.}%
\label{Mill}%
\end{figure}

It is convenient to describe the symmetry of PDF with respect to the axes
$y=0$ by antisymmetric component, $\mathcal{P}^{a}\left(  x,y\right)  =\left[
\mathcal{P}\left(  x,y\right)  -\mathcal{P}\left(  x,-y\right)  \right]  /2$.
In Fig.~\ref{Mill} a) we plot equiprobability levels of positive part of this
function, $(\mathcal{P}^{a}\left(  x,y\right)  +\left\vert \mathcal{P}%
^{a}\left(  x,y\right)  \right\vert )/2$. 

For small $\left\vert \varepsilon\right\vert \ll\phi$ the plot demonstrates
four--blade mill--like pattern (the ``market mill''\ pattern), that was
observed first time in Ref. \cite{LeTr-06}, see Fig.~\ref{Mill} c). To analyze
these pictures it is convenient to divide the push-response plane $\left(
x,y\right)  $ into sectors numbered counterclockwise from I to VIII. In
agreement with empirical data at $\varepsilon>0$ the blades in II and IV
quadrants of the $\left(  x,y\right)  $ plane are thinner than their
counterparts, which extend out of I and III quadrants. The situation is
reversed at $\varepsilon<0$. With the rise of $\left\vert \varepsilon
\right\vert $ the market mill pattern becomes distorted and only two
corresponding blades of the mill pattern left well expressed.
\begin{figure}[tb]
\begin{center}
\includegraphics[
height=3.033in,
width=3.163in
]{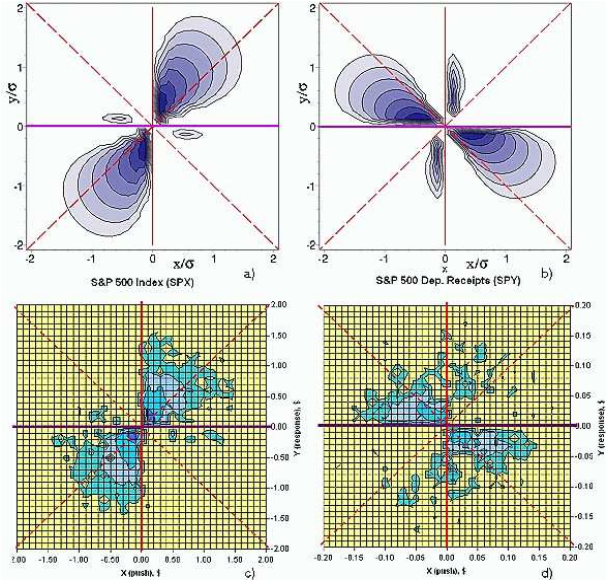}
\end{center}
\caption{Changing of asymmetry with respect to $y=0$ axes with parameter
$\varepsilon=\phi_{+}-\phi_{-}$. We use $\nu=0.95$ and $\phi_{+}=8^{\circ}$
and vary the angle $\phi_{-}=14^{\circ}$ a), $\phi_{-}=6^{\circ}$ b). For
comparison sake we show typical patterns observed for different
stocks\cite{Le3-06} c) and d).}%
\label{Mill_px}%
\end{figure}

In Figs.~\ref{Mill_px} a) and b) we show how the market mill pattern is
deformed with variation of $\varepsilon$. Varying the angle $\phi_{-}$ for
fixed $\sigma,\nu$ and $\phi_{+}$, we get good qualitative agreement with
observable patterns, shown in Figs.~\ref{Mill_px} c) and d). We conclude that
the theory allows to explain all the variety of basic patterns for different
stocks\cite{Le3-06}, and may be considered as the basis for their quantitative
classification: Fig.~\ref{Mill} with $\phi_{-}\simeq\phi_{+}>0$ ($\varepsilon
\simeq0$) corresponds to the mill pattern (MILL), Fig.~\ref{Mill_px} a) with
$\phi_{-}>\phi_{+}$ ($\varepsilon<0$) corresponds to negative autocorrelation
(ACOR), and Fig.~\ref{Mill_px} b) with $\phi_{-}<\phi_{+}$ ($\varepsilon>0$)
corresponds to positive autocorrelation (COR). Anti-mill pattern (AMILL) with
$\phi_{-}\simeq\phi_{+}<0$ was never observed in Ref.\cite{Le3-06}.

Similar patterns are obtained for symmetry properties of the bivariate PDF
$\mathcal{P}\left(  x,y\right)  $ with respect to different axes $y=x,$ $x=0$,
or $y=-x$. As example we show in Fig.~\ref{Mill} b) equiprobability levels of
positive part of the function~$\mathcal{P}^{a}\left(  x,y\right)  =\left[
\mathcal{P}\left(  x,y\right)  -\mathcal{P}\left(  y,x\right)  \right]  /2$.
The blades of this market mill are more symmetric than those in
Fig.~\ref{Mill} a), in agreement with empirical pictures in Figs.~\ref{Mill}
c) and d).

An attempt to explain market mill patterns for the asymmetry with respect to
the axis $y=0$ was made in Ref.\cite{Le-07}, where ``hand-made''\ analytical
ansatz for conditional PDF was proposed. It was explicitly assumed, that the
response $y$ depends only on push $x$ at previous time, and no long-range
correlations were taken into account. We do not think, that such Markovian
model can give adequate description of real market with extremely wide
spectrum of relaxation times.

\subsubsection{Univariate PDF}

\begin{figure}[tb]
\begin{center}
\includegraphics[
height=1.2769in,
width=3.3949in
]{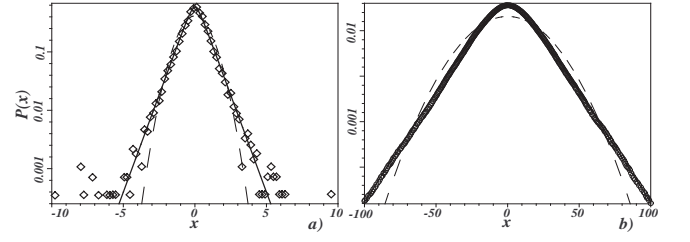}
\end{center}
\caption{Probability distribution function for the S\&P500. Daily data from
31/1/1950 to 18/7/2003\cite{BaTh-04} a) 5 minute increments for
1991-1995\cite{Co-97} b) We show the best fit by Eq.~(\ref{Pxt}) with
$\theta=0.3$ $a)$ and $\theta=0.4$ $b)$, for comparison we show by dotted
lines the best fit by Gaussian PDF.}%
\label{S&P500}%
\end{figure}

Now we calculate one-point PDF, Eq.~(\ref{Pint0}), of Double Gaussian model.
Expression~(\ref{G-1}) of Appendix~\ref{PDF} for the Fourier component
$G\left(  k,0\right)  $ can be represented in the form%
\begin{align}
G\left(  k,0\right)   &  =\left(  1+\sigma^{2}\alpha_{1}^{2}k^{2}/2\right)
^{-1}\left(  1+\sigma^{2}\alpha_{2}^{2}k^{2}/2\right)  ^{-1},\label{Gk0}\\
\alpha_{1}  &  =\cos\theta,\qquad\alpha_{2}=\sin\theta\nonumber
\end{align}
with the angle $\theta$ defined by
\begin{equation}
\sin\left(  2\theta\right)  =\sqrt{1-\nu^{2}}\sin\left(  2\phi\right)  .
\label{theta}%
\end{equation}
Calculating the integral over $k$ in Eq.~(\ref{Pyx}) with this function
$G\left(  k,0\right)  $ we find one-point PDF%
\begin{equation}
\mathcal{P}\left(  x\right)  =\allowbreak\frac{\alpha_{1}e_{1}\left(
x\right)  -\alpha_{2}e_{2}\left(  x\right)  }{\sqrt{2}\sigma\left(  \alpha
_{1}^{2}-\alpha_{2}^{2}\right)  },\quad e_{i}\left(  x\right)  =e^{-\sqrt
{2}\left\vert x\right\vert /\left(  \alpha_{i}\sigma\right)  }. \label{Pxt}%
\end{equation}
As one can see from Fig.~\ref{S&P500} this distribution is in good agreement
with observable PDF of the Standard\&Poor 500 (S\&P500) index, that is one of
the most widely used benchmarks for U.S. equity performance.

\subsubsection{Conditional response}

\begin{figure}[tb]
\begin{center}
\includegraphics[
height=1.701in,
width=2.981in
]{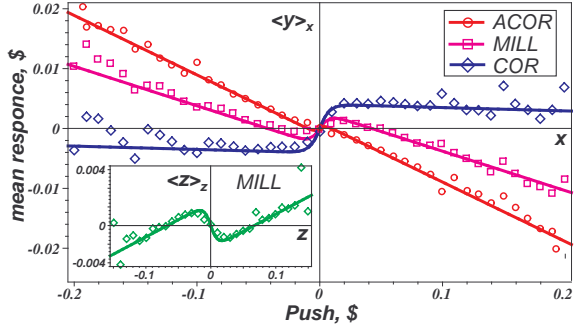}
\end{center}
\caption{The dependence of mean conditional responce on push $x$ for different
angles $\phi_{-}$. We use $\sigma=\$0.04$ and$\ \phi_{+}=8$, $\nu=0.97$ and
$\phi_{-}=12.5^{\circ}$ (ACOR),, $\nu=0.9$ and $\phi_{-}=9^{\circ}$ (MILL),
and $\nu=0.8$ and $\phi_{-}=4.5^{\circ}$ (COR). Fitting parameters were taken
to reproduce empirical dependences presented for some stocks\cite{Le3-06}.
Conditional responce $\left\langle z\right\rangle _{\bar{z}}$ for $\nu=0.95,$
$\phi_{+}=8^{\circ}$ and $\phi_{-}=7.7^{\circ}$ in line with corresponding
empirical data are shown in Insert. }%
\label{Mean2}%
\end{figure}

Calculating the integral~(\ref{yav}), we find the mean conditional response
\begin{align}
\left\langle y\right\rangle _{x}  &  =-sign\left(  x\right)  \sqrt{2}%
\sigma\dfrac{2\varepsilon\alpha_{1}^{2}\alpha_{2}^{2}-A}{\left(  \alpha
_{1}^{2}-\alpha_{2}^{2}\right)  ^{2}}\dfrac{e_{1}\left(  x\right)
-e_{2}\left(  x\right)  }{\alpha_{1}e_{1}\left(  x\right)  -\alpha_{2}%
e_{2}\left(  x\right)  }\allowbreak\nonumber\\
+  &  x\dfrac{\left(  \varepsilon\alpha_{1}-A/\alpha_{1}\right)  e_{1}\left(
x\right)  +\left(  \varepsilon\alpha_{2}-A/\alpha_{2}\right)  e_{2}\left(
x\right)  }{\left(  \alpha_{1}^{2}-\alpha_{2}^{2}\right)  \left[  \alpha
_{1}e_{1}\left(  x\right)  -\alpha_{2}e_{2}\left(  x\right)  \right]  },
\label{yx}%
\end{align}
where $A=\alpha_{1}\alpha_{2}\sqrt{\left(  \alpha_{1}^{2}-\alpha_{2}%
^{2}\right)  ^{2}-\nu^{2}}$, $\alpha_{i}$ and $e_{i}\left(  x\right)  $ are
defined in Eqs.~(\ref{Gk0}) and~(\ref{theta}). In Fig.~\ref{Mean2} a) we show
how the dependence~(\ref{yx}) of mean conditional response on push $x$ depends
on the angle $\phi_{\_}$. This dependence has zigzag structure for MILL group
($\varepsilon\simeq0$), it is almost monotonic for ACOR group ($\varepsilon
<0$), with linear limiting dependence~(\ref{ACOR})), and is essentially
nonlinear for COR group ($\varepsilon>0$). Similar calculations of conditional
mean absolute response $\left\langle \left\vert y\right\vert \right\rangle
_{x}$ (which shows how the response volatility is grow with the amplitude of
the given push $x$) show that in the case $\varepsilon=0$ to a good accuracy
it is linear in the absolute value of the push $\left\vert x\right\vert $,
$\left\langle \left\vert y\right\vert \right\rangle _{x}\simeq c_{0}%
+c_{1}\left\vert x\right\vert $, in good agreement with empirical
data\cite{Le1-06}. 

To analyze the asymmetry of PDF $\mathcal{P}\left(  x,y\right)  $ with respect
to time reversion in the case of MILL group ($\varepsilon=0$), it is
convenient to introduce the total increment of price during the two time
intervals, and also the difference of these increments $z=2^{-1/2}\left(
x+y\right)  $ and $\bar{z}=2^{-1/2}\left(  y-x\right)  $. PDF $\mathcal{P}%
\left(  z,\bar{z}\right)  $ of these random variables takes the form of
Eq.~(\ref{Pppp}) with the substitution $\phi\rightarrow\varphi=\pi/4-\phi$,
describing the rotation of the push-response plane by the angle $\pi/4$.
Therefore, both PDF and all conditional averages are given by above
expressions under the substitution $\left\langle y\right\rangle _{x}%
\rightarrow-\left\langle \bar{z}\right\rangle _{z}$ and $\theta\rightarrow
\theta^{\prime}$, where
\[
\sin\left(  2\theta^{\prime}\right)  =\sqrt{1-\allowbreak\nu^{2}}\sin\left(
2\varphi\right)  =\sqrt{1-\nu^{2}}\cos\left(  2\phi\right)  .
\]
Conditional response $\left\langle \bar{z}\right\rangle _{z}$ also has
$z$-shaped structure, and it is shown in Insert in Fig.~\ref{Mean2} in
comparison with empirical data.

Nonvanishing of average responses $\left\langle y\right\rangle _{x}$ and
$\left\langle \bar{z}\right\rangle _{z}$ (Fig.~\ref{Mean2}) allows one to make
some ``nonlinear''\ predictions~(\ref{yx}) about future price changes on the
market, which can not be obtained from the knowledge of only linear
correlations: the response $y$ in the next time interval is correlated with
initial increment of the price $x$ at small $\left\vert x\right\vert
\lesssim\sigma$, and is anticorrelated with it at large $\left\vert
x\right\vert \gtrsim\sigma$. The order of price increments is also important
for given total increment $\sqrt{2}z$: for small $z<z_{0}\sim\sigma$ the
average initial variation $x$ is larger than next one, $y$, and the situation
is reverted at large $z$.

\subsubsection{Conditional double dynamics\label{DOUBLE}}

\begin{figure}[tb]
\begin{center}
\includegraphics[
height=1.222in,
width=3.124in
]{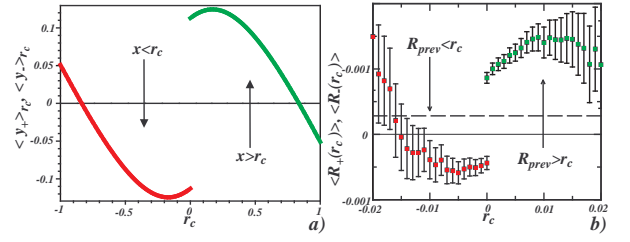}
\end{center}
\caption{Average daily return given that the previous day had a return greater
than $r_{c}$ (right) and given that the previous day had a return smaller than
$r_{c}$ (left). Prediction of Gaussian model at $\theta=0.2$ a) and empirical
data\cite{BoMa-06} b).}%
\label{PM}%
\end{figure}

In this section we discuss the hypothesis of Ref.\cite{BoMa-06}, that the
average return is actually the result of composition of two independent
signals with Markovian statistics: one of them positive, and another one
negative. It is proposed to characterize this effect by average daily returns
$\left\langle y_{-}\right\rangle _{r_{c}}$ and $\left\langle y_{+}%
\right\rangle _{r_{c}}$ given that the previous day had a return greater than
$r_{c}$ and smaller than $r_{c}$:%
\begin{align}
\left\langle y_{-}\right\rangle _{r_{c}}  &  =\left.  \int_{-\infty}^{r_{c}%
}dx\left\langle y\right\rangle _{x}\mathcal{P}\left(  x\right)  \right/
\int_{-\infty}^{r_{c}}dx\mathcal{P}\left(  x\right)  ,\label{yp}\\
\left\langle y_{+}\right\rangle _{r_{c}}  &  =\left.  \int_{r_{c}}^{\infty
}dx\left\langle y\right\rangle _{x}\mathcal{P}\left(  x\right)  \right/
\int_{r_{c}}^{\infty}dx\mathcal{P}\left(  x\right)  . \label{ym}%
\end{align}
Calculating integrals~(\ref{yp}) and~(\ref{ym}) for Double Gaussian model in
MILL case $\varepsilon=0$, we find at $r_{c}>0$:%
\begin{equation}%
\begin{array}
[c]{c}%
\left\langle y_{-}\right\rangle _{r_{c}}=-\left\langle y_{+}\right\rangle
_{r_{c}}=\dfrac{\sigma}{\sqrt{2}\left(  \alpha_{1}^{2}-\alpha_{2}^{2}\right)
}\times\\
\allowbreak\dfrac{A}{\alpha_{1}^{2}\allowbreak e_{1}\left(  r_{c}\right)
-\alpha_{2}^{2}e_{2}\left(  r_{c}\right)  }\left\{  \dfrac{\alpha_{1}%
e_{1}\left(  r_{c}\right)  +\alpha_{2}e_{2}\left(  r_{c}\right)  }{2}-\right.
\\
\left.  -\dfrac{\alpha_{1}e_{1}\left(  r_{c}\right)  -\alpha_{2}%
e_{2}\allowbreak\left(  r_{c}\right)  }{\alpha_{1}^{2}-\alpha_{2}^{2}}%
-\dfrac{r_{c}}{\sqrt{2}\sigma}\left[  e_{1}\left(  r_{c}\right)  +e_{2}\left(
r_{c}\right)  \right]  \right\}  .
\end{array}
\label{ypm}%
\end{equation}
This function is shown in Fig.~\ref{PM} in line with empirical data. As one
can expect from Fig.~\ref{PM} a) the average response $\left\langle
y_{+}\right\rangle _{r_{c}}$ is correlated with $r_{c}$ at small
$r_{c}\lesssim\sigma$ and is anticorrelated with it at larger $r_{c}$. We
added horizontal dotted line in Fig.~\ref{PM} b), shifting the $y$-axis by
unconditional average return\cite{BoMa-06} $\overline{\left\langle
y\right\rangle }=0.00025$. This shift and remaining difference in shape
between Figs.~\ref{PM} a) and b) is related to the buy/sale asymmetry,
discussed below. 

At $r_{c}=0$ the average daily return for given sign of the previous day
return is finite, reproducing the effect of ``double dynamics''\ of the
market, attributed in Ref.\cite{BoMa-06} to ``propagation''\ of two
independent signals in ``Markovian''\ market.

In fact, the market is not Markovian, but the sign of price increments is
determined only by the noise $\xi$. Markovian ``double dynamics''\ of signs is
direct consequence of Markovian statistics of noise correlations, see
section~\ref{FAT}. Anticorrelations between the noise and the amplitude are
responsible for small systematic trend of the price, $\overline{\left\langle
y\right\rangle }\simeq\sqrt{2}\alpha\zeta$~(\ref{Shift}), reproducing
empirical data\cite{BoMa-06} for $\zeta\simeq0.02$. This positive trend leads
to corresponding increase of the probability to have positive price increment,
$p_{+}=1/2+c_{1}\zeta$ with $c_{1}\simeq1$. Conditional probabilities of the
two-state model\cite{BoMa-06} can be expressed through the angle $\phi
\simeq\phi_{-}\simeq\phi_{+}$, which determines the amplitude of the response
$\left\langle y\right\rangle _{x}$ on previous price increment $x$ ($c_{2}%
\sim1$):%
\begin{align*}
p_{++}  &  =1/2+c_{1}\zeta+c_{2}\phi,\\
p_{--}  &  =1/2-c_{1}\zeta+c_{2}\phi.
\end{align*}

Empirical observation of ``double dynamics''\ may be considered as direct
confirmation of Markovian statistics of noise fluctuations, but not of the
whole market, as conjectured in Ref.\cite{BoMa-06}. We show in
section~\ref{BFT}, that multifractal evolution of the amplitude is strongly
non-Markovian. But consideration of only signs of returns
``erases''\ information about the amplitude from the time series.

\subsubsection{Skewness}

\begin{figure}[tb]
\begin{center}
\includegraphics[
height=2.35in,
width=2.73in
]{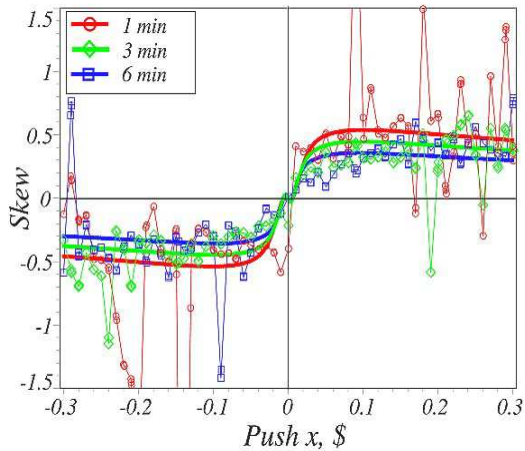}
\end{center}
\caption{Theoretical dependence of the skewness $\rho_{x}$ on $x$ for double
Gaussian model at $\phi=0.2$, $\varepsilon=0$ and $\nu=0.9$ and empirically
observable dependence\cite{Le2-06}. }%
\label{Skew}%
\end{figure}

Asymmetry of the conditional distribution $\mathcal{P}\left(  y|x\right)  $
with respect to the average~(\ref{yav}) is characterized by the skewness of
the conditional response:%
\[
\rho_{x}=\frac{1}{\sigma_{x}^{3}}\int_{-\infty}^{\infty}dy\left(
y-\left\langle y\right\rangle _{x}\right)  ^{3}\mathcal{P}\left(  y|x\right)
.
\]
The conditional mean-square deviation $\sigma_{x}$ is defined in
Eq.~(\ref{scond}). Positive value of $\rho_{x}$ indicates that only few agents
perform great profits, while many of them have small losses with respect to
the mean. A negative $\rho_{x}$ describes a complementary case. As one can see
from Fig.~\ref{Skew}, the skewness has the sign of initial push $x$ in
accordance with the empirical dependence. Notice that although the skewness is
very sensitive characteristic of PDF, our theory reproduces both observed
shape and values of $\rho_{x}$. 

In this section we show, that separation of hot and cold degrees of freedoms
allows to reproduce numerous empirical data, known for high-frequency
fluctuations on financial markets. For Gaussian statistics of all degrees of
freedom this model captures main features of all groups of stocks, including
``market mill''\ patterns, ``dependence-induced volatility smile'', $z$-shaped
response function and so on. Correlations between hot and cold variables are
responsible for observable double dynamics of the market, mixing propagating
signals of opposite signs and providing systematic positive trend of prices.

\subsection{Results and restrictions\label{CONCL2}}

In this section we demonstrate, that the idea of hot and cold variables allows
to capture main features of price fluctuations, which can be described by
Double Gaussian model -- a generalization of the random walk model for the
case of multiscale fluctuations. For different sets of parameters the
analytical solution of this model reproduces the behavior of all kinds of
stocks on financial market, as well as the market as a whole.

Consider some restrictions of this approach. Using coarse-grained description
of price fluctuations at times $\tau>\tau_{k}\lesssim1$ min we loose
information about sale/buy mechanisms\cite{Ro-01,WeRo-03}. This knowledge
(see, for example, Minority and Majority Games\cite{Minor}) is important to
derive parameters of our models for particular markets. We also considered
only uni- and bivariate distribution functions, while market dynamics is
described by the whole family of $n$-point correlation functions. In next
section we present alternative description of the market at different time
scales using ideas of renormalization group approach.

\section{Multiscale dynamics of the market\label{RG}}

Standard thermodynamics can describe only ergodic systems, while the market is
the system with ``restricted ergodicity'': during the time $\tau$ it can
explore only small part of the total configuration space near current local
equilibrium. Increasing the time $t$, this equilibrium is shifted because of
long-time variations. The resulting multi-time dynamics of fluctuations on the
market is not ergodic, and can only be described by continuous set of Langevin
or Fokker-Planck-type equations for all time scales.

Note, that this is not exclusive, but standard behavior of complex physical
systems. As we will show, different local equilibriums on the market are
organized into tree cascades (``self-organized criticality''), by analogy with
``hot spots''\ in Quantum Chromodynamics\cite{QCD}, and dynamics of unergodic
spin-glasses, which is government by continuous set of Fokker-Planck
equations\cite{SG}. Continuous set of equilibriums was also predicted for
incomplete markets\cite{BC-89}.

The behavior of the market at the trade by trade level was studied in many
details\cite{BiHiSp-95,BoMePo-02,MaMi-01}. At larger times collective effects
become important, and financial time series display long-time nonlinear
correlations\cite{Ma-St-99,DaGeMu-01,Co-01}, which puzzle many
researchers\cite{BoCo-98,Interm,KrHoHe-02}. Different models have been
proposed in order to reproduce some ``stylized facts''\cite{Styl}\ of
empirical time series. L\'{e}vy flight processes\cite{Levi} were used to model
jumping character of price variations. Volatility (the amplitude of
fluctuations) clustering effects have been studied in frameworks of stochastic
volatility models\cite{Ta-86} and GARCH-type models\cite{GARCH}. Mixed effect
of jumps and stochastic volatility was taken into account in some
models\cite{CGM-03}. A key to study multiscale properties of price
fluctuations is provided by phenomenological multifractal models, see Refs.
\cite{CaFi-04,FiCaMa-97}. Renormgroup approach, describing evolution of
multiscale systems, was first proposed for glass systems\cite{Renorm,Do-85},
and in present work we extend it to the market.

The key question is the source of price fluctuations. Assumption about totally
random activity of traders\cite{BaPaSh-97-Ph,DaFaIo-03} leads to Brownian
motion of prices. Although random trading model predicts many qualitative and
quantitative properties of the order books\cite{BoMePo-02,SmFaGi}, it can not
describe existing correlations in price fluctuations. An alternative
``efficient market hypothesis''\ assumes, that the price can be changed only
because of unanticipated and totally unpredictable news. This hypothesis lays
in the basis of the model of fully rational agents\cite{Fa-70}, which also
predicts Brownian walk statistics of prices. Observed volatility of the market
is too high to be compatible with the idea of fully rational
pricing\cite{Sc-AER-81}, and can only be reproduced by introducing an
artificial random source -- ``sunspots''\cite{CS-83}. In addition, the
analysis of Ref.\cite{GLGB-08} shows, that most of large fluctuations in the
market are due to trading activity, independently of real news.

\begin{figure}[tb]
\begin{center}
\includegraphics[
height=1.6777in,
width=3.2474in
]{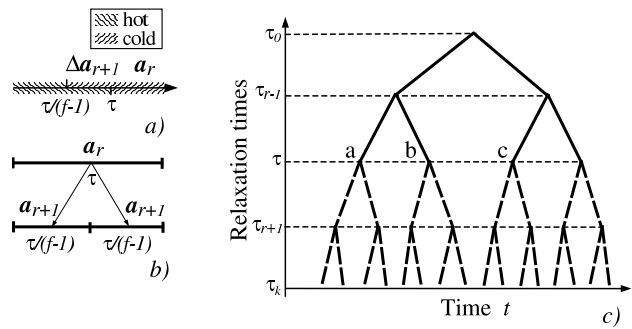}
\end{center}
\caption{Hot and cold degrees of freedom have, respectively, times small and
large with respect to coarse graining time $\tau$. The scale of relaxation
times for two observation times $\tau$ and $\tau/\left(  f-1\right)  $, Fig.
a). Elementary step of renorm group transformation corresponds to division of
``parent''\ time interval $\tau$ into $\left(  f-1\right)  $ ``child''\ time
intervals $\tau/\left(  f-1\right)  $, see Fig. b). Hierarhical tree in
ultrametric time ``space''\ is shown in Fig. c). For given observation time
$\tau$ upper part of the tree, shown by solid lines, corresponds to
``cold''\ degrees of freedom, and lower (dotted) part corresponds to
``hot''\ degrees of freedom.}%
\label{Tree}%
\end{figure}

The main idea of this paper is that market activity can be described as random
trading at all time horizons $\tau_{p}$\ from a minute to tenths years. The
market tends to reach equilibriums on extremely wide baseband of time scales,
and all these equilibriums are continuously changed both because of long-time
modes and external events. Multiple local equilibriums can be represented by
an hierarchical tree, see Fig.~\ref{Tree}. Each generation $r$ of this tree is
characterized by its relaxation time $\tau_{r}$. For any observation time
interval $\tau=\tau_{r}$, all states with times smaller than $\tau_{r}$ are in
equilibrium, and fluctuations near this equilibrium are described by hot
degrees of freedoms. The states with times larger $\tau_{r}$ are frozen, and
are described by cold degrees of freedom. 

In this section we derive an analog of renormgroup equations for the market,
which\ relate fluctuations at different coarse grained time scales $\tau$.
With decrease of the time interval $\tau$, which can be thought of as
effective temperature, the market experiences a cascade of dynamic phase
transitions of broken ergodicity, when some hot degrees of freedom become
frozen. This cascade can be graphically shown as the hierarchical tree, each
branching point of which represents ``phase transition''\ to a state with
frozen degrees of freedom with relaxation times $\tau_{r}>\tau$, see
Fig.~\ref{Tree}. We show in section~\ref{ULTRA}, that the topology of this
tree reflects ultrametricity of the time ``space''.

In section~\ref{RECUR} we demonstrate, that fluctuations at given time scale
$\tau$ are determined by contributions of all ``parent''\ time scales of the
hierarchical tree in Fig.~\ref{Tree}, what is the reason of non-Markovian
dynamics of the market. Cumulative contribution of all time scales allows to
explain extremely high volatility of the market (section~\ref{ACCES}), and is
responsible for power low decay of correlation functions (section~\ref{COREL})
and their multifractal properties. In section~\ref{COREL} we formulate
self-consistency condition, under which the hierarchical tree in
Fig.~\ref{Tree} describes coarse-graining dynamics at all levels of the coarse
graining time $\tau$, and find the $\tau$-dependence of parameters of our
Double Gaussian model, section~\ref{ASYMM}.

In section~\ref{MULTI} we derive a set of Langevin equations, describing
dynamics of the market with extremely wide range of characteristic times --
from minute to tenths years, and calculate the price shift in the response on
imbalance of trading volumes (section~\ref{RESP}).

We also calculate PDF of volatility (section~\ref{PROBAB}) and show, that it
has fat tail with stable exponent $\mu=3$ for stock jumps and $\mu=2$ for news
jumps. We derive, that coarse grained dynamics of the market can be reduced to
the multifractal random walk model\cite{MuDeBa-00,BaDeMu-01}, which determines
multifractal properties of price fluctuations, related to the ultrametric
structure of the tree in Fig.~\ref{Tree}. We calculate volatility patters
after news and stock jumps, and find their conditional probabilities. In
section~\ref{CGM} we show, that the price $P\left(  t\right)  $ behaves as
fractional Brownian motion. We demonstrate in section~\ref{BFT}, that Brownian
motion, sub- and super-diffisive regimes change each other at the long-time
scale. The knowledge of history can be used to estimate the tendency and risks
of future price variations.

Main results of this section are summarized in section~\ref{CONCL3}. In
Appendix~\ref{VOLAT} we show details of calculations of the volatility distribution.

\subsection{Renormalization group transformation}

Consider statistics of price increments (returns)
\begin{equation}
\Delta_{\tau}P\left(  t\right)  =P\left(  t+\tau\right)  -P\left(  t\right)  ,
\label{DP}%
\end{equation}
as the function of the coarse-graining time interval $\tau$. Here $P\left(
t\right)  $ is the price or its logarithm at time $t$. For definiteness sake,
we consider only ACOR group of stocks\cite{Le3-06}, when $\Delta_{\tau
}P\left(  t\right)  $ can be represented as scalar product of complex
amplitude $\mathbf{a}\left(  t\right)  $ and complex noise $\mathbf{\xi
}\left(  t\right)  $, Eq.~(\ref{dP}). By analogy with renormgroup
consideration, cold variable $\mathbf{a}\left(  t\right)  $ slowly varies at
time scale $\tau$, while hot variable $\mathbf{\xi}\left(  t\right)  $ quickly
fluctuates at this scale, see Fig.~\ref{Tree} a).

\subsubsection{Ultrametricity and restricted ergodicity\label{ULTRA}}

In order to establish an analog of renormgroup transformation for the market
we first introduce corresponding partitioning of the time ``space''. At
elementary step of the renormgroup each ``parent''\ interval $\tau_{r}$\ of
time axis can be divided into $f-1$ ``child''\ time interval $\tau_{r+1}%
=\tau_{r}/\left(  f-1\right)  $, see Fig.~\ref{Tree} b). Repeating such
division, we arrive to the hierarchical tree with functionality $f$, shown
schematically in Fig.~\ref{Tree} c). For this tree the time
\begin{equation}
\tau_{r}=\tau_{0}e^{-\kappa r},\qquad\kappa=\ln\left(  f-1\right)
\label{taur}%
\end{equation}
depends exponentially on the current rank $r$, $\tau_{0}$ is maximal
relaxation time. Minimal time $\tau_{k}=\tau_{0}e^{-\kappa k}$ ($k$ is the
number of generations of the tree) is about average time between trades.
Typically, $\tau_{0}$ is about several years and $\tau_{k}$ is about minute
for liquid markets, and so $\kappa k\gtrsim13$.

We define the ``distance''\ $z$ between events at times $t$ and $t^{\prime}$
by the condition $\tau_{r-z}=\left\vert t-t^{\prime}\right\vert $:
\begin{equation}
z\left(  t-t^{\prime}\right)  =\frac{1}{\kappa}\ln\frac{\left\vert
t-t^{\prime}\right\vert }{\tau}\quad\text{at\quad}\left\vert t-t^{\prime
}\right\vert \gg\tau=\tau_{r}, \label{ztt}%
\end{equation}
which can be identified with the distance (number of generations) along the
tree in Fig.~\ref{Tree} c) between these points. One can show, that for three
different times $t,t^{\prime},t^{\prime\prime}$
\[
z\left(  t-t^{\prime\prime}\right)  \simeq\max\left[  z\left(  t-t^{\prime
}\right)  ,z\left(  t^{\prime}-t^{\prime\prime}\right)  \right]  ,
\]
and, therefore, the metric~(\ref{ztt}) generates ultrametric time
``space''\ (with only isosceles and equilateral triangles), which can be
really mapped to the tree. For example, in Fig.~\ref{Tree} c) the distance
between points $a$ and $b$ is $z_{ab}=1$, $z_{bc}=2$, and $z_{ac}=\max\left(
z_{ab},z_{bc}\right)  =2$.

Note, that Eq.~(\ref{taur}) gives standard relation between time scales of
discrete wavelet transformation. The tree in Fig.~\ref{Tree} can be thought of
as a skeleton of the wavelet transformation of time series. We turn to wavelet
interpretation of our approach in section~\ref{MULTI}.

Each of horizontal levels at the time $\tau=\tau_{r}$ on the tree in
Fig.~\ref{Tree} c) corresponds to course-grained description of fluctuations
at the time scale $\tau$. Hot degrees of freedom are ``melted'', and described
by complex noise $\mathbf{\xi}\left(  t\right)  $ with continuum spectrum of
relaxation times extended from $\tau_{k}$ through $\tau$. Cold degrees of
freedom are characterized by complex amplitude $\mathbf{a}=\mathbf{a}_{r}$ of
the noise, which is frozen at the time $\tau$, see Eq.~(\ref{dP}).

By analogy with glasses, the states of real market are highly degenerated,
what is reflected in the presence of gauge transformation~(\ref{calibr_1}) of
complex noise and amplitude variables, which do not affect price variations
$\Delta_{\tau}P\left(  t\right)  $, Eq.~(\ref{dP}). The degeneracy is the
reason of high sensibility of the market to external events. Recall, that in
spin glasses any observable are not affected by ``non-serious''\ part of
disorder, which can be removed by gauge transformation of glass degrees of freedoms.

Following this analogy, the time $\tau$ plays the role of the temperature $T$.
With decrease of the temperature $T\sim\tau$ from $\tau=\tau_{0}$ the market
experiences a cascade of dynamic phase transitions of broken ergodicity, when
some hot degrees of freedom become frozen (the system is unergodic if its
fluctuations can not explore the whole configuration space). This cascade
proceeds continuously down to the time $\tau_{k}$, and can be graphically
shown as hierarchical tree, each branching point of which represents phase
transition to a state with frozen degrees of freedom with relaxation times
$\tau_{r}>\tau$, see Fig.~\ref{Tree}. The parameter $\kappa\ll1$ determines
the probabilities of such transitions, which are relatively rare for real markets.

In this sense at any $\tau<\tau_{0}$ the market is just at the point of
dynamic phase transition of broken ergodicity, and has, therefore, increased
amplitude of fluctuations -- the volatility. This observation supports the
idea that the market is always operating at a critical point as the result of
competition between two populations of traders: ``liquidity providers'', and
``liquidity takers''\cite{Bo-Ch-05,Bak}. Liquidity providers correspond to hot
degrees of freedom, creating antipersistence in price changes, whereas
liquidity takers correspond to cold degrees of freedom, and they lead to the
long range persistence in prices.

Since such separation of the market into hot and cold degrees of freedom takes
place at any time scale, $\tau_{k}\ll\tau\ll\tau_{0}$, there could not be any
unique classification of traders, which can be divided also into ``positive
feedback''\ traders and ``fundamentalists''\cite{Positive},
``contrarian''\ traders and ``trend followers''\cite{Ma-01} and so on. There
is, however, important difference between market and spin-glass hierarchical
trees: while the states of the glass are not ordered, there is strong time
ordering of all ``points''\ of the market tree at any level $r$ of the hierarchy.

\subsubsection{Recurrence relation\label{RECUR}}

General recurrence relation between amplitudes $\mathbf{a}_{r}$ and
$\mathbf{a}_{r+1}$ at levels $r$ and $r+1$ of the hierarchical tree can be
written through the random transition matrix $\mathbf{u}_{r}$:%
\begin{equation}
\mathbf{a}_{r+1}=\mathbf{u}_{r+1}\mathbf{a}_{r}+\Delta\mathbf{a}_{r+1}.
\label{recur}%
\end{equation}
In general, there could be a term $\sim\left(  \mathbf{a}_{r}\right)  ^{\ast}$
in the rhs of this equation, but it is not invariant with respect to gauge
transformation, Eq.~(\ref{calibr_1}), and should be dropped. The term
$\sim\mathbf{u}_{r+1}$ describes the inheritance of the amplitude
$\mathbf{a}_{r}$ of the ``parent''\ levels of the hierarchy, and
$\Delta\mathbf{a}_{r+1}$ gives the contribution of ``newborn''\ during the
transition $r\rightarrow r+1$ unfrozen degrees of freedom to the
``child''\ amplitude $\mathbf{a}_{r+1}$.

The recurrence relation~(\ref{recur}) can also be rewritten in the
multiplicative form, introducing relative increment $\Delta_{r+1}%
=\Delta\mathbf{a}_{r+1}/\mathbf{a}_{r}$:
\begin{equation}
\mathbf{a}_{r+1}=e^{\omega_{r+1}}\mathbf{a}_{r},\quad e^{\omega_{r+1}}%
\simeq1+\omega_{r+1}=\mathbf{u}_{r+1}+\Delta_{r+1} \label{aar}%
\end{equation}
Random variables $\mathbf{u}_{r+1}$ and $\Delta\mathbf{a}_{r+1}$
($\Delta_{r+1}$) are determined by degrees of freedom with characteristic
times $\tau_{r}<\tau<\tau_{r+1}$, while $\mathbf{a}_{r}$ is formed by degrees
of freedom with times larger $\tau_{r}$. Assuming independence of fluctuations
of different time scales, $\mathbf{a}_{r}$ do not depend on $\mathbf{u}_{r+1}$
and $\Delta\mathbf{a}_{r+1}$. We estimate the mean squared amplitudes of
fluctuations of $\Delta\mathbf{a}_{r}$ and $\mathbf{u}_{r}$ as
\begin{equation}
\overline{\left(  \Delta\mathbf{a}_{r}\right)  ^{2}}=D_{0}\tau_{r}%
,\qquad\overline{\mathbf{u}_{r}^{2}}=u^{2}, \label{dat}%
\end{equation}
There is important difference of Eq.~(\ref{dat}) from the case of usual
diffusion, when $\left\langle r^{2}\right\rangle =Dt$ is the consequence of
independence of fluctuations at \emph{different times} $t$. In contrast,
diffusion-like\ dependence~(\ref{dat}) with effective coefficient $D_{0}$ is
the consequence of independence of fluctuations of \emph{different time
scales} $\tau_{r}$. The time $t$-dependence of price fluctuations is strongly non-diffusive.

\subsection{Amplitude of fluctuations}

\subsubsection{Excess of volatility\label{ACCES}}

\begin{figure}[tb]
\begin{center}
\includegraphics[
height=1.9285in,
width=3.3512in
]{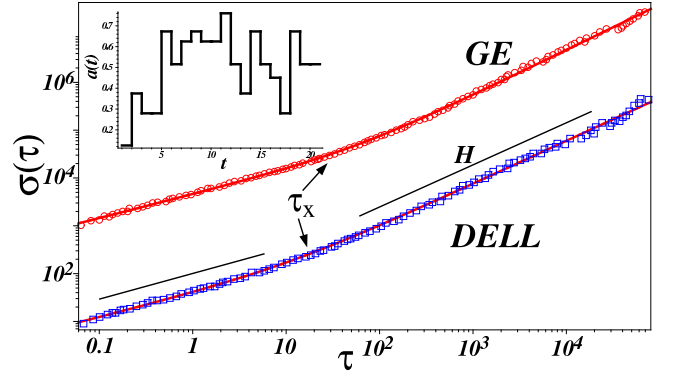}
\end{center}
\caption{Empirical dependence\cite{Eisler} of dispersion $\sigma\left(
\tau\right)  $ on time interval $\tau$, and its fitting by Eq.~(\ref{sigt}),
$\lambda_{0}^{2}=0.9$ for DELL and $\lambda_{0}^{2}=0.8$ for General Electric.
The effective Hurst exponent $H$ is different at $\tau<\tau_{x}$ and
$\tau>\tau_{x}$. In Insert -- computer simulation of random amplitude,
Eq.~(\ref{ak}).}%
\label{Sigma}%
\end{figure}

In the mean field approximation we neglect fluctuations of $\mathbf{u}_{r}=u$,
and find the solution of Eq.~(\ref{recur}) in the form of the sum of
independent random signals $\Delta\mathbf{a}_{r-k}$ from time intervals
$\tau\left(  f-1\right)  ^{k}$, obtained by multiplicative merging of $\left(
f-1\right)  ^{k}$ previous time intervals $\tau$: \
\begin{equation}
\mathbf{a}_{r}\left(  t\right)  =\sum\nolimits_{k>0}u^{k}\Delta\mathbf{a}%
_{r-k}\left(  t\right)  . \label{ak}%
\end{equation}
Weights of these signals exponentially fall with the distance $k$ in time
hierarchy from the current rank $r$. Simulated time series~(\ref{ak}) for the
amplitude $a\left(  t\right)  $ in the model with random $\Delta\mathbf{a}%
_{r}=\pm\sqrt{D_{0}\tau_{r}}$ are shown in Insert in Fig.~\ref{Sigma}. This
picture demonstrates multiscale character of resulting price fluctuations.

Averaging the square of the recurrence relation~(\ref{recur}), we find
difference equation
\begin{equation}
\sigma^{2}\left(  \tau_{r+1}\right)  =u^{2}\sigma^{2}\left(  \tau_{r}\right)
+D_{0}\tau_{0}e^{-\kappa\left(  r+1\right)  } \label{diff}%
\end{equation}
for the dispersion $\sigma\left(  \tau_{r+1}\right)  $ of the amplitude
$\mathbf{a}_{r}$, which has the solution%
\begin{equation}
\sigma^{2}\left(  \tau_{r}\right)  =D\tau_{r}+Lu^{2r}, \label{sr}%
\end{equation}
where\ $L$ is the constant of integration and $D$ is the effective diffusion
coefficient:
\begin{equation}
D=\dfrac{D_{0}}{1-e^{-\kappa\lambda_{0}^{2}}}\gg D_{0},\quad e^{-\kappa
\lambda_{0}^{2}}\equiv u^{2}e^{\kappa}. \label{DD0}%
\end{equation}

From Eqs.~(\ref{sr}) and~(\ref{taur}) we find the dependence of the dispersion
of price increments on the coarse-graining time $\tau$:
\begin{equation}
\sigma^{2}\left(  \tau\right)  =D\tau+L\left(  \tau/\tau_{0}\right)
^{1+\lambda_{0}^{2}}. \label{sigt}%
\end{equation}
The dependence~(\ref{sigt}) for different stocks is in good agreement with
empirical data, see Fig.~\ref{Sigma}. Although at small $\tau\ll\tau_{x}$,
\begin{equation}
\tau_{x}=\tau_{0}\left(  D\tau_{0}/L\right)  ^{1/\lambda_{0}^{2}}, \label{tx}%
\end{equation}
it looks like diffusion with apparent diffusion coefficient $D$~(\ref{DD0}),
price fluctuations do not really have diffusive behavior. As the sign of it,
the amplitude of price fluctuations is anomalously large due to the presence
of a big prefactor for $\kappa\ll1$ in Eq.~(\ref{DD0}). It was shown by
Schiller\cite{Sh-00}, that even accounting the volatility of
dividends\cite{Sh-81} leaves the empirical volatility at least a factor $5$
too large with respect to the random walk model. Such anomalous ``excess of
volatility'',\ $D/D_{0}\sim10$, originates from the superposition of signals
from all time scales, see Eq.~(\ref{ak}). Similar effect (by 10 orders of
value) is well known in spin glasses, where the parameter $\kappa$ in
Eq.~(\ref{ztt}) is extremely small.

Eq.~(\ref{sr}) can be used to estimate the amplitude of the transition matrix
in Eq.~(\ref{aar}). From Eqs.~(\ref{dat}) and~(\ref{sr}) with $L=0$ (at
$\tau\ll\tau_{x}$) we find $\overline{\Delta_{r+1}^{2}}=e^{-\kappa}\left(
1-e^{-\kappa\lambda_{0}^{2}}\right)  $, and get%
\begin{equation}
\overline{e^{2\omega_{r}}}=e^{-\kappa}. \label{omr}%
\end{equation}

\subsubsection{Cross-over time and Hurst exponents\label{TIME}}

The term $\sim L$ in Eq.~(\ref{sigt}) appears as a constant of integration of
the recurrence equation~(\ref{diff}), which is determined by the ``boundary
condition''\ at trading time $\tau_{k}$. Therefore, $L$ is not universal and
determined by microstructure of the market.

At small $\tau<\tau_{x}$ the first term in Eq.~(\ref{sigt}) gives the main
contribution, $\sigma\left(  \tau\right)  \sim\tau^{H}$, with effective Hurst
exponent close to $1/2$. At large $\tau>$ $\tau_{x}$ the second term in
Eq.~(\ref{sigt}) gives dominating contribution to the Hurst exponent:
\begin{equation}
H=(1+\lambda_{0}^{2})/2. \label{H2}%
\end{equation}
Such behavior with different exponents $H$ at $\tau<\tau_{x}$ and $\tau>$
$\tau_{x}$ was really observed for S\&P 500 stock index
(1984-1996)\cite{LGCMPS-99} with different values of the cross-over time
$\tau_{x}$ for individual companies, see Fig.~\ref{Sigma}.

The removal\cite{LGCMPS-99} of the largest 5 and 10\% events kills
correlations of the noise $\xi\left(  t\right)  $ at small time scales,
reducing the constant $L$. According to the prediction~(\ref{tx}) of our
theory, it shifts $\sigma\left(  \tau\right)  $ to lower values, and strongly
increases $\tau_{x}$. Excluding the shift of $L$ from variations of both
$\sigma$ and $\tau_{x}$, we find linear relation between two these shifts:
$\Delta\ln\sigma\simeq-\left(  \lambda_{0}^{2}/4\right)  \Delta\ln\tau_{x}$.
Comparison with empirical data\cite{LGCMPS-99} gives the estimation
$\lambda_{0}^{2}\simeq1$, in good agreement with observable exponent~$H\simeq
0.93$ for the regime $\tau>\tau_{x}$, see Eq. (\ref{H2}). Similar behavior is
observed for different stocks with typical transition times $\tau_{x}$ about
several days.

\subsubsection{Parameters of Double Gaussian model\label{PARAM}}

Let us show, that in order to represent coarse grained dynamics of price
fluctuations for the time interval $\tau=\tau_{r}$, noise variables
$\mathbf{\xi}_{r+1}^{n}$ at different ``child''\ subintervals $n=1,\ldots,f-1$
of the same ``parent''\ interval should be (anti)correlated. According to the
idea of the coarse grained description, the price increments for the time
interval $\tau_{r}$ is the sum
\begin{equation}
\Delta_{\tau_{r}}P\left(  t\right)  =\sum\nolimits_{n=1}^{f-1}\Delta
_{\tau_{r+1}}P\left(  t+n\tau_{r+1}\right)  \label{cgr}%
\end{equation}
of price increments for all $f-1$ adjacent time intervals $\tau_{r+1}$.
Substituting Eq.~(\ref{dP}) into Eq.~(\ref{cgr}), this last equation can be
rewritten in the form%
\begin{equation}%
\begin{array}
[c]{c}%
\left(  \mathbf{a}_{r},\mathbf{\xi}_{r}\right)  =%
{\displaystyle\sum_{n=1}^{f-1}}
\left(  \mathbf{a}_{r},\mathbf{\xi}_{r+1}^{n}\right) \\
=\left(  \mathbf{u}_{r}\mathbf{a}_{r},%
{\displaystyle\sum_{n=1}^{f-1}}
\mathbf{\xi}_{r+1}^{n}\right)  +%
{\displaystyle\sum_{n=1}^{f-1}}
\left(  \Delta\mathbf{a}_{r+1}^{n},\mathbf{\xi}_{r+1}^{n}\right)  ,
\end{array}
\label{axi}%
\end{equation}
where we substituted Eq.~(\ref{recur}) for the amplitude $\mathbf{a}_{r+1}%
^{n}$ to the right hand side of Eq.~(\ref{axi}). Calculating the average (both
quenched and annealed) of the square of this equation, we find the
self-consistency equation%
\[
\sigma^{2}\left(  \tau_{r}\right)  =u^{2}\sigma^{2}\left(  \tau_{r}\right)
\left(  f-1\right)  \left(  1+\varepsilon\right)  +D_{0}\tau_{r}\left(
f-1\right)  ,
\]
where $\varepsilon$ is the noise correlator at neighboring time intervals,
\begin{align*}
\varepsilon &  \equiv\left.  \overline{\left\langle \Delta_{\tau_{r+1}%
}P\left(  t\right)  \Delta_{\tau_{r+1}}P\left(  t+\tau_{r+1}\right)
\right\rangle }\right/  \sigma^{2}\left(  \tau_{r+1}\right) \\
&  =\left\langle \left(  \mathbf{\xi}_{\alpha}^{r+1},\mathbf{\xi}_{\beta
}^{r+1}\right)  \right\rangle ,
\end{align*}
Substituting here $u^{2}=e^{-\kappa(1+\lambda_{0}^{2})}$~(\ref{sr}) with
$e^{\kappa}=f-1$ we find in the case $L=0$%
\[
\varepsilon\simeq-\left(  e^{\kappa}-1\right)  \kappa\lambda_{0}^{2}%
\simeq-\kappa^{2}\lambda_{0}^{2}.
\]
This value is always negative and small at $\kappa\ll1$.

\subsubsection{Time and size dependence of fluctuations\label{SIZE}}

\begin{figure}[tb]
\begin{center}
\includegraphics[
height=1.8in,
width=3.384in
]{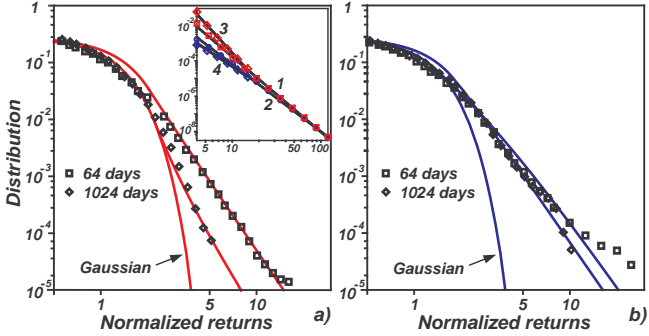}
\end{center}
\caption{Empirically observed positive a) and negative b) tails\cite{PGMNS-99}
of the probability distribution, and their matching by shifted Gaussian and
power tail $\left(  x-\bar{x}\right)  ^{-4}$, $\bar{x}=0.1\sigma$ for time
interval $\tau=64$ days and $\bar{x}=1.1\sigma$ for $\tau=1024$ days. In
Insert we show matching of $x^{-4}$ by $\left(  x-\bar{x}\right)
^{-1-\mu_{\pm}}$ with $\mu_{+}=3.25$ (1)$,\mu_{-}=2.8$ (2) at $\bar{x}=1$, and
$\mu_{+}=4.5$ (3)$,\mu_{-}=2.3$ (4) at $\bar{x}=2$.}%
\label{Stanl}%
\end{figure}

The effect of noise/amplitude anticorrelations, studied in section~\ref{FAT},
is small in the parameter $\zeta\ll1$, and leads to the asymmetry of the tails
of probability distributions (see Fig.~\ref{Gr2}), observed for PDF of
individual companies in Ref.\cite{PGMNS-99}. It is also responsible for
different apparent exponents for positive ($\mu_{+}>3$) and negative ($\mu
_{-}<3$) power tails. This effect is illustrated in Insert in Fig.~\ref{Stanl}%
, where we show, that the function $\left(  x-\bar{x}\right)  ^{-4}$ at
$\bar{x}=1$ is indistinguishable at about two decades in $x$ from power tails
$\left\vert x\right\vert ^{-1-\mu_{\pm}}$ with exponents $\mu_{+}=3.25$ and
$\mu_{-}=2.8$. Increasing $\bar{x}\sim\zeta$ leads to larger deviations of
these exponents from the universal value $3$, see Insert. 

Simple analytical expression for the whole distribution $\mathcal{P}\left(
x\right)  $ for general $\tau$ can be easily constructed by consecutive
matching Gaussian, exponential and power distributions, $\mathcal{P}\left(
x\right)  \sim x^{-4}$, at some points $x_{\pm}$ and $y_{\pm}$, respectively.
The resulting expression well reproduces most of empirical data for PDF
$\mathcal{P}\left(  x\right)  $. Below we use this expression to explain main
features of PDF at $\tau\ll\tau_{x}$ and $\tau\gg\tau_{x}$:

We can estimate the effective exponents $\mu_{\pm}$ re-expanding $x^{-4}%
\sim\left(  x-\bar{x}\right)  ^{-1-\mu_{\pm}}$ about correlation induced
systematic shift $\bar{x}\simeq\sqrt{2}\zeta\alpha\simeq\sqrt{2}\zeta\sigma$
(see Eq.~\ref{Shift}), and find $\mu_{\pm}\simeq3+4\bar{x}/x_{\pm}$ near
cross-over points $x=x_{\pm}$. At small $\tau\ll\tau_{x}$ the central part of
the distribution has exponential shape~(\ref{Pxpm}) with maximum at $\bar{x}$.
Matching it at $x=x_{\pm}$ with power tails $A_{\pm}x^{-4}$ we find $x_{\pm
}=\pm2\sqrt{2}\sigma_{\pm}$ and $\mu_{\pm}\simeq3\pm2\zeta$, in good agreement
with Ref.\cite{PGMNS-99}.

At large $\tau\gg\tau_{x}$ the central part has Gaussian shape. Matching it at
$x=y_{\pm}$ with power tails, we get $y_{\pm}=\bar{x}/2\pm\sqrt{\bar{x}%
^{2}/4+4\sigma^{2}}$. With increase of $\zeta$ (at $\bar{x}\simeq\sqrt{2}%
\zeta\sigma\gtrsim\sigma$) Gaussian region of the positive distribution is
progressively extended, while negative distribution remains fat-tailed,
explaining corresponding mysterious behavior of empirical data\cite{PGMNS-99},
see Fig.~\ref{Stanl}.

The size dependence of the volatility was studied for individual companies in
Ref.\cite{PGMNS-99}. It is shown, that Eq.~(\ref{dr}), $\sigma\sim G^{-\beta}%
$, well describes the dependence of dispersion of returns on market
capitalization. For $\tau=1$ day $\beta\simeq0.2$, while it gradually
decreases with the rise of $\tau$, approaching the value $\beta\simeq0.09$ for
$\tau=1000$ days. This effect supports our self-similar model of companies
(see section~\ref{MODEL}), when the index $\beta=1/\left(  2n\right)  $ is
determined by the number $n$ of generations of the hierarchical tree. The
effective number $n$ of tree generations logarithmically depends on relaxation
time $\tau$, Eq.~(\ref{taur}), and for $\tau_{k}\ll\tau\ll\tau_{0}$ the
dependence $\beta$ on $\tau$ can be approximated by%
\begin{equation}
\beta\simeq\beta_{0}-\beta_{1}\ln\tau. \label{beta}%
\end{equation}
From equation $\sigma\sim G^{-\beta}\sim\tau^{H}$ we find that the Hurst
exponent~(\ref{H2}) should grow logarithmically,
\[
H=H_{0}+\beta_{1}\ln G,
\]
with market capitalization $G$, in good agreement with empirical
data\cite{Eisler}.

\subsection{Nonlinear dynamics of fluctuations}

\subsubsection{Correlation functions: multifractality\label{COREL}}

From Eqs.~(\ref{recur}) and~(\ref{sr}) we find simple analytical expression
for the correlation function of amplitudes:%
\begin{equation}
\overline{\left(  \mathbf{a}\left(  t\right)  ,\mathbf{a}\left(  t^{\prime
}\right)  \right)  }=D\tau_{0}e^{-\kappa r-\kappa\lambda_{0}^{2}z\left(
t,t^{\prime}\right)  }+Lu^{2r}, \label{qz}%
\end{equation}
where $z$ is the logarithmic distance~(\ref{ztt}) in the ultrametric space,
see Fig.~\ref{Tree}. Therefore, observed power autocorrelations in the time
series are the consequence of the self-similiarity of the hierarchical tree in
Fig.~\ref{Tree}. Neglecting the term $L$ at large enough $r$ (small
coarse-graining time $\tau<\tau_{\times}$) in Eq.~(\ref{qz}) we find that
amplitude correlation function decays as the power of the time%
\[
\overline{\left(  \mathbf{a}\left(  t\right)  ,\mathbf{a}\left(  t^{\prime
}\right)  \right)  }\sim\exp\left(  -\lambda_{0}^{2}\ln\left\vert t-t^{\prime
}\right\vert /\tau\right)  ,\ \left\vert t-t^{\prime}\right\vert \gg\tau.
\]

Now consider fluctuations of the modulus $a\left(  t\right)  $ of the
amplitude $\mathbf{a}\left(  t\right)  $. The solution of the multiplicative
recurrent relation~(\ref{aar}) for the coarse graining time $\tau=\tau_{r}$ is%
\begin{equation}
a\left(  t\right)  \simeq\sigma_{0}e^{\omega\left(  t\right)  },\qquad
\omega\left(  t\right)  =\sum\nolimits_{p\leqslant r}\omega_{p}\left(
t\right)  . \label{om1}%
\end{equation}
From expansion~(\ref{om1}) we find%
\begin{equation}
\overline{a^{q}\left(  t\right)  a^{q}\left(  t^{\prime}\right)  }\sim\left(
\dfrac{\tau}{|t-t^{\prime}|}\right)  ^{\tau\left(  q\right)  },\ \tau\left(
q\right)  =\dfrac{g\left(  2q\right)  -2g\left(  q\right)  }{\kappa}
\label{aqaq}%
\end{equation}
with%
\[
g\left(  q\right)  \equiv\ln\overline{e^{q\omega_{r}}}=-q\kappa/2+\kappa
\lambda^{2}\left(  q^{2}/2-q\right)  +...,
\]
where we expanded $g\left(  q\right)  $ over irreducible correlators of
$\omega_{r}=\bar{\omega}_{r}+\Delta\omega_{r}$, $\kappa\lambda^{2}%
=\overline{\Delta\omega_{r}^{2}}$, and used Eq.~(\ref{omr}) to find the linear
in $q$ term. For Gaussian $\omega_{r}$ there are only two first terms in this
expansion, and we get $\tau\left(  q\right)  =\lambda^{2}q^{2}$. We also find%
\[
\overline{a^{q}\left(  t\right)  }\sim\tau^{q\tilde{H}\left(  q\right)
},\qquad q\tilde{H}\left(  q\right)  =-g\left(  q\right)  /\kappa.
\]
In the case of Gaussian $\omega_{p}$ this gives us the generalized Hurst
exponent $\tilde{H}\left(  q\right)  =1/2+\lambda^{2}-\lambda^{2}q/2$, see
also Ref.\cite{Sh-03}. The intermittence parameter $\lambda^{2}$ characterizes
the uncertainty on the market, and we expect, that $\lambda^{2} $ is
relatively large for emerging markets with large uncertainty, and small for
well-developed markets (see section~\ref{MULTI}).

We conclude, that hierarchical structure of market times, see Fig.~\ref{Tree},
generates multifractal time series with $q$-dependent generalized Hurst
exponent. The amplitude $a$ of fluctuations is randomly renewed with time $t$:
with the probability $p_{0}\sim\tau_{0}^{-1}$ for the root of the hierarchical
tree in Fig.~\ref{Tree}, ..., and with the probability $p_{k}\sim\tau_{k}%
^{-1}\gg\gamma_{r}$ for maximum rank $i=k$ of the tree. This random process
generalizes the Markov-Switching Multi-Fractal process\cite{CaFi-04} with
$a^{2}=\sigma^{2}\prod\nolimits_{r=1}^{k}M^{(r)}$. The multiplier $M^{\left(
r\right)  }$ is renewed with probability $p_{r}$ exponentially depending on
its rank $r$ within the hierarchy of multipliers.

\subsubsection{Volume statistics\label{VOLUME}}

In this section we introduce an analog of canonical action-angle variables, in
which coarse-grained market dynamics can be described by a set of linear
Langeven equations for all time scales $\tau_{p}$ in the market. The
``thermodynamic force''\ of price variations is the imbalance of trading
volumes, $V\left(  t\right)  $, which can be considered as random function of
time (volume time series). The increment of the volume at the time interval
$\tau=\tau_{r}$ can be written by analogy with price increment (Eq.~\ref{om1})
in the form:%
\begin{equation}
\Delta_{\tau}V\left(  t\right)  \simeq\sigma_{0}e^{v\left(  t\right)  }%
\eta\left(  t\right)  ,\quad v\left(  t\right)  =\sum\nolimits_{p\leqslant
r}v_{p}\left(  t\right)  . \label{dVV0}%
\end{equation}
Normalized random noise $\eta\left(  t\right)  $ is proportional to the sign
of the increment $\Delta_{\tau}V\left(  t\right)  $. Gaussian random variable
$v\left(  t\right)  $ slowly varies at the time scale $\tau$, and can be
expanded over modes $p$ covering the frequency band from $\tau_{p}^{-1}$ to
$\tau_{p+1}^{-1}$. Explicit expression for $v_{p}\left(  t\right)  $ can be
obtained expanding its variation $\Delta v\left(  t\right)  =v_{p}\left(
t\right)  -\bar{v}_{p}$ over normalized wavelet functions $\psi$ with
expansion coefficients $\hat{v}_{p}\left(  t^{\prime}\right)  $ ($\bar{v}_{p}$
describes regular trend):
\begin{equation}%
\begin{array}
[c]{c}%
\Delta v_{p}\left(  t\right)  =\int\tau_{p}^{-1/2}\psi\left[  \left(
t-t^{\prime}\right)  /\tau_{p}\right]  \hat{v}_{p}\left(  t^{\prime}\right)
dt^{\prime},\\
\hat{v}_{p}\left(  t^{\prime}\right)  =\int\tau_{p}^{-3/2}\psi^{\ast}\left[
\left(  t-t^{\prime}\right)  /\tau_{p}\right]  \Delta v\left(  t\right)  dt.
\end{array}
\label{Wd}%
\end{equation}
Similar equations relate modes $\omega_{p}\left(  t\right)  $~(\ref{om1}) with
the volatility $\omega\left(  t\right)  $. In the case of random activity of
traders $\hat{v}_{p}\left(  t^{\prime}\right)  $ at all time horizons
$\tau_{p}$ should be considered as uncorrelated random values:%
\begin{equation}
\overline{\hat{v}_{p}\left(  t\right)  \hat{v}_{p^{\prime}}\left(  t^{\prime
}\right)  }\simeq\lambda^{2}\kappa\delta_{pp^{\prime}}\delta\left(
t-t^{\prime}\right)  . \label{vvp}%
\end{equation}

The noise function $\eta\left(  t\right)  $ in Eq.~(\ref{dVV0}) can be
presented in the form $\eta\left(  t\right)  \simeq\cos\phi\left(  t\right)
$, where $\phi\left(  t\right)  $ is the phase of corresponding complex
amplitude (see Eq.~\ref{dP}). Random function $\phi\left(  t\right)  $ can be
expanded over wavelet modes (Eq.~\ref{Wd}). Assuming that corresponding
expansion coefficients $\hat{\phi}_{p}\left(  t^{\prime}\right)  $ are
independent random values at all time horizons,%
\begin{equation}
\overline{\hat{\phi}_{p}\left(  t\right)  \hat{\phi}_{p^{\prime}}\left(
t^{\prime}\right)  }\simeq\gamma\kappa\delta_{pp^{\prime}}\delta\left(
t-t^{\prime}\right)  , \label{ffp}%
\end{equation}
we find:%
\[%
\begin{array}
[c]{c}%
\overline{\eta\left(  t\right)  \eta\left(  t^{\prime}\right)  }%
-\overline{\eta\left(  t\right)  }^{2}\simeq\exp\left[  -%
{\displaystyle\sum\limits_{\tau_{k}<\tau_{p}<\left\vert t-t^{\prime
}\right\vert }}
\overline{\phi_{p}\left(  t\right)  \phi_{p}\left(  t^{\prime}\right)
}\right] \\
\simeq\exp\left[  -%
{\displaystyle\int_{\tau_{k}}^{\left\vert t-t^{\prime}\right\vert }}
\dfrac{d\tau_{p}}{\kappa\tau_{p}}\left(  \gamma\kappa\right)  \right]  .
\end{array}
\]
Calculating the integral over $\tau_{p}$ we arrive to%
\begin{equation}
\overline{\eta\left(  t\right)  \eta\left(  t^{\prime}\right)  }%
-\overline{\eta\left(  t\right)  }^{2}\sim\left\vert t-t^{\prime}\right\vert
^{-\gamma}. \label{etet}%
\end{equation}

From above equations we find that the amplitude of volume increments
$\left\vert \Delta_{\tau}V\right\vert $ is log-normally distributed and
uncorrelated with the sign of $\Delta_{\tau}V\left(  t\right)  $. These
predictions are supported by empirical data\cite{BJPW-03}, which also show,
that signs $\eta\left(  t\right)  $ of trade volumes (and, therefore, the very
$\eta\left(  t\right)  $) have long-range power correlations, Eq.~(\ref{etet}%
), with stock dependent exponent $\gamma<1$. We conclude, that observed
power-low correlations in signs of volume are the consequence of the
self-similiarity of price fluctuations at different time scales, which lead to
scale invariant intermittence parameter $\lambda^{2}$ (Eq.~\ref{vvp}) and the
amplitude of fluctuations $\gamma$ (Eq.~\ref{ffp}). Such long range
correlations are usually considered as the result of cutting of large trades
into small chunks (see Ref.\cite{BJPW-03}).

\subsubsection{Langeven equations and market entropy\label{MULTI}}

The key idea of Langeven formulation of multi-time market dynamics is that
fluctuations at different time scales $\tau_{p}$ are statistically
independent. Therefore, the logarithmic volatility $\omega_{p}\left(
t\right)  $ of the mode with relaxation time $\tau_{p}$ is induced only by
corresponding volume mode $v_{p}\left(  t\right)  $~(\ref{Wd}). Since both
$\omega_{p}\left(  t\right)  $ and $v_{p}\left(  t\right)  $ have Gaussian
statistics, general Langeven equations are linear (different choice of
fluctuation modes makes these equations highly nonlinear):%
\begin{equation}
\tau_{p}\dfrac{\partial\omega_{p}\left(  t\right)  }{\partial t}+\omega
_{p}\left(  t\right)  =v_{p}\left(  t\right)  \label{Lang}%
\end{equation}
with $\delta$-correlated noise:
\begin{equation}
\overline{\Delta v_{p}\left(  t\right)  \Delta v_{p^{\prime}}\left(
t^{\prime}\right)  }=2\delta_{pp^{\prime}}\kappa\lambda^{2}\tau_{p}%
\delta\left(  t-t^{\prime}\right)  . \label{vv}%
\end{equation}
At time scale $\tau_{p}$ this equation is in agreement with Eq.~(\ref{vvp}).
After standard calculations we find correlation function of volatility modes%
\begin{equation}
\overline{\Delta\omega_{p}\left(  t\right)  \Delta\omega_{p^{\prime}}\left(
t^{\prime}\right)  }=\delta_{pp^{\prime}}\kappa\lambda^{2}e^{-\left\vert
t-t^{\prime}\right\vert /\tau_{p}}, \label{dwdw}%
\end{equation}
and fluctuations of logarithmic volatility for the coarse graining time
$\tau=\tau_{r}$:%
\begin{equation}%
\begin{array}
[c]{c}%
G\left(  t-t^{\prime}\right)  =\overline{\Delta\omega\left(  t\right)
\Delta\omega\left(  t^{\prime}\right)  }=\sum_{p\leqslant r}\overline
{\Delta\omega_{p}\left(  t\right)  \Delta\omega_{p}\left(  t^{\prime}\right)
}\\
\simeq\lambda^{2}\ln\left(  \tau_{0}/\left\vert t-t^{\prime}\right\vert
\right)  ,\quad\tau\ll\left\vert t-t^{\prime}\right\vert \ll\tau_{0}.
\end{array}
\label{dw}%
\end{equation}
This expression lays in the basis of multifractal random walk
model\cite{MuDeBa-00,BaDeMu-01}, and reproduces Eq.~(\ref{aqaq}):%
\[
\overline{a^{q}\left(  t\right)  a^{q}\left(  t^{\prime}\right)  }%
\sim\overline{\exp q\left[  \omega\left(  t\right)  +\omega\left(  t^{\prime
}\right)  \right]  }\sim\left(  \tau/\left\vert t-t^{\prime}\right\vert
\right)  ^{\tau\left(  q\right)  }.
\]

Eqs.~(\ref{om1}),~(\ref{Lang}) and~(\ref{vv}) present Langeven formulation of
multifractal market dynamics. Using standard transformations, they can be
rewritten in the form of Smoluchovski equations for probability function
$\Psi\left\{  \omega_{p}\right\}  $. The importance of this function is that
it defines rigor entropy of the market%
\[
S\left[  \Psi\right]  =\int D\omega_{p}\Psi\left\{  \omega_{p}\right\}
\ln\Psi\left\{  \omega_{p}\right\}  ,
\]
which can only increase with time. The entropy $S\left[  \Psi\right]  $
characterizes informational content of the market.

From Eq.~(\ref{Lang}) we find the relation between averages: $\bar{v}_{p}%
=\bar{\omega}_{p}\sim-\kappa$ (see Eq.~\ref{omr}), which allows one to express
parameters $\kappa$ and $\lambda^{2}$ of our theory through corresponding
moments of the trade volume $V$:%
\begin{equation}
\kappa\sim\dfrac{2}{k}\overline{\ln\dfrac{|V|}{V_{k}}},\quad\lambda^{2}%
\sim\dfrac{\overline{\left(  \Delta\ln V\right)  ^{2}}}{k\kappa}\sim
\dfrac{\overline{\left(  \Delta\ln V\right)  ^{2}}}{\overline{\ln|V/V_{k}|}},
\label{param}%
\end{equation}
where $V_{k}^{2}\simeq\sigma_{0}^{2}\tau_{k}/\tau_{0}$ is about squared
bid-ask spread.

\subsubsection{Response functions\label{RESP}}

From Langeven equation~(\ref{Lang}) we find the response of the mode $p$ on
volume imbalance $v_{p}\left(  t\right)  $:%
\begin{align}
\omega_{p}\left(  t\right)   &  =\int_{-\infty}^{t}\chi_{p}\left(
t-t^{\prime}\right)  v_{p}\left(  t^{\prime}\right)  dt^{\prime},\nonumber\\
\chi_{p}\left(  t-t^{\prime}\right)   &  =\dfrac{\kappa}{\tau_{p}}e^{-\left(
t-t^{\prime}\right)  /\tau_{p}}\theta\left(  t-t^{\prime}\right)  .
\label{chir}%
\end{align}
Using Eq.~(\ref{Wd}) one can check, that the response function\ $\chi$ of
volatility on volume imbalance at the coarse graining time $\tau_{r}=\tau$ is
determined by the sum of contributions of modes $p>r$:%
\begin{align}
\omega\left(  t\right)   &  \simeq\int_{-\infty}^{t}\chi\left(  t-t^{\prime
}|\tau\right)  v\left(  t^{\prime}\right)  dt^{\prime},\nonumber\\
\chi\left(  t-t^{\prime}|\tau\right)   &  =\sum\nolimits_{p=0}^{r}\chi
_{p}\left(  t-t^{\prime}\right)  \simeq\dfrac{\theta\left(  t-t^{\prime
}\right)  }{t-t^{\prime}}. \label{chi}%
\end{align}

The average price shift because of a single trade at the time $t=0$ of the
volume%
\[
\left\vert \Delta V\right\vert =V_{k}\left(  e^{\Delta v}-1\right)
\]
($-1$ is only important at small $\left\vert \Delta V\right\vert \sim\Delta v
$) is determined by all modes with times from $\tau_{k}$ through $\tau_{0}$
(the noise $\xi_{k}\simeq1$ at $\tau_{k}$):
\begin{align}
\Delta P\left(  t\right)   &  \simeq\sigma_{k}\left[  e^{\Delta\omega\left(
t\right)  }-1\right]  \simeq\sigma_{k}\Delta\omega\left(  t\right)
\label{dtP}\\
&  \simeq\sigma_{k}\chi\left(  t|\tau_{k}\right)  \tau_{k}\Delta v.\nonumber
\end{align}
The dispersion $\sigma_{k}$ at time interval $\tau=\tau_{k}$ is obtained by
averaging over fluctuations of random variables~(\ref{vv}), describing
variations of the liquidity of the market. In general, the liquidity (at
physical language, susceptibility) strongly depends on the history: small
volumes can initiate large jumps or make almost no effect. Similar
``aging''\ effect exists for spin glasses, where the susceptibility\ strongly
depends on the history of temperature and magnetic field changes.

Using Eq.~(\ref{dVV0}) we get%
\begin{align}
\Delta P\left(  t\right)   &  =G_{0}\left(  t\right)  sign(\Delta V)\ln\left(
1+\left\vert \Delta V\right\vert /V_{k}\right)  ,\label{dPt}\\
G_{0}\left(  t\right)   &  \simeq\sigma_{k}\theta\left(  t\right)  \tau
_{k}\left(  t+\tau_{k}\right)  ^{-1}.\nonumber
\end{align}
In general, $1$ under logarithm is out of accuracy of our calculations, and we
hold it to reproduce expected linear response $\Delta P\sim\Delta V$ at
extremely small $\Delta V$. The result $G_{0}\left(  \tau_{k}\right)
\simeq\sigma_{k}=\sigma\left(  \tau_{k}\right)  $ extremely well supported by
data\cite{BJPW-03}. Weak logarithmic dependence of average price shift $\Delta
P$ on the trade volume $\Delta V$ is related to multi-time character of volume
fluctuations, described by Eq.~(\ref{dVV0}). \begin{figure}[tb]
\begin{center}
\includegraphics[
height=2.0686in,
width=2.9412in
]{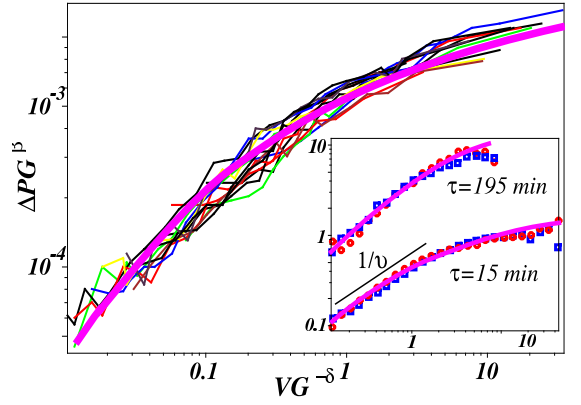}
\end{center}
\caption{The price shift $\Delta P$ per trade vs. transaction size $V$, for
buy orders in 1996, renormalized by powers of market capitalization
$G$.\cite{LiFaMa-02} Theoretical prediction, Eq.~(\ref{dPt}), is shown by
solid line. Results for 1995, 1997 and 1998 are very similar. In Insert:
normalized $\Delta_{\tau}P$ for different $\tau$ as function of $V$, $V>0$
($\bigcirc$) and $V<0$ ($\square$), data from Ref.\cite{PGGS-08}. Solid lines
show fitting by Eq.~(\ref{dtPV}). }%
\label{Impact}%
\end{figure}

The dispersion, $\sigma_{k}\sim G^{-\beta}$, is inversely correlated to the
capitalization $G$ of the market, see Eq.~(\ref{dr}). The exponent
$\beta\simeq0.3$ (Eq.~\ref{beta} at $\tau=\tau_{k}$) is smaller than the
Gaussian value $1/2$, because of hierarchical structure of financial markets,
see Ref.~\cite{Ma-EPJ-99}. It is shown in Fig.~\ref{Impact} that price impact
curves for $1000$ stocks are collapsed very well by Eq.~(\ref{dPt}) with
$V_{k}\sim G^{\delta}$ and $\delta\simeq0.3-0.4$.

The average price shift during the time interval $\tau>\tau_{k}$ can be
estimated considering several trades as one large trade of summary volume
$\Delta V$, and renormalizing minimal relaxation time $\tau_{k}\rightarrow
\tau$ in Eq.~(\ref{dPt}):
\begin{equation}
\Delta_{\tau}P\simeq\sigma sign\left(  \Delta V\right)  \ln\left(
1+\left\vert \Delta V\right\vert /V_{\tau}\right)  , \label{dtPV}%
\end{equation}
with $V_{\tau}^{2}=V_{k}^{2}\tau/\tau_{k}$. We show in Insert in
Fig.~\ref{Impact}, that the above dependence $\Delta_{\tau}P$ well agrees with
empirical data at different $\tau$. Eq.~(\ref{dtPV}) can be approximated by
power dependence $\Delta P\simeq h\Delta V^{1/\upsilon}$, with time, volume
and capitalization dependent effective exponent (see Insert in
Fig.~\ref{Impact}):
\[
\upsilon\equiv\dfrac{d\ln V}{d\ln\Delta_{\tau}P}=\left(  1+\dfrac{V_{\tau}%
}{\left\vert \Delta V\right\vert }\right)  \ln\left(  1+\dfrac{\left\vert
\Delta V\right\vert }{V_{\tau}}\right)  .
\]
At small $\Delta V$ the apparent exponent $\upsilon$ is large at small $\tau$,
and $\upsilon\simeq1$ at large $\tau$ ($\upsilon\simeq3$ for $\tau=5\min$ and
$\upsilon\simeq1$ for $\tau=195\min,$ see Ref.\cite{PGGS-08} and Insert in
Fig.~\ref{Impact}). At large $\Delta V$ typical value $\upsilon\simeq2$, see
Ref.\cite{GGPS-03}, while $1/\upsilon$ slowly decreases\cite{PI-03} with
$\Delta V$, and\cite{PB-03} $1/\upsilon\rightarrow0$ for very large $\Delta V
$.

\begin{figure}[tb]
\begin{center}
\includegraphics[
height=3.0407in,
width=3.0848in
]{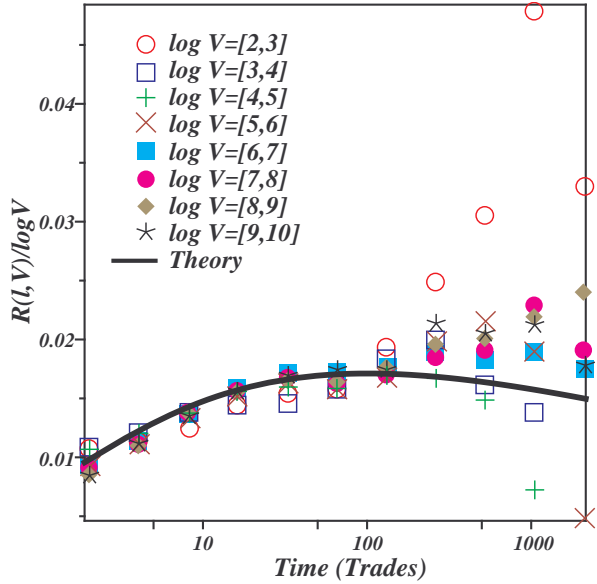}
\end{center}
\caption{Response function $R\left(  l,V\right)  $, conditioned to a certain
volume $V$, as a function of dimensionless time $l$. Data for
France-Telecom\cite{BJPW-03}. The thick line is the prediction of
Eq.~(\ref{RlV}).}%
\label{resp_v0}%
\end{figure}

In the end of this section we estimate introduced in Ref.\cite{BJPW-03}
response function\ conditioned to a volume $V$:%
\begin{align*}
\mathcal{R}\left(  l,V\right)   &  \simeq\left.  \overline{\sum\nolimits_{n<l}%
\Delta P\left[  \left(  l-n\right)  \tau_{k}\right]  \eta\left(  0\right)
}\right\vert _{\Delta V=V}\\
&  \simeq\int_{0}^{l}\left\vert \Delta P\left[  \left(  l-n\right)  \tau
_{k}\right]  \right\vert \overline{\eta\left(  n\tau_{k}\right)  \eta\left(
0\right)  }dn.
\end{align*}
Here $\left\vert \Delta P\left[  \left(  l-n\right)  \tau_{k}\right]
\right\vert $ is the value of average price shift at time $l\tau_{k}$ because
of a trade at time $n\tau_{k}$. It is important, that $\ln\left\vert
V\right\vert $ at time $l\tau_{k}$ for given value of $\ln\left\vert
V\right\vert $ at time $n\tau_{k}$ logarithmically weakly depends on time
interval $\left(  l-n\right)  \tau_{k}$ because of multi-time relaxation of
this value (see Eq.~\ref{wt} below as an example of calculation of such
conditional average). Substituting Eqs.~(\ref{dPt}) and~(\ref{etet}) we find
with logarithmic accuracy%
\begin{equation}%
\begin{array}
[c]{c}%
\mathcal{R}\left(  l,V\right)  \simeq\mathcal{R}\left(  l\right)
\ln\left\vert V/V_{k}\right\vert ,\\
\mathcal{R}\left(  l\right)  \sim%
{\displaystyle\int_{0}^{l}}
\dfrac{1}{l-n+1}\dfrac{dn}{n^{\gamma}}\simeq\dfrac{\ln\left(  1+l\right)
}{l^{\gamma}},
\end{array}
\label{RlV}%
\end{equation}
Plotted in Fig.~\ref{resp_v0} function $\mathcal{R}\left(  l\right)  $ is in
good agreement with empirical data for France-Telecom\cite{BJPW-03} with the
sole parameter $\gamma=1/5$. In general, $\mathcal{R}\left(  l\right)  $
initially grows, reaching maximum at certain $l^{\ast}\simeq e^{1/\gamma}$,
and than decreasing back with dimensionless time $l>l^{\ast}$.

Notice, that Eqs.~(\ref{dPt}),~(\ref{dtPV}) and~(\ref{RlV}) were obtained by
pre-averaging over fluctuations of the liquidity of market, and can not be
applied to find higher moments of price increments, like dispersion
$\sigma\left(  \tau\right)  $. Multifractality changes power dependences of
these values: in contrast to the first moment, Eq.~(\ref{RlV}), when the
intermittency effect is not important, it gives a major contribution to higher
moments. We show in section~\ref{CGM}, that the liquidity fluctuations lead to
strong variations of the virtual trading time, the rate of which is determined
by local time between trades. Therefore correlation function $\mathcal{R}%
\left(  l,V\right)  $ with pre-averaged time intervals $\tau_{k}$ carries no
information about dispersion $\sigma\left(  \tau\right)  $. The relation
between $\mathcal{R}\left(  l,V\right)  $ and $\sigma^{2}\left(  \tau\right)
\simeq D\tau$ was used in Ref.\cite{BJPW-03} to demonstrate a very delicate
balance between liquidity takers and liquidity providers to put the market at
the border between sub- and super-diffusive behavior. In section~\ref{ACCES}
we show, that apparent diffusive behavior $\sigma^{2}\left(  \tau\right)
\simeq D\tau$ is really a result of random trader activity at all time scales.

\subsubsection{Stock and news jumps\label{PROBAB}}

In this section we study volatility patterns of large price jumps in the
market. The volatility variable $\omega\left(  t\right)  $~(\ref{om1}) can be
measured empirically as the average over $n\gg1$ time intervals $\tau$ of the
logarithmic modulus of price increments:%
\[
\omega\left(  t\right)  =\dfrac{1}{n}\sum\nolimits_{k=1}^{n}\ln\left\vert
\Delta P_{k}\left(  t\right)  \right\vert ,\ \Delta P_{k}\left(  t\right)
=\Delta_{\tau}P\left(  t-k\tau\right)  .
\]
In the limit $n\rightarrow\infty$ $\omega\left(  t\right)  $ can be considered
as asymptotically Gaussian random variable. To prove this it is instructive to
define generalized volatility%
\begin{equation}
V_{q}\left(  t\right)  =\dfrac{1}{n}\sum\nolimits_{k=1}^{n}\left\vert \Delta
P_{k}\left(  t\right)  \right\vert ^{q}, \label{Volat}%
\end{equation}
which turns to standard definition of volatility at $q=1$, while in the limit
$q\rightarrow0$ we have%
\begin{equation}
\omega\left(  t\right)  =\left.  \dfrac{dV_{q}\left(  t\right)  }%
{dq}\right\vert _{q=0}. \label{Vq0}%
\end{equation}

In Appendix~\ref{VOLAT} we show, that at large $n\gg1$ PDF of volatility
converges to universal function, which depends only on $q$, exponent $\mu=3$
of the fat tail of PDF, $\mathcal{P}\left(  \Delta P\right)  \sim\left\vert
\Delta P\right\vert ^{-1-\mu}$,~and non-universal constant $c>0$ (will be
calculated later):%
\begin{equation}
P\left(  V_{q}\right)  =\dfrac{x^{-1-\mu/q}}{c\Gamma\left(  c\mu/q\right)
V_{m}}e^{-x^{-1/c}},\quad x=\dfrac{V_{q}}{V_{m}}. \label{pVu}%
\end{equation}
The maximum of this distribution is reached at the point $V_{\max}=\left[
c\left(  1+\mu/q\right)  \right]  ^{-c}V_{m}$.

In Fig.~\ref{distvolatil} we show that expression~(\ref{Pn}) of
Appendix~\ref{VOLAT} for $q=1$ with $\mu=3$ and $c_{1}=2/3$ is in excellent
agreement with known empirical data. Usually this dependence is fitted by
log-normal, Eq.~(\ref{lnn}) of Appendix~\ref{VOLAT}, or inverse gamma
distributions\cite{Volat} (Eq.~(\ref{pVu}) with $c=1$) with extremely high
exponent $\mu=5-7$. Our calculations show, that the distribution $P\left(
V_{q}\right)  $ has the same tail exponent $\mu=3$ as PDF of price incemants.
This result clearly demonstrates the absence of the self-averaging of price
fluctuations: large variations of the coarse-grained ``volatility''\ variable
$V_{q}\left(  t\right)  $~(\ref{Volat}) are induced by large short time jumps,
the contribution of which is dominated even after averaging over
$n\tau\longrightarrow\infty$ time interval. \begin{figure}[tb]
\begin{center}
\includegraphics[
height=2.06in,
width=2.8746in
]{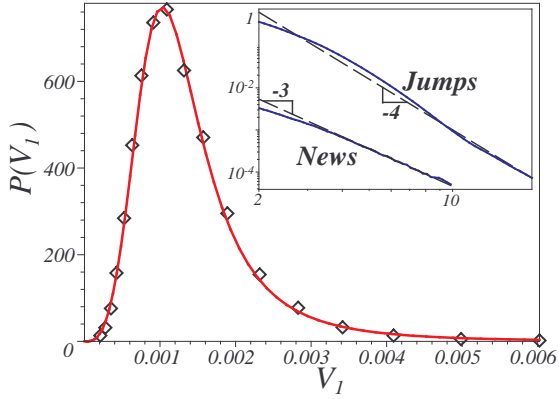}
\end{center}
\caption{Comparison of theoretical (with $\mu=3$ and $c=2/3$) and
empirical\cite{LGCMPS-99} volatility distributions for $\tau=30$ min,
$n\tau=120$ min. In insert we show the tail distribution of volatility of
jumps and news\cite{GLGB-08}.}%
\label{distvolatil}%
\end{figure}

In Insert in Fig.~\ref{distvolatil} we show the probability of large
volatility fluctuations\cite{GLGB-08}. As one can see, the probability of
large ``stock jumps''\ has power tail, $P\left(  V\right)  \sim V^{-1-\mu}$
with $\mu=3$, while $\mu=2$ for ``news jumps'', induced by independent
macroeconomic events. We show in section~\ref{FAT} that $\mu$ equals to the
number of essential degrees of freedom of the noise: complex uncorrelated
noise of ``news jumps''\ has $\mu=2$ components, while there is additional
component of price at previous time interval for Markovian\ noise of
stock\ jumps. The prediction $\mu=3$ is strongly supported by the analysis of
distinct databases with extremely large number of records\cite{GPNM-98} for
the interval $\tau$ from a minute through several months.

Both predictions for the tail exponent of PDF, $\mu=2$ and $\mu=3$, are quite
general. For example, they describe two major universal classes of city grow
(discovered from empirical data in Ref.\cite{CP-03}), because of adding new
street lines. The PDF describes the distribution of lengths of these lines.
New lines are created randomly for cities with $\mu=2$, while there are strong
local correlations in line creation for cities, characterized by the exponent
$\mu=3$. Similar Gutenberg-Richter power law describes earthquakes of a given strength.

Since $\mu/q\rightarrow\infty$ at $q\rightarrow0$, the distribution of $V_{q}%
$~(\ref{pVu}) in this limit becomes asymptotically Gaussian, and random
variable $\omega\left(  t\right)  $ at large $n\gg1$ has Gaussian statistics
with the probability%
\begin{equation}%
\begin{array}
[c]{c}%
P\left\{  \omega\right\}  \sim e^{-H\left[  \omega\right]  },\\
H\left\{  \omega\right\}  =\dfrac{1}{2}\iint dtdt^{\prime}\omega\left(
t\right)  \omega\left(  t^{\prime}\right)  G^{-1}\left(  t-t^{\prime}\right)
,
\end{array}
\ \label{Pw}%
\end{equation}
where $G^{-1}$ is an inverse to the kernel $G$, Eq.~(\ref{dw}) (explicit
expression for $G^{-1}$ is given in section~\ref{CONCL3}). We checked Gaussian
character of $\omega\left(  t\right)  $ by numerical simulations in the model
of Eq.~(\ref{ak}). Eqs.~(\ref{om1}) and~(\ref{dw}) lay in the basis of the
famous Multifractal Random Walk model\cite{MuDeBa-00,BaDeMu-01}.

Minimizing $H\left\{  \omega\right\}  $~(\ref{Pw}) under the condition of
fixed $\omega\left(  t_{0}\right)  $, we find deterministic component of
$\omega\left(  t\right)  $ at $t-t_{0}>\tau$:%
\begin{align}
\omega\left(  t\right)   &  =\Lambda G\left(  t\right)  =\omega\left(
t_{0}\right)  h\left(  t-t_{0}\right)  ,\label{wt}\\
h\left(  t\right)   &  \equiv\varepsilon\ln\left(  \tau_{0}/\left\vert
t\right\vert \right)  ,\qquad\epsilon=1/\ln\left(  \tau_{0}/\tau\right)  ,
\label{ht}%
\end{align}
where we expressed the Lagrange multiplier $\Lambda=\epsilon\omega\left(
t_{0}\right)  /(2\lambda^{2})$ through $\omega\left(  t_{0}\right)  $.

Eq.~(\ref{wt}) describes the result of trading activity, while news coming at
time $t=0$ induce additional volatility, see Eq.~(\ref{chi}):%
\begin{equation}
\omega_{n}\left(  t\right)  =\omega_{0}\tau/t,\qquad t>\tau, \label{wn}%
\end{equation}
$\omega_{0}$ is the amplitude of the news jump. Combining both contributions,
Eq.~(\ref{wt}) and~(\ref{wn}), we find differential equation for the most
probable $\omega\left(  t\right)  $ after a news jump $d\omega=-\epsilon\omega
d\ln(t/\tau)+d\omega_{n}$, with the solution
\begin{equation}
\omega\left(  t\right)  =\dfrac{\omega_{0}}{1-\epsilon}\left[  \dfrac{\tau}%
{t}-\epsilon\left(  \dfrac{\tau}{t}\right)  ^{\epsilon}\right]  ,\quad t>\tau.
\label{wt1}%
\end{equation}
The resulting volatility pattern, $a\left(  t\right)  =a_{0}e^{\omega\left(
t\right)  }$, can be measured by averaging over all significant news.

\begin{figure}[tb]
\begin{center}
\includegraphics[
height=2.1724in,
width=3.1609in
]{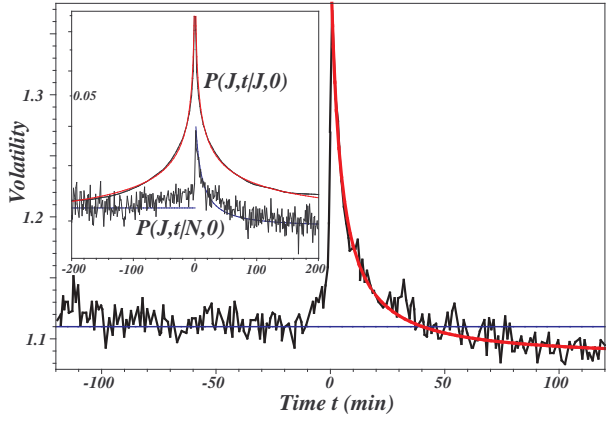}
\end{center}
\caption{The volatility pattern before and after news jump, given by
Eq.~(\ref{wt1}) with $\tau=3,\tau_{0}=10$ years, in comparison with empirical
data\cite{GLGB-08}. The amplitude $\omega_{0}$ is choosed from the condition
of best fitting. In Insert we show, that the probabilities of stock jump after
jump, $P\left(  J,t|J,0\right)  $, decays to jump probability as $\left\vert
t\right\vert ^{-1/2}$, as predicted by our theory. We also show the
probability to observe a jump after news, $P\left(  J,t|N,0\right)  $,
Eq.~(\ref{PJN}), which is increased at small times and decreased at
intermediate times.}%
\label{NJump}%
\end{figure}

Last term in Eq.~(\ref{wt1}) appears because of long memory effect, known as
``aging''\ effect in spin glasses. It can also be interpreted\cite{GLGB-08} as
the reduction of measure of uncertainty after news, since some previously
unknown information becomes available. In Fig.~\ref{NJump} we show the
prediction of Eq.~(\ref{wt1}) in comparison with empirical data\cite{GLGB-08}.
Initial increase of $\omega\left(  t\right)  $ at $t<0$ is related to finite
waiting time $\tau_{w}$, see Eq.~(\ref{wt1}).

Although stock jumps have Markovian statistics, correlations quickly decay at
the coarse graining time $\tau$, which is slightly longer than the time of
returns, and at larger times these events can be described by uncorrelated
random variable $\eta\left(  t\right)  $. The volatility, Eq.~(\ref{dw}), can
be rewritten through $\eta\left(  t\right)  $ as\cite{SoMaMu-02}%
\begin{equation}
\Delta\omega\left(  t\right)  =\int_{-\infty}^{t}\dfrac{dt^{\prime}\eta\left(
t^{\prime}\right)  }{\sqrt{t-t^{\prime}}},\quad\overline{\eta\left(  t\right)
\eta\left(  t^{\prime}\right)  }=\lambda^{2}\delta\left(  t-t^{\prime}\right)
. \label{ome}%
\end{equation}
Substituting $\eta\left(  t\right)  =\omega_{0}\delta\left(  t\right)  $ in
Eq.~(\ref{ome}), we find, that volatility after stock jump at $t=0$ is
relaxing as $\omega\left(  t\right)  =\omega_{0}t^{-1/2}$, in good agreement
with empirical observations\cite{GLGB-08}. Stock jumps can be interpreted as
stochastic resonance of different fluctuation modes $p$ in Eq.~(\ref{om1})
because of their random concurrence. Such resonance is usually happened
because of delaying the jump until it will be anchored by jumps of larger time
scales. The amplitude of the resulting stock jump is significantly increased,
and slowly decays with time. The model predicts diffusion dependence
$\omega_{0}^{2}\sim\lambda^{2}t_{w}$ of the amplitude of a jump on the waiting
time $t_{w}$ between neighboring stock jumps.

The central part of PDF of the volatility fluctuation $V_{1}\left(
t_{0}\right)  =a_{0}e^{\omega\left(  t_{0}\right)  }$, averaged over all
fluctuations, can be found by substituting Eq.~(\ref{wt}) into~(\ref{Pw}):%
\begin{equation}
P\left(  V_{1}\right)  \sim\exp\left[  -G\left(  0\right)  \dfrac{\Lambda^{2}%
}{2}\right]  \simeq\exp\left[  -\allowbreak\dfrac{\epsilon}{4\lambda^{2}%
}\left(  \ln\dfrac{V_{1}}{a_{0}}\right)  ^{2}\right]  . \label{P1V}%
\end{equation}
The log-normal of this distribution is supported by numerous empirical
data\cite{Volat}. Comparing this expression with Eq.~(\ref{lnn}) of
Appendix~\ref{VOLAT}, we find the value of constant $c\simeq\lambda^{2}\left(
\mu+1\right)  /\epsilon$ in Eq.~(\ref{pVu}) for $q=1$.

The volatility pattern of a news jump at time $t=0$ followed by a stock jump
at time $t$ can be presented as the sum of corresponding volatility patterns,
Eq.~(\ref{wt1}) and~(\ref{wt}). Substituting it into Eq.~(\ref{Pw}) we find
the probability of this pattern:%
\begin{equation}
P\left(  J,t|N,0\right)  =P_{n}\left(  \omega_{0}\right)  P\left(
V_{1}\right)  \exp\left[  -\dfrac{\epsilon\omega\left(  t\right)  }%
{2\lambda^{2}}\ln\dfrac{V_{1}}{a_{0}}\right]  . \label{PJN}%
\end{equation}
Here $P_{n}$ and $P$ are probabilities of news and stock jumps, and function
$\omega\left(  t\right)  $ is calculated in Eq.~(\ref{wt1}). We conclude, that
at small time interval $t$ there is asymmetric increase of the probability to
see a jump induced by a news, following by the decrease of this probability at
intermediate times. Similar expression~(\ref{PJN}) with $\omega\left(
t\right)  =\omega_{0}t^{-1/2}$ can be found for the probability $P\left(
J,t|J,0\right)  $ to find a jump at time $t$ after a jump at time $t=0 $.
These predictions are in good agreement with empirical data, see Insert in
Fig.~\ref{NJump}.

\subsubsection{Virtual trading time\label{CGM}}

In this section we show, that fluctuations of the liquidity of the market lead
to corresponding variations of the virtual trading time, $\Theta\left(
t\right)  $, which is proportional to the number of trades per given time
interval. Logarithmic volatility $\omega\left(  t\right)  $, Eq.~(\ref{wt}),
gives the deterministic part of time dependence of the amplitude at
$t-t_{0}>\tau$:
\begin{equation}
a\left(  t\right)  \simeq a_{0}e^{\omega\left(  t\right)  }\simeq a\left(
t_{0}\right)  \left[  \left(  t-t_{0}\right)  /\tau\right]  ^{\alpha},
\label{Vt}%
\end{equation}
with the ``feedback parameter''%
\begin{equation}
\alpha=-\epsilon\omega\left(  t_{0}\right)  . \label{alpha}%
\end{equation}

In Multifractal models\cite{Ma-63} the (logarithmic) price is assumed to
follow%
\begin{equation}
P\left(  t\right)  =B\left[  \Theta\left(  t\right)  \right]  , \label{PX}%
\end{equation}
where $B\left(  t\right)  $ is Brownian motion and $\Theta\left(  t\right)  $
is the random trading time, which is an increasing function of $t$.
Differentiating Eq.~(\ref{PX}) over $t$, we can represent the increment of
price in the form~$\Delta_{\tau}P\left(  t\right)  =a\left(  t\right)
\hat{\xi}\left(  t\right)  $ with $a\left(  t\right)  \sim\Theta^{\prime
}\left(  t\right)  $. Substituting ``classical trajectory''\ $a\left(
t\right)  $ from Eq.~(\ref{Vt}), we find
\begin{equation}
\Theta\left(  t\right)  -\Theta\left(  t_{0}\right)  \sim\left(
t-t_{0}\right)  ^{1+\alpha}, \label{vt}%
\end{equation}
and the mean square increment of the price is%
\begin{equation}
\overline{\left\langle \left[  P\left(  t\right)  -P\left(  t_{0}\right)
\right]  ^{2}\right\rangle }\sim\Theta\left(  t\right)  -\Theta\left(
t_{0}\right)  \sim\left(  t-t_{0}\right)  ^{2H} \label{Hrst}%
\end{equation}
with the local Hurst exponent
\begin{equation}
H=\left(  1+\alpha\right)  /2. \label{Hurst}%
\end{equation}
Expression~(\ref{Hrst}) is valid for any $t_{0}$ with current $H(t_{0})$.

\begin{figure}[tb]
\begin{center}
\includegraphics[
height=2.1378in,
width=2.7155in
]{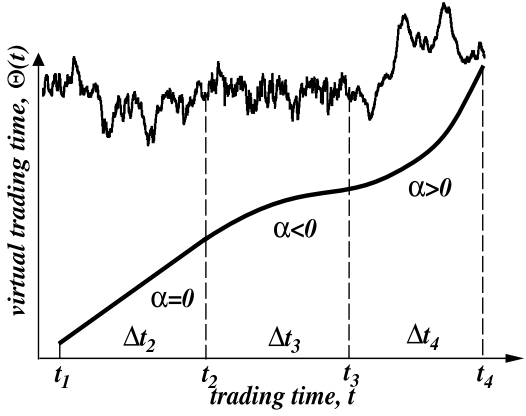}
\end{center}
\caption{Price fluctuations and corresponding virtual trading time
$\Theta\left(  t\right)  $ as functions of real time $t$ for Brownian motion
($\alpha=0$), sub- ($\alpha<0$) and super-diffusive behavior ($\alpha>0$). }%
\label{VTime}%
\end{figure}

The price $P\left(  t\right)  $ experiences different types of fractional
Brownian motion in time intervals $\Delta t_{i}$ with different feedback index
$\alpha$, which randomly change each other, see Fig.~\ref{VTime}. The case
$H\simeq1/2$ ($\alpha\simeq0$) describes usual Brownian motion. A Hurst
exponent value $0<H<1/2$ ($\alpha<0$) will exist for a time series with
sub-diffusive (anti-persistent) behavior. A Hurst exponent value from the
interval $1/2<H<1$ ($\alpha>0$) indicates super-diffusive\ (persistent)
behavior. $H\neq1/2$ can be interpreted as the result of local unbalance
between the competing liquidity providers and liquidity takers\cite{BJPW-03}.

The fractal dimension of the fractional Brownian motion is $D_{ext}=2-H$. At
large $H$ the motion becomes more regular ($D_{ext}\rightarrow1$), with large
up- and downturns, while at small $H$ it quickly fluctuates, trying to covers
the whole plane ($D_{ext}\rightarrow2$). Therefore, establishing of a super-
and sub-diffusive behavior leads to significant suppression/creation of
short-time fluctuations, see Fig.~\ref{VTime}. This effect was really
observed\cite{DSD-04}, and may be used as an indicator of establishment of
large volatility at long time scales, which is hard to detect at short time intervals.

\subsubsection{Brownian motion, sub- and super-diffusion\label{BFT}}

Switching of fluctuation regimes between Brownian motion ($\alpha\simeq0$),
sub- ($\alpha<0$) and super-diffusion ($\alpha>0$) is happened randomly at
``frustration times''\ (with equal probability of different choices) with the
probability (see Eq.~(\ref{P1V}))%
\begin{equation}
p\left(  \alpha\right)  =\dfrac{1}{\sqrt{2\pi}\sigma_{0}}\exp\left(
-\dfrac{\allowbreak\alpha^{2}}{2\sigma_{0}^{2}}\right)  ,\quad\sigma_{0}%
^{2}=2\epsilon\lambda^{2}. \label{s0}%
\end{equation}
Because of multifractality, such change of fluctuation regimes is happened at
all time scales $\tau$.

We also define a multivariate PDF
\begin{equation}
p\left(  \alpha_{0},\cdots,\alpha_{k}\right)  \equiv\overline{\prod
\nolimits_{l=0}^{k}\delta\left[  \alpha_{l}+\epsilon\omega\left(
t_{l}\right)  \right]  }, \label{p0k}%
\end{equation}
which determines information entropy conditional to the set of indexes
$\alpha_{0},\cdots,\alpha_{k}$:
\begin{equation}
S\left(  \alpha_{0},\cdots,\alpha_{k}\right)  \simeq\ln p\left(  \alpha
_{0},\cdots,\alpha_{k}\right)  . \label{S}%
\end{equation}
The entropy, Eqs.~(\ref{S}) and~(\ref{s0}), is maximal for Brownian motion,
$\alpha=0$, sub- and super-diffusion lower the entropy production, since the
market behavior is more predictable in these regimes.

Conditional dynamics of mode switching can be described by the probability
$p\left(  \alpha_{1}|\alpha_{0}\right)  =p\left(  \alpha_{0},\alpha
_{1}\right)  /p\left(  \alpha_{0}\right)  $ to find given value of the index
$\alpha_{1}$ at time $t_{1}=t_{0}+\Delta t_{1}$ under the condition that it
was $\alpha_{0}$ at previous time $t_{0}$. Calculating the average~(\ref{p0k})
at $k=1$ with Gaussian distribution function, Eq.~(\ref{Pw}), we find:
\begin{align}
p\left(  \alpha_{1}|\alpha_{0}\right)   &  =\dfrac{1}{\sqrt{2\pi}\sigma_{1}%
}\exp\left[  -\allowbreak\dfrac{\left(  \alpha_{1}-\bar{\alpha}_{1}\right)
^{2}}{2\sigma_{1}^{2}}\right]  ,\label{paa}\\
\bar{\alpha}_{1}  &  =\alpha_{0}h\left(  \Delta t_{1}\right)  ,\ \sigma
_{1}^{2}=\sigma_{0}^{2}\left[  1-h^{2}\left(  \Delta t_{1}\right)  \right]
\label{ac0}%
\end{align}
Conditional average $\bar{\alpha}_{1}$ decreases with time $\Delta t_{1}$,
while the conditional dispersion $\sigma_{1}$ grows with this time. The
transition to a new state is happened in average when two these amplitudes
become of the same order:%
\begin{equation}
\Delta t_{1}\simeq\tau_{0}\left(  \dfrac{\tau}{\tau_{0}}\right)
^{1/\sqrt{1+z}},\quad z=\dfrac{\alpha_{0}^{2}}{4\epsilon^{2}\lambda^{2}}.
\label{dt1}%
\end{equation}
It is surprising, that the length of the time interval $\Delta t_{1}$ grows
with the rise of $\left\vert \alpha_{0}\right\vert $ (although the probability
of large initial $\left\vert \alpha_{0}\right\vert $, Eq.~(\ref{s0}), is
small). Such counter-intuitive behavior is related to the absence of any
``restoring force''\ to $\alpha=0$. \begin{figure}[tb]
\begin{center}
\includegraphics[
height=2.0003in,
width=3in
]{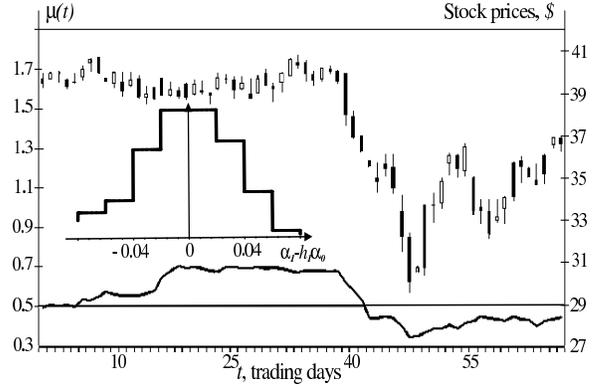}
\end{center}
\caption{Daily series of Exxon Mobil Corporation and corresponding $\mu\left(
t\right)  $\cite{DSD-04}. Probability distribution of $\alpha_{1}-h_{1}%
\alpha_{0}$ with $h_{1}=0.9$ at $\Delta t_{1}=16$ days is shown in Insert.}%
\label{Mu}%
\end{figure}

The characteristic time of sub- and super-diffusion behavior can be roughly
estimated substituting in the above expression the most probable value
$\alpha_{0}^{2}\simeq\sigma_{0}^{2}$ from Eq.~(\ref{s0}), giving
$z\simeq\frac{1}{2}\ln\left(  \tau_{0}/\tau\right)  $. For $\tau=1$ day,
$\tau_{0}\simeq10^{3}$ days and $\lambda^{2}\simeq0.1$ this gives $\Delta
t_{1}$ about a month, in agreement with empirical observations, see
Fig~\ref{Mu} and Ref.\cite{DSD-04}. The effect of sub- and super-diffusion
switching can be hardly detectable at small time intervals $\tau$, because of
small $\alpha$ (see Eq.~(\ref{alpha})), but it is well pronounced for daily
time intervals $\tau$.

The feedback index $\alpha$ is related to ``variation index''\ $\mu=\left(
1-\alpha\right)  /2$, introduced in Ref.\cite{DSD-04} from the fractal
analysis of empirical data (local extension of the R/S analysis\cite{Chaos} of
the Hurst exponent $H$). In Insert in Fig.~\ref{Mu} we plot estimated
probability distribution of $\alpha_{1}-h_{1}\alpha_{0}$ for $\Delta t_{1}=16
$ days, which is proportional to the conditional probability~(\ref{paa}). The
value $h_{1}$ is chosen to get a maximum at the beginning, and it decreases
with the rise of the time interval $\Delta t_{1}$ from $0.99$ at $\Delta
t_{1}=1$ to $0.9$ at $\Delta t_{1}=16$ days. The dispersion of this
distribution $\sigma_{0}\simeq0.2$, and from Eq.~(\ref{s0}) we get reasonable
estimation $\lambda^{2}\simeq0.08$. From second Eq.~(\ref{ac0}) we find the
dispersion $\sigma_{1}=\sigma_{0}\sqrt{1-h_{1}^{2}}\simeq0.07$, close to
observed value.

We can also calculate the probability to find the feedback index $\alpha_{k}$
at time $t_{k}$ under the condition that it was $\alpha_{k-1}$ at time
$t_{k-1}$, $\alpha_{k-2}$ at time $t_{k-2}$, and so on:%
\begin{align}
&  p\left(  \alpha_{k}|\alpha_{0},\cdots,\alpha_{k-1}\right) \nonumber\\
&  \equiv p\left(  \alpha_{0},\cdots,\alpha_{k}\right)  /p\left(  \alpha
_{0},\cdots,\alpha_{k-1}\right) \label{pk}\\
&  =\dfrac{1}{\sqrt{2\pi}\sigma_{k}}\exp\left[  -\allowbreak\dfrac{\left(
\alpha_{k}-\bar{\alpha}_{k}\right)  ^{2}}{2\sigma_{k}^{2}}\right]  ,\nonumber
\end{align}
The logarithm of this probability determines the entropy lowering because of
the knowledge about previous events $\alpha_{0},\cdots,\alpha_{k}$. The
conditional average $\bar{\alpha}_{k}$ corresponds to the maximum of entropy
production, Eq.~(\ref{S}). It has the meaning of average response of $\alpha$
on previous values $\alpha_{0},\cdots,\alpha_{k-1}$:
\begin{equation}
\bar{\alpha}_{k}=\sum\nolimits_{i=0}^{k-1}K_{ki}\alpha_{i},\quad
K_{ki}=-a_{ki}/a_{kk}, \label{ac}%
\end{equation}
$a_{kj}$ are adjoints of the matrix $\mathbf{h}$ with elements $h\left(
t_{i}-t_{j}\right)  $. Conditional dispersion
\begin{equation}
\sigma_{k}^{2}=\sigma_{0}^{2}\det\mathbf{h/}a_{kk} \label{sk}%
\end{equation}
estimates the accuracy of the prediction~(\ref{ac}). Probabilities~(\ref{pk})
depend not only on index $\alpha_{k-1}$ at previous time $t_{k-1}$, but also
on all $\alpha_{0},\cdots,\alpha_{k-1}$ -- the random process is not
Markovian. As the consequence, the probability to have the same value of all
three indexes $\alpha_{2}=\alpha_{1}=\alpha_{0}$ (continuation of a sub- and
super-diffusive regimes) grows with increase of the initial time interval
$\Delta t_{1}$.

\subsubsection{Fluctuation corrections\label{TREND}}

The ``classical trajectory''~(\ref{Vt}) predicts that the feedback index
$\alpha$~(\ref{alpha}) can be changed only because of fluctuations. In this
section we demonstrate, that fluctuations lead to an additive shift of
$\alpha$ in the super-diffusion direction. Calculating the average of
$V_{q}\left(  t\right)  $ with Gaussian weight~(\ref{Pw}) under the condition
$\omega\left(  0\right)  =\omega_{0}$, we find:%
\begin{align}
\left\langle V_{q}\left(  t\right)  \right\rangle _{\omega_{0}}  &
\simeq\left\langle a^{q}\left(  t\right)  \right\rangle _{\omega_{0}}%
\simeq\left\langle V_{q}\left(  0\right)  \right\rangle _{\omega_{0}}\left(
t/\tau\right)  ^{q\alpha+q^{2}\beta\left(  t\right)  },\nonumber\\
\beta\left(  t\right)   &  =\lambda^{2}\left[  1+h\left(  t\right)  \right]
/2, \label{fluct}%
\end{align}
where $h\left(  t\right)  $ is defined in Eq.~(\ref{ht}). The $\beta$-term
describes deviation from fractional Brownian motion because of multifractal
behavior. At $q=2$ this expression can be interpreted as the response on
``endogenous event''\ $\omega\left(  0\right)  $\cite{SoMaMu-02}, while at
$q=1 $ it gives fluctuation correction to the feedback index $\alpha$.

Empirical study of year correlations shows\cite{WB-03}, that financial market
is really ``locked''\ in sub- and super-diffusive or Brownian motion states at
extremely long periods (conventions can persist up to 30 years). The change in
convention can be rather smooth, like during the second part of the century,
or occur suddenly, triggered by an extreme event, like it did after 1929. From
the data, it was also observed a systematic bias towards the persistent
following convention.

\subsection{Universality of fluctuations\label{CONCL3}}

Non-universal properties of market at trading times $\lesssim\tau_{k}$ can
only be described by models of agent-based strategies. In this section we
consider only universal properties of price fluctuations at time scales
$\tau>\tau_{k}\simeq1\min$, see Ref.\cite{NP-06}. At qualitative level the
presence of universality is known for a long time as ``stylized
facts''\cite{Styl}. We show, that our approach captures such stylized facts
and gives explanation for many others empirical observations.

The universality is related to self-similiarity of price fluctuations at
different time scales\cite{Intro-01,MaAsDa-03,BaDeMu-01}: the change of time
interval $\tau$ corresponds to the change of characteristic scale along the
hierarchical tree in Fig.~\ref{Tree}. We demonstrate, that resulting time
series have complex non-periodic behavior with chaotic changes of usual
Brownian motion, sub- and super-diffusion, reflecting cyclic dynamics of the
market. We show, that the impact function of the market logarithmically
depends on volume imbalance.

Fluctuations on financial market have unexpected physical interpretation,
reflecting the unified nature of physics. The effective Hamiltonian~(\ref{Pw})
can be rewritten as%
\[
H\left\{  \omega\right\}  =\dfrac{\eta}{2\pi}\iint\left(  \dfrac{\omega\left(
t\right)  -\omega\left(  t^{\prime}\right)  }{t-t^{\prime}}\right)
^{2}dtdt^{\prime}.
\]
This expression describes diffusion of quantum Brownian particle with the
coordinate $\omega\left(  t\right)  $ and the coefficient of linear friction
$\eta=1/\left(  2\pi\lambda^{2}\right)  $.

A microscopic model of quantum diffusion is based on coupling to a termostat
-- the reservoir of harmonic oscillators\cite{CaLe-83}, presenting the
``army''\ of traders in the case of the market. The resulting dynamics is
intrinsically non-Markovian in that the evolution depends on history rather
than just on present state\cite{HPZ-92}. Brownian particle can respond to a
very wide range of reservoir frequencies, and this is the origin of
time-irreversive behavior and slow relaxation after fluctuation cast, see
Eq.~(\ref{wt}). The production of information entropy (see Eq.~(\ref{S}) and
Appendix~\ref{THERMO}) is related to enviroment-induced decoherence of the
quantum particle\cite{PZ-86}, and it is at the peak of many recent studies.

\section{Conclusion\label{CONCL}}

Many concepts of equilibrium macroeconomic (resources, unemployers, different
firm dynamics at small- and long- time horizons, taxes and so on) naturally
enter into proposed coalescent theory, which integrates both physical and
economic concepts of essentially non-equilibrium market in one unique
approach. We also developed new approach to study fluctuations on the market,
well describing empirical data of both firm grow rates and price increments on
financial markets. We propose the set of Langeven equations, describing
multi-time dynamics of price and volume fluctuations at different time scales
on the market. Using these equations, we derived analytically equations of
multifractal random walk model.

In the end we discuss physical meaning of our theory:

a) What is physics of extreme events on the market (problem of fat tails)?

There are two sources of fluctuations: macroeconomic events -- news, and
traders activity because of uncertainty of equilibrium prices at different
time scales, which generate two types of large price jumps: News jumps are
created by the inflow of news, while stock jumps are generated during random
concurrence of different fluctuation modes. We show, that the decay of
volatility observed after news jumps is related to the effect, similar to
``aging''\ effect in spin glasses.

Fluctuations on the market are characterized by the normalized noise
$\mathbf{\xi}$ and its amplitude $\mathbf{a}$ (volatility). New key idea of
our approach is that $\mathbf{\xi}$ and $\mathbf{a}$ are independent complex
random variables, separated on time scale: the noise is generated by hot
degrees of freedom on times small with respect to observation time interval
$\tau$, while evolution of the amplitude is determined by cold degrees of
freedom on times large with respect to $\tau$.

In stochastic volatility and multifractal models jumps are predicted as the
result of volatility fluctuations, and characterized by large non-universal
tail exponent $\mu\gg1$, while the noise is assumed to be Gaussian
uncorrelated random variable. In fact, the noise is strongly correlated, and
can experience large non Gaussian jumps. We calculate the contribution of such
jumps to PDF and show, that the distribution of stock jumps is characterized
by the tail exponent $\mu=3$, while the distribution of news jumps has tail
exponent $\mu=2$. The exponent $\mu$ remains stable with the rise of $\tau$
(recall, that Levy distribution with $\mu>2$ is unstable).

b) Why market dynamics is so complex: can it be described by simple Markovian
or Gaussian processes?

We show, that local equilibriums on the market are self-organized in the
hierarchical tree, according to their relaxation times. The amplitude
$\mathbf{a}$ of the noise at given time scale is determined by cumulative
signal from all ``parent''\ time scales, and its dynamics is complex
multifractal process. But the information about the amplitude can be
``erased''\ from time series considering only signs of price fluctuations. The
resulting Markovian process describes propagation of positive and negative
signals, and determines conditional double dynamics of the market.

Typical noise $\mathbf{\xi}$ and amplitude $\mathbf{a}$ are determined by
signals from large number of, respectively, short and long (with respect to
$\tau$) time scales, and they have asymptotically Gaussian statistics. We
propose and solve Double Gaussian model of market fluctuations, and show good
agreement with empirical data for different groups of stocks.

c) What physics stands behind ``random trading time''\ in Multifractal
models\cite{Ma-63}?

At large time intervals the price randomly cycles between Brownian motion,
sub- and super-diffusive regimes, which change each other because of liquidity
fluctuations. The virtual trading time is proportional to the real time for
Brownian motion and experiences time shifts in sub- and super-diffusive
regimes. The theory predicts systematic bias to persistent behavior, observed
for many markets and exchange rates.

d) And finally, can price behavior be described by universal physical lows or
it is dictated only by the zoo of microstructures of markets (see
Refs.\cite{Ho-97,GiBo-03})?

The universality of price fluctuations on financial markets was demonstrated
at time scales from a minute to tenths years in many studies, see for example,
Refs.\cite{SGPS-02,BoMa-06,Le3-06}. We show, that it is related to the
self-similarity of the underlying hierarchical tree of amplitudes, see
Fig.~\ref{Tree} (we do not give here lists of all stocks, used for comparison
with our theory, since they are shown in corresponding references). In
contrast, statistics of trades and volumes is not universal, and strongly
depends on details of market microstructure.

Our theory can also be used to study other time series, such as variations of
cloudiness, temperature, earthquake frequencies, rate of traffic flow and so
on. It looks attractive to apply analytical approach of this paper for the
description of social processes, which are driven by frustrations at turning
points of the mankind history. Events between these points support the social
activity, but do not change the state of the society.

\medskip\textbf{Acknowledgement} \emph{I would like to thank Andrey Leonidov
for attracting my attention to the problem of market fluctuations and helpful
notes, M. Dubovikov for discussion of some results, and J.-P. Bouchaud for
critical comments.}

\appendix{}

\section{Entropy formulation\label{THERMO}}

In order to reveal the economic meaning of Master equation~(\ref{dxdt}), it is
convenient to rewrite it in the form%
\begin{equation}
\dfrac{dG}{dt}=p_{c}\left(  G\right)  -p_{d}\left(  G\right)  , \label{pqG}%
\end{equation}
where $p_{c}\left(  G\right)  $ and $p_{d}\left(  G\right)  $ are
probabilities of job creation and destruction per unit time in the firm of $G$
people. In the absence of any external supply, $U=0$, the probability of job
creation is zero. In the main order in ``concentration''\ $U$ (we use physical
term to emphasize the analogy with coalescence) $p_{c}\left(  G\right)  $ is
proportional to $U$, while the probability of job destruction $p_{d}\left(
G\right)  $ is determined mainly by internal firm structure, and do not depend
of $U$. Comparing Eqs.~(\ref{pqG}) and~(\ref{dxdt}) we find explicit
expressions for these probabilities for our model
\begin{equation}
p_{c}\left(  G\right)  =qUG,\qquad p_{d}\left(  G\right)  =qU_{\ast
}G+pG^{1-\beta}. \label{pdg}%
\end{equation}

We define the ``entropy''\ $S\left(  G\right)  $ of the firm of size $G$ as
logarithm of equilibrium distribution function of firms over their sizes,
$f_{eq}\left(  G\right)  $. Since equilibrium values do not depend of a way
how the system is assembling, consider the process when the size $G$ is
varying by one. In this case $f_{eq}\left(  G\right)  $ is determined by the
detailed balance condition, $p_{c}\left(  G\right)  f_{eq}\left(  G\right)
=p_{d}\left(  G+1\right)  f_{eq}\left(  G+1\right)  $. The solution of this
equation at $G\gg1 $ relates firm entropy with probabilities of job creation
and destruction:%
\begin{equation}
f_{eq}\left(  G\right)  =e^{S\left(  G\right)  },\qquad S\left(  G\right)
=\int dG\ln\frac{p_{c}\left(  G\right)  }{p_{d}\left(  G\right)  }. \label{SG}%
\end{equation}
As one can naively expect, the entropy of the firm increases with the rise of
the probability to get a job and decreases with the rise of the probability to
loose it. In the case of overheated market $U<U_{\ast}$ the entropy $S\left(
G\right)  $ monotonically grows with the firm size $G$, while in the
``supersaturated''\ case $U>U_{\ast}$ it initially decreases with $G$,
reaching its minimum for firms of critical size, $G=G_{c}$, Eq.~(\ref{Gc}).

Calculating this integral~(\ref{SG}) with functions~(\ref{pdg}), we find%
\[
S\left(  G\right)  \simeq\mu G-G\ln\left[  U_{\ast}/U_{0}+\left(  G/e\right)
^{-\beta}\right]  +const,
\]
where $U_{0}=p/g$ and $\mu=\ln\left(  U/U_{0}\right)  $ has the meaning of
chemical potential. The entropy of the whole market
\begin{equation}
S=-U\ln\frac{U}{eU_{0}}+\int dGS\left(  G\right)  f\left(  G,t\right)
-\mu\left(  Q-U\right)  \label{SUG}%
\end{equation}
is the sum of the entropy of the ``ideal gas''\ of unemployments, the entropy
of all firms and the term $\sim\mu$, which takes into account the supply of
external resources~(\ref{Qf}). By analogy with thermodynamics, there is
maximum principle for the entropy: maximizing it with respect to $U$ we
reproduce the balance condition~(\ref{Qf}). Using Eqs.~(\ref{SG})
--~(\ref{SUG}) one can check that in the case when $Q\left(  t\right)  $ is
not (quickly) decreasing function, the market entropy always increases with
time%
\begin{equation}
dS/dt>0. \label{Sp}%
\end{equation}
Since the variation of entropy $\Delta S$ is opposite to the variation of
information, $\Delta I=-\Delta S$, Eq.~(\ref{Sp}) means, that the activity of
the market leads to ``erasing''\ of\ initial information -- the effect, well
known for some ``laundering''\ schemes.

\section{Solution of coalescence equations\label{SLEZOV}}

To find PDF of coalescent model~(\ref{dxdt}) we use the method of
Ref.\cite{LP}. We define dimensionless time $\tau$ and introduce the function
$u\left(  \tau\right)  $:%
\[
\tau=\ln\left[  G_{c}\left(  t\right)  /G_{c}\left(  t_{0}\right)  \right]
,\qquad u\left(  \tau\right)  =G\left(  t\right)  /G_{c}\left(  t\right)  ,
\]
where $t_{0}$ is coalescent time. In new variables the Master
equation~(\ref{dxdt}) takes the form%
\begin{equation}
du/d\tau=v\left(  u\right)  =\gamma\left(  \tau\right)  \left(  u-u^{1-\beta
}\right)  -u, \label{vu}%
\end{equation}
where%
\begin{equation}
\gamma\left(  \tau\right)  =p\frac{dt}{G_{c}^{\beta-1}dG_{c}}. \label{gam}%
\end{equation}
The balance equation~(\ref{Qt}) can only be satisfied if the plot of function
$v\left(  u\right)  $ lays below the axis $u$, and touches it at one point
$u=u_{0}$. Such locking point $u=u_{0}$, $\gamma=\gamma_{0}$, for
Eq.~(\ref{vu}) is determined by equations%
\[
v\left(  u_{0}\right)  =0,\ dv\left(  u_{0}\right)  /du_{0}=0,\ d^{2}v\left(
u_{0}\right)  /du_{0}^{2}<0,
\]
and we find that $u_{0}\rightarrow\infty$ and $\gamma_{0}=1$.

From Eq.~(\ref{gam}) we get that the critical size $G_{c}\left(  t\right)  $
grows and ``supersaturation''\ $\Delta\left(  t\right)  $ (we use physical
terms here) decreases with time as
\begin{equation}
G_{c}\left(  t\right)  =\left(  \dfrac{\beta qt}{\gamma_{0}}\right)
^{1/\beta},\qquad\Delta\left(  t\right)  =\frac{\gamma_{0}}{q\beta t}.
\label{dt}%
\end{equation}

PDF of firms can be rewritten through PDF of variables $u$ and $\tau$:
$f\left(  G,t\right)  =\varphi\left(  u,\tau\right)  /G_{c}\left(  t\right)
$, and neglecting the diffusion inflow of new firms we find%
\begin{equation}
\frac{\partial\varphi}{\partial\tau}+\frac{\partial}{\partial u}\left[
v_{0}\left(  u\right)  \varphi\right]  =0, \label{df}%
\end{equation}
where the velocity $v_{0}\left(  u\right)  =du/d\tau=-u^{1-\beta}$ is given by
Eq.~(\ref{vu}) with $\gamma=\gamma_{0}$.

General solution of Eq.~(\ref{df}) is%
\begin{equation}
\varphi\left(  u,\tau\right)  =u^{\beta-1}\chi\left[  \tau-\tau\left(
u\right)  \right]  ,\qquad\tau\left(  u\right)  =-u^{\beta}/\beta\label{phi}%
\end{equation}
with arbitrary function $\chi\left(  \tau\right)  $. To find $\chi\left(
\tau\right)  $ substitute this expression into the balance equation~(\ref{Qf})
and~(\ref{Qt}) with $Q\left(  t\right)  \gg U\left(  t\right)  $:%
\begin{equation}
Q_{0}\left(  \frac{\gamma_{0}}{\beta h}\right)  ^{m}G_{c}^{m-1}\left(
t_{0}\right)  e^{\tau\left(  \beta m-1\right)  }=\int_{0}^{\infty}%
duu\varphi\left(  u,\tau\right)  . \label{nrm}%
\end{equation}

This condition can be satisfied only if the function $\chi$ has the form
$\chi\lbrack\tau-\tau\left(  u\right)  ]=Ae^{\left(  \beta m-1\right)  \left[
\tau-\tau\left(  u\right)  \right]  }$. Substituting this function into
Eq.~(\ref{phi}), we find%
\begin{equation}
\varphi\left(  u,\tau\right)  =A\left(  1-\beta m\right)  ^{-1}e^{\left(
\beta m-1\right)  \tau}dF\left(  u\right)  /du, \label{phi1}%
\end{equation}
with%
\begin{equation}
F\left(  u\right)  =e^{-\left(  1/\beta-m\right)  u^{\beta}},\quad A\simeq
Q_{0}\left(  \frac{\gamma_{0}}{\beta p}\right)  ^{m}G_{c}^{m-1}\left(
t_{0}\right)  . \label{Pu1}%
\end{equation}

\section{Macroeconomic interpretation\label{MACRO}}

To get better understanding of coalescent model, consider its microeconomic
interpretation. Optimal firm size, $G=G_{c}$, is determined from the maximum
of the profit function%
\begin{equation}
\pi\left(  G\right)  =Py\left(  G\right)  -wG, \label{pi}%
\end{equation}
where $y\left(  G\right)  $ is the number of units produced by firm of $G$
peoples, $P$ is the price of one unit, and $w$ is the average wage per one
man. The technology is usually characterized by the standard Cobb-Douglas
function $y\left(  G\right)  =K^{\eta}G^{1-\eta}$, where $K$ is the firm
capital and the exponent $\eta>0$. Both the price $P$~and the capital $K$~are
reduced to initial time. Maximizing the profit function~(\ref{pi}) we get
$G_{c}\sim w^{-1/\eta}$.

Variation of wages $w\left(  t\right)  $ with the time is determined by the
Fillips low\cite{Ph-E-58}:%
\begin{equation}
\frac{1}{w\left(  t\right)  }\frac{dw\left(  t\right)  }{dt}\simeq a\left[
U_{\ast}-U\left(  t\right)  \right]  =-a\Delta\left(  t\right)  , \label{Fil}%
\end{equation}
with positive constant $a>0$. Substituting expression~(\ref{dt}) for
$\Delta\left(  t\right)  $ in Eq.~(\ref{Fil}), we find its solution $w\left(
t\right)  =const\times t^{-\zeta}$ with $\zeta=a/\left(  \beta q\right)  $.
Substituting this function into $G_{c}\sim w^{-1/\eta}$, we get the optimal
firm size $G_{c}\sim t^{\zeta/\eta}$. Comparing this dependence with
Eqs.~(\ref{Gc}) and~(\ref{dt}), we find the Fillips parameter in
Eq.~(\ref{Fil}): $a=\eta q$. We conclude, that coalescent approach is
consistent with the maximum profit principle and the Fillips low.

Notice that while the parameter $q=a/\eta$ of Master equation~(\ref{pqG}%
),~(\ref{pdg}) is determined by technology ($\eta$) and market structure
($a$), economic analysis do not impose any restrictions on the second
parameter $p$ of Master equation. Therefore, $p$ can depend on management
ability of firm head, relations between firm staff, industry shocks and so on,
and can experience strong random fluctuations $\Delta p$. This observation
explains the empirical fact, that there is much more variance in job
destruction than in job creation time series\cite{DaHaSh-96} (as was noted by
Lev Tolstoy: all fortunate families are happy alike -- each unfortunate family
is unhappy in own way).

\section{PDF of Double Gaussian model\label{PDF}}

It is convenient in Eq.~(\ref{b}) to use instead of $\left\{  \mathbf{a}%
_{i}^{0}\right\}  $ Gaussian random variables $\left\{  \mathbf{\alpha}%
_{i}\right\}  $:%
\[
\mathbf{a}_{i}^{0}=c_{i}\mathbf{\alpha}_{i},\quad\mathbf{\varepsilon}_{i}%
=\sum\nolimits_{j}c_{ij}\mathbf{\alpha}_{j},
\]
such as $\overline{\alpha_{i}^{2}}=1,\overline{\left(  \mathbf{\alpha}%
_{1},\mathbf{\alpha}_{2}\right)  }=\nu$. Averaging over fluctuating Gaussian
variables $\mathbf{\xi}_{i}^{0}$ and $\mathbf{\alpha}_{i}$, we get general
expression for inverse Fourier component of PDF:%
\begin{align*}
G^{-1}\left(  k,p\right)   &  =1+k^{2}\sigma_{11}/2+p^{2}\sigma_{22}%
/2+kp\sigma_{12}+\\
&  \left(  1-\nu^{2}\right)  \left(  \varkappa_{11}k^{2}+\varkappa_{22}%
p^{2}+\varkappa_{12}kp\right)  ^{2}%
\end{align*}
with%
\begin{align*}
\sigma_{11}  &  =c_{1}^{2}+c_{12}^{2}+c_{11}^{2}+2\nu c_{11}c_{12},\\
\sigma_{22}  &  =c_{2}^{2}+c_{21}^{2}+c_{22}^{2}+2\nu c_{22}c_{21},\\
\sigma_{12}  &  =c_{1}c_{21}+c_{2}c_{12}+\nu\left(  c_{1}c_{22}+c_{2}%
c_{11}\right)  ,\\
\varkappa_{11}  &  =c_{1}c_{12},\quad\varkappa_{22}=c_{2}c_{21},\\
\varkappa_{12}  &  =c_{1}c_{2}-c_{11}c_{22}+c_{12}c_{21}.
\end{align*}
An important relation between elements of matrixes $\mathbf{\sigma}$ and
$\mathbf{\varkappa}$ follows from the condition of stationarity of PDF, which
leads to physical constraint $G\left(  k,0\right)  =G\left(  0,k\right)  $ on
Fourier components of univariate PDFs of $\Delta_{\tau}P\left(  t\right)  $
and $\Delta_{\tau}P\left(  t+\tau\right)  $, and we get $\sigma_{11}%
=\sigma_{22}=\sigma^{2},\varkappa_{11}^{2}=\varkappa_{22}^{2}$. Here $\sigma$
is the dispersion of price fluctuations,%
\begin{align}
\overline{\left\langle \Delta_{\tau}P^{2}\left(  t\right)  \right\rangle }  &
=\overline{\left\langle \Delta_{\tau}P^{2}\left(  t+\tau\right)  \right\rangle
}=\sigma^{2},\nonumber\\
\overline{\left\langle \Delta_{\tau}P\left(  t\right)  \Delta_{\tau}P\left(
t+\tau\right)  \right\rangle }  &  =\sigma_{12}=\varepsilon\sigma^{2}.
\label{ppe}%
\end{align}
We conclude, that in addition to $\sigma$ and $\varepsilon$, our model is
characterized by dimensionless constant $\nu$ and the angle $\phi$:%
\begin{equation}%
\begin{array}
[c]{c}%
G^{-1}\left(  k,p\right)  =1+\left[  \left(  k^{2}+p^{2}\right)
/2+\varepsilon kp\right]  \sigma^{2}+\left(  1-\nu^{2}\right)  \times\\
\dfrac{\sigma^{4}}{4}\left[  \dfrac{k^{2}-p^{2}}{2}\sin\left(  2\phi\right)
/2-kp\cos\left(  2\phi\right)  \right]  ^{2}.
\end{array}
\label{G-1}%
\end{equation}

The quadratic part of this expression can be diagonalized by changing
variables, $K=k\cos\psi-p\sin\psi^{\prime},P=k\sin\psi+p\cos\psi^{\prime}$,
with $\psi^{\prime}=\phi-\frac{1}{2}\arcsin\varepsilon,\psi=\phi+\frac{1}%
{2}\arcsin\varepsilon$. In new variables Eq.~(\ref{G-1}) takes the form%
\[%
\begin{array}
[c]{c}%
G^{-1}\left(  k,p\right)  =1+\dfrac{\sigma^{2}}{2}\left(  K^{2}+P^{2}\right)
+\\
\dfrac{1-\nu^{2}}{(1-\varepsilon^{2})^{2}}\dfrac{\sigma^{4}}{4}\left[
KP-\tau\left(  K,P\right)  \right]  ^{2},
\end{array}
\]
with%
\[%
\begin{array}
[c]{c}%
\tau\left(  K,P\right)  =\dfrac{\varepsilon}{2}\cos\left(  2\phi\right)
\left(  K^{2}+P^{2}\right)  +\dfrac{1}{2}\left(  1-\sqrt{1-\varepsilon^{2}%
}\right) \\
\times\sin\left(  2\phi\right)  \left[  2KP\sin\left(  2\phi\right)  +\left(
K^{2}-P^{2}\right)  \cos\left(  2\phi\right)  \right]  \allowbreak.
\end{array}
\]
Eq.~(\ref{G-1}) can be simplified if we note, that correlations of price
increments, Eq.~(\ref{ppe}), are always very small, $\left\vert \varepsilon
\right\vert \ll1$. In polar coordinates $K=\left\vert \mathbf{K}\right\vert
\cos\varphi,P=\left\vert \mathbf{K}\right\vert \sin\varphi$ we have in the
main order in $\varepsilon$%
\[%
\begin{array}
[c]{c}%
\left[  KP-\tau\left(  K,P\right)  \right]  ^{2}\simeq\frac{1}{4}\left\vert
\mathbf{K}\right\vert ^{4}\left[  \sin\left(  2\varphi\right)  -\varepsilon
\cos\left(  2\phi\right)  \right]  ^{2}\\
\simeq\frac{1}{4}\left\vert \mathbf{K}\right\vert ^{4}\sin^{2}\left[  2\left(
\varphi-\frac{\varepsilon}{2}\cos\left(  2\phi\right)  \right)  \right]
=\left(  K^{\prime}P^{\prime}\right)  ^{2},
\end{array}
\]
where we introduced new orthogonal rotated coordinate system $K\simeq
K^{\prime}-\left(  \varepsilon/2\right)  P^{\prime}\cos\left(  2\phi\right)
,P\simeq P^{\prime}+\left(  \varepsilon/2\right)  K^{\prime}\cos\left(
2\phi\right)  $. Changing integration variables $\left(  k,p\right)
\rightarrow\left(  K^{\prime},P^{\prime}\right)  $ in the integral~(\ref{Pint}%
) in the main order in the small parameter $\varepsilon$ we get
Eq.~(\ref{Pppp}), where angles $\phi_{+}$ and $\phi_{-}$ are defined by
$\phi_{-}=\phi-\varepsilon\cos^{2}\phi,\phi_{+}=\phi+\varepsilon\sin^{2}\phi$,
and functions $\mathcal{P}_{l}\left(  x\right)  $ are determined by
Eqs.~(\ref{Pl}) and~(\ref{Pxt}).

\section{PDF of volatility fluctuations\label{VOLAT}}

PDF of the volatility variable $V_{1}\left(  t\right)  $~(\ref{Volat}) at
$q=1$ can be expressed through the $n$-point PDF $\mathcal{P}\left\{  \Delta
P_{k}\right\}  $ of variables $\Delta P_{k}$,%
\begin{equation}
P_{n}\left(  V_{1}\right)  \equiv\prod_{k=1}^{n}\int\limits_{0}^{\infty
}d\Delta P_{k}\delta\left(  V_{1}-\dfrac{1}{n}\sum_{k=1}^{n}\left\vert \Delta
P_{k}\right\vert \right)  \mathcal{P}\left\{  \Delta P_{k}\right\}  \label{pV}%
\end{equation}
Asymptotes of $P_{n}\left(  V_{1}\right)  $ can be found both for $V_{1}%
\ll\sigma$ and for $V_{1}\gg\sigma$. At $V_{1}\ll\sigma$ only small
$\left\vert \Delta P_{k}\right\vert \ll\sigma$ contribute to the
integral~(\ref{pV}) and we have $P_{n}\left(  V_{1}\right)  \sim V_{1}^{n-1}$.
In the opposite case of large $V_{1}\gg\sigma$ the integral is dominated by
the power tail of PDF $\mathcal{P}\left\{  \Delta P_{k}\right\}  $ with
typical $\left\vert \Delta P_{k}\right\vert \sim V_{1}\gg\sigma$. From
dimension consideration we find for such $\Delta P_{k}$ that $\mathcal{P}%
\left\{  \Delta P_{k}\right\}  \sim V_{1}^{-n-\mu}$, where $\mu$ is the
exponent of one-point PDF, Eq.~(\ref{pow}), and the integral~(\ref{pV}) is
estimated as$\mathcal{\ }$ $P_{n}\left(  V_{1}\right)  \sim V_{1}^{-1-\mu}$.

Both these limits are matched by the function%
\begin{align}
P_{n}\left(  V_{1}\right)   &  =V_{0}^{-1}N_{0}^{-1}f\left(  V_{1}%
/V_{0}\right)  ,\label{Pn}\\
f\left(  z\right)   &  =z^{-1}\left(  z^{-n/s}+z^{\mu/s}\right)
^{-s}\nonumber
\end{align}
with $V_{0}\sim\sigma$. The dependence of a new parameter $s>0$ on $n$ will be
found later from the condition that at large $n$ the distribution
$P_{n}\left(  V_{1}\right)  $ should not depend of $n$. Momentums of this
distribution $\left\langle V_{1}^{k}\right\rangle =V_{0}^{k}N_{k}/N_{0}$ are
determined by normalization integrals,%
\[%
\begin{array}
[c]{c}%
N_{k}=\int\limits_{0}^{\infty}z^{k}f\left(  z\right)  dz=mB\left[  m\left(
n+k\right)  ,m\left(  \mu-k\right)  \right]  ,\\
m=s/\left(  n+\mu\right)  ,
\end{array}
\]
where $B$ is the Beta-function.

The function $f\left(  z\right)  $~(\ref{Pn}) reaches its maximum at $z_{\max
}=\left(  n-1\right)  ^{m}/\left(  \mu+1\right)  ^{m}$. The central part of
the distribution is obtained by expanding the probability $P_{n}\left(
V_{1}\right)  $ over $\ln\left(  V_{1}/V_{\max}\right)  $ near its maximum at
$V_{\max}=z_{\max}V_{0}$, and it has log-normal form:%
\begin{equation}
\ln P_{n}\left(  V_{1}\right)  =const-\dfrac{1}{2}\dfrac{n-1}{s}\left(
\mu+1\right)  \allowbreak\left(  \ln\dfrac{V_{1}}{V_{\max}}\right)  ^{2}.
\label{lnn}%
\end{equation}
Since the distribution $P_{n}\left(  V_{1}\right)  $~(\ref{lnn}) should not
depend on $n$ at large $n$, we find $s=c\left(  n-1\right)  $ with certain
constant $c$. In the limit $n\rightarrow\infty$ $P_{n}\left(  V_{1}\right)  $
becomes universal function of $V_{1}/V_{\max}$, Eq.~(\ref{pVu}) with $q=1$.
Repeating our calculations for general $q>0$, we find that it is given by the
substitution $\mu\rightarrow\mu/q$ and $n\rightarrow$ $n/q$ in the above expressions.


\begin{thebibliography}{999}                                                                                              %


\bibitem {Gibrat}R. Gibrat, Lesin' egalit'es' economiques, Sirey, Paris (1931).

\bibitem {Pe-PRE-96}R. Perline, \emph{Phys. Rev. E} \textbf{54,} 220 (1996).

\bibitem {LS-96}M. Levy, S. Solomon, \emph{Int. J. Mod. Phys. C} \textbf{7},
595 (1996); D. Sornette, R. Cont, \emph{J. Phys. I} (France) \textbf{7}, 431 (1997).

\bibitem {Am-PRL-98}L.A.N. Amaral, S.V. Buldyrev, S. Havlin, M.A. Salinger,
and H.E. Stanley, \emph{Phys. Rev. Lett.} \textbf{80}, 1385 (1998).

\bibitem {Le-PRL-98}Y. Lee, L.A.N. Amaral, D. Canning, M. Meyer, and H.E.
Stanley, \emph{Phys. Rev. Lett.} \textbf{81}, 3275 (1998).

\bibitem {Su-Ph-02}G. Sutton, \emph{Physica A,} \textbf{312}, 577 (2002).

\bibitem {Sa-80}P.A. Samuelson, \emph{Economics.} Mc-Graw Hill Int., Auckland (1980).

\bibitem {Cha-95}B.K. Chakrabarti, S. Marjit, Indian \emph{J. Phys.} \emph{B
}\textbf{69}, 681 (1995); S. Ispolatov, P.L. Krapivsky, S. Redner, \emph{Eur.
Phys. J.} \textbf{B2}, 267 (1998).

\bibitem {Dra-00}A.A. Dr\u{a}gulescu, V.M. Yakovenko, \emph{Eur. Phys.
J.,}\textbf{\ B17}, 723 (2000).

\bibitem {Cha-04}A. Chatterjee, B.K. Chakrabarti, S.S. Manna, \emph{Physica}
\emph{A, }\textbf{335}, 155 (2004).

\bibitem {Sl-04}F. Slanina, \emph{Phys. Rev.} \emph{E,}\textbf{\ 69}, 046102 (2004).

\bibitem {Ra-Ph-00}J.J. Ramsden and Gy. Kiss-Hayp'al, \emph{Physica}
\emph{A,}\textbf{\ 277,} 220 (2000).

\bibitem {DY-01}A. Dr\u{a}gulescu, V.M. Yakovenko, \emph{Physica A}
\textbf{299,} 213 (2001); V.M. Yakovenko,\texttt{\ arXiv:physics/0709.3662v3} [physics.soc-ph].

\bibitem {St-JPh-97}L.A.N. Amaral, S.V. Buldyrev, S. Havlin, H. Leschhorn, P.
Maass, M. A. Salinger, H. E. Stanley, and M. H. R. Stanley, \emph{J. Phys. I
France,} \textbf{7}, 621 (1997); S. V. Buldyrev, L.A.N. Amaral, S. Havlin, H.
Leschhorn, P. Maass, M.A. Salinger, H.E. Stanley, and M.H.R. Stanley, \emph{J.
Phys. I France,} \textbf{7}, 635 (1997).

\bibitem {Fr--68}M. Friedman, \emph{Amer. Econ. Rev.}, \textbf{58,} 8 (1968).

\bibitem {Ax-01}R. Axtell, \emph{Science}, \textbf{293}, 1818 (2001).

\bibitem {Ga-03}X. Gabaix, P. Gopikrishnan, V. Plerou and H.E. Stanley,
\emph{Nature}, \textbf{423}, 267 (2003).

\bibitem {PuHa-04}D. Pushkin and A. Hassan, \emph{Physica A}, \textbf{336},
571 (2004).

\bibitem {Soo-02}Soo, K.T., Zipf's law for cities: a cross country
investigation. London School of Economics, preprint (2002).

\bibitem {St-Na-96}M.H.R. Stanley, L.A.N. Amaral, S. Buldyrev, S. Havlin, H.
Leschorn, P. Maass, M.A. Salinger, H. E. Stanley, \emph{Nature}, \textbf{319,}
804 (1996).

\bibitem {La-98}J. Laberr\`{e}re, D. Sornette, \emph{Eur. Phys. J. B}
\textbf{2}, 525 (1998).

\bibitem {Da-02}J.A. Davies, \emph{Eur. Phys. J. B} \textbf{27}, 445 (2002).

\bibitem {Ne-01}M.E.J. Newman, \emph{Phys. Rev. E} \textbf{64}, 016131 (2001).

\bibitem {KN-01}T. Knudsen, \emph{Amer. J. of Economics and Sociology}
\textbf{60,} 123, (2001).

\bibitem {GU-EB-03}C.Di Guilmi, E. Gaffeo, and M. Gallegati, \emph{Econom.
Bull.}, \textbf{15}, No. 6, 1 (2003).

\bibitem {SY-05}A.C. Silva and V.M. Yakovenko, \emph{Europhys. Lett.},
\textbf{69,} 304 (2005).

\bibitem {PeAxTe-06}R. Perline, R. Axtell and D. Teitelbaum, \emph{Small
Business Research Summary}, N285 (2006).

\bibitem {SGPS-02}H.E. Stanley, L.A.N Amaral1, P. Gopikrishnan, V. Plerou. and
M.A. Salinger, \emph{J. Phys.: Condens. Matter,} \textbf{14,} 2121 (2002).

\bibitem {TeAx-SB-05}D. Teitelbaum and R. Axtell, \emph{Small Business
Research Summary}, N247 (2005).

\bibitem {ARCH}T. Bollerslev, R.F. Engle, D.B. Nelson, ARCH models, Handbook
of Econometrics, vol 4, ch. 49, R.F Engle and D.I. McFadden Edts,
North-Holland, Amsterdam, 1994.

\bibitem {BoBo-05}L. Borland, J.-Ph.
Bouchaud,\texttt{\ arXiv:physics/0507073v1} [physics.soc-ph].

\bibitem {Turbul}R.N. Mantegna and E Stanley, \emph{Physica A}\textbf{\ 239,}
255 (1997).

\bibitem {MRW}E. Bacry, J. Delour and J.F. Muzy, \emph{Phys. Rev. E,
}\textbf{64} 26103 (2001).

\bibitem {MaHu-04}B. Mandelbrot, R.L. Hudson, ``The (Mis)behavior of Prices: A
Fractal View of Risk, Ruin, and Reward''. New York: Basic Books; London:
Profile Books, 2004.

\bibitem {CL-96}T.F. Crack, O. Ledoit, \emph{Journal of Finance,} \textbf{51,}
751 (1996).

\bibitem {Le3-06}A. Leonidov, V. Trainin, S. Zaitsev, A. Zaitsev,
\texttt{\ arXiv:physics/0605138}.

\bibitem {BoMa-06}M. Bogu\~{n}\'{a} and J. Masoliver, \emph{Eur. Phys. J. B}
\textbf{40}, 347 (2004).

\bibitem {LM-01}F. Lillo and R.N. Mantegna, \emph{Physica A}, \textbf{299},
161, (2001).

\bibitem {Le2-06}A. Leonidov, V. Trainin, S. Zaitsev, A. Zaitsev,
\texttt{\ arXiv:physics/0601098}.

\bibitem {Styl}R. Cont, \emph{Quant. Finance}, \textbf{1,} 223 (2001).

\bibitem {SPY-04}A.C. Silva, R.E. Prange, V.M. Yakovenko, \emph{Physica A}
\textbf{344,} 227 (2004).

\bibitem {Le1-06}A. Leonidov, V. Trainin, S. Zaitsev, A. Zaitsev,
\texttt{\ arXiv:physics/0603103}.

\bibitem {LeTr-06}A. Leonidov, V. Trainin, A. Zaitsev,
\texttt{\ arXiv:physics/0506072}.

\bibitem {Leverage}J. Perello, J. Masoliver, \emph{Phys. Rev. E,} \textbf{67},
037102 (2003).

\bibitem {Le-07}A. Leonidov, V. Trainin, S. Zaitsev, A. Zaitsev,
\texttt{\ arXiv:physics/0701158}; \emph{Physica A}, \textbf{386}, 240 (2007).

\bibitem {BaTh-04}M. Bartolozzi and A.W. Thomas, \emph{Physical Review}
\emph{E} \textbf{69,} 046112 (2004).

\bibitem {Co-97}R. Cont, \texttt{\ arXiv:cond-mat/9705075v3}.

\bibitem {Ro-01}B. Rosenow, \emph{Int. J. Mod. Phys. C,} \textbf{13}, 419 (2002).

\bibitem {WeRo-03}P. Weber and B. Rosenow, \emph{Quant. Finance}, \textbf{5},
357 (2005).

\bibitem {Minor}See web page: \url{
http://www.unifr.ch/econophysics/minority/}

\bibitem {QCD}R. Peschanski, \texttt{\ arXiv:hep-ph/0610019v1}.

\bibitem {SG}K. Binder and A.P. Young, \emph{Rev. Mod. Phys.} \textbf{58,} 801 (1986).

\bibitem {BC-89}Y. Balasko and D. Cass, \emph{Econometrica}, \textbf{57}, 135 (1989).

\bibitem {BiHiSp-95}B. Biais, P. Hilton, C. Spatt, \emph{Journal of Finance},
\textbf{50}, 1655 (1995).

\bibitem {BoMePo-02}J.-P. Bouchaud, M. Mezard, M. Potters, \emph{Quantitative
Finance}, \textbf{2}, 251 (2002).

\bibitem {MaMi-01}S. Maslov, M. Millis, \emph{Physica A}, \textbf{299}, 234 (2001).

\bibitem {Ma-St-99}R. Mantegna, H. E. Stanley, An Introduction to
Econophysics, Cambridge University Press, Cambridge, 1999.

\bibitem {DaGeMu-01}M.M. Dacorogna, R. Gencay, U.A. Muller, R.B. Olsen, O.V.
Pictet, An Introduction to High-Frequency Finance, Academic Press, San Diego, 2001.

\bibitem {Co-01}R. Cont, \emph{Quantitative Finance}, \textbf{1}, 223 (2001).

\bibitem {BoCo-98}J.-P. Bouchaud and R. Cont, \emph{Eur. Phys. J.,}
\textbf{B6}, 543 (1998).

\bibitem {Interm}I. Giardina and J-P. Bouchaud, \emph{Eur. Phys. J. B,}
\textbf{31,} 421 (2003).

\bibitem {KrHoHe-02}A. Krawiecki, J.A. Ho\l yst, and D. Helbing, \emph{Phys.
Rev. Lett.}, \textbf{89, }158701 (2002).

\bibitem {Levi}R.N. Mantegna, H.E. Stanley, \emph{Nature} \textbf{376,} 46 (1995).

\bibitem {Ta-86}S.J. Taylor. Modelling Financial Time Series. Chichester,
Wiley (1986).

\bibitem {GARCH}R. Engle, \emph{Econometrica}, \textbf{50,} 987 (1982).

\bibitem {CGM-03}P. Carr, H. Geman, D. Madan, M. Yor, \emph{Math. Finance},
\textbf{13,} 345 (2003).

\bibitem {CaFi-04}L. Calvet and A. Fisher, \emph{J. Financ. Econometrics},
\textbf{2}, 49 (2004).

\bibitem {FiCaMa-97}B. Mandelbrot, A. Fisher, L. Calvet, Cowles Foundation
Disc. Paper, 1164 (1997).

\bibitem {Renorm}V.S. Dotsenko, \emph{J. Phys. C }\textbf{20}, 5473 (1987);
\emph{J. Phys.: Condens. Matter.,} \textbf{2}, 2721 (1990).

\bibitem {Do-85}V.S. Dotsenko, \emph{J. Phys.} \emph{C }\textbf{18}, 6023 (1985).

\bibitem {BaPaSh-97-Ph}P. Bak, M. Paczuski, and M. Shubik, \emph{Physica A,}
\textbf{246,} 430 (1997).

\bibitem {DaFaIo-03}M.G. Daniels, J. D. Farmer, G. Iori, E. Smith, \emph{Phys.
Rev. Lett.} \textbf{90,} 108102 (2003).

\bibitem {SmFaGi}E. Smith, J. D. Farmer, L. Gillemot, S. Krishnamurthy,
\emph{Quant. Finance,} \textbf{3,} 481, (2003).

\bibitem {Fa-70}E.F. Fama, \emph{J. of Finance} \textbf{25,} 383 (1970).

\bibitem {Sc-AER-81}R.J. Schiller, \emph{Am. Econ. Rev.,} \textbf{71,} 421 (1981).

\bibitem {CS-83}D. Cass and K. Shell, \emph{J. Polit. Econ.,} \textbf{92}, 193 (1983).

\bibitem {GLGB-08}A. Joulin, A. Lefevre, D. Grunberg, J-P. Bouchaud,
\texttt{\ arXiv:0803.1769v1} [physics.soc-ph].

\bibitem {MuDeBa-00}J.F. Muzy, J. Delour and E. Bacry, \emph{Eur. Phys. J.} B
\textbf{17,} 537 (2000).

\bibitem {BaDeMu-01}E. Bacry, J. Delour and J.F. Muzy, \emph{Phys. Rev. E
}\textbf{64,} 026103 (2001).

\bibitem {Bo-Ch-05}J.-P. Bouchaud, \emph{Chaos}, \textbf{15}, 026104 (2005).

\bibitem {Bak}P. Bak, How Nature Works: The Science of Self-Organized
Criticality, Copernicus Springer, New York, (1996).

\bibitem {Positive}M. Youssefmir, B.A. Huberman and T. Hogg, \emph{Comp.
Econ.} \textbf{12,} 97 (1998).

\bibitem {Ma-01}M. Marsili, \emph{Physica A,} \textbf{299,} 93 (2001).

\bibitem {Eisler}Z. Eisler, Ph.D. thesis, ``Fluctuation Phenomena on the Stock
Market'', Budapest, (2007).

\bibitem {Sh-00}A. Shleifer, Inefficient Markets, An Introduction to
Behavioral Finance, Oxford University Press (2000).

\bibitem {Sh-81}R. J. Schiller, \emph{Amer. Econ. Rev.}, \textbf{71,} 421 (1981).

\bibitem {LGCMPS-99}Y.I. Liu, P. Gopikrishnan, P. Cizeau, M. Meyer, C-K. Peng,
and H.E. Stanley, \emph{Phys. Rev.} \emph{E, }\textbf{60,} 1390 (1999).

\bibitem {PGMNS-99}V. Plerou, P. Gopikrishnan, L.A.N. Amaral, M. Meyer and
H.E. Stanley, \emph{Phys. Rev. E} \textbf{60,} 6519 (1999).

\bibitem {Sh-03}F. Schmitt, \emph{Eur. J. Phys. B, }\textbf{34,} 85 (2003).

\bibitem {BJPW-03}J.-P. Bouchaud, Y. Gefen, M. Potters, M. Wyart, \emph{Quant.
Finance}, \textbf{4,} 176 (2004).

\bibitem {LiFaMa-02}F. Lillo, J.D. Farmer, R.N. Mantegna, \emph{Nature,}
\textbf{421}, 123 (2003).

\bibitem {PGGS-08}V. Plerou, P. Gopikrishnan, X. Gabaix and H.E. Stanley,
\emph{Phys. Rev. E} \textbf{66}, 027104 (2002).

\bibitem {Ma-EPJ-99}R.N. Mantegna, \emph{Europ. Phys. J. B }\textbf{11}, 193 (1999).

\bibitem {GGPS-03}X. Gabaix, P. Gopikrishnan, V. Plerou and H.E. Stanley,
\emph{Physica A} \textbf{324,} 1 (2003).

\bibitem {PI-03}F. Lillo, J.D. Farmer and R.N. Mantegna, \emph{Nature}
\textbf{421,} 129 (2003).

\bibitem {PB-03}M. Potters and J.-P. Bouchaud, \emph{Physica A}\textbf{\ 324,}
133 (2003).

\bibitem {Volat}P. Cizeau, Y. Liu, M. Meyer, Ci-K. Peng, H.E. Stanley,
\emph{Physica} \emph{A,} \textbf{245,} 441 (1997).

\bibitem {GPNM-98}P. Gopikrishnan, V. Plerou1, L.A.N. Amaral1, M. Meyer1, and
H.E. Stanley, \emph{Phys. Rev. E} \textbf{60}, 5305 (1999).

\bibitem {CP-03}R. Carvalho and A. Penn, \emph{Physica A,} \textbf{332,} 539 (2003).

\bibitem {SoMaMu-02}D. Sornette, Y. Malevergne and J.-F. Muzy,
\texttt{\ arXiv:cond-mat/0204626v1}.

\bibitem {Ma-63}B. Mandelbrot, \emph{J. of Business}, \textbf{36,} 307 (1963).

\bibitem {DSD-04}M.M. Dubovikov, N.V. Starchenko, M.S. Dubovikov,
\emph{Physica A,} \textbf{339,} 591 (2004).

\bibitem {Chaos}Jens Feder, Fractals. New York, NY: Plenum Press, 1988.

\bibitem {WB-03}M. Wyart and J.-P. Bouchaud,
\texttt{\ arXiv:cond-mat/0303584v2}.

\bibitem {NP-06}A.P. Nawroth and J. Peinke, \emph{Eur. Phys. J. B,}
\textbf{50,} 147 (2006).

\bibitem {Intro-01}M.M. Dacorogna, R. Gen\c{c}ay, U.A. M\H{u}ller, R.B. Olsen,
O.V. Pictet, An Introduction to High Frequency Finance Academic Press, San
Diego, (2001).

\bibitem {MaAsDa-03}T. Di Matteo, T. Aste, and M. Dacorogna, \emph{Physica A,}
\textbf{324,} 183 (2003).

\bibitem {CaLe-83}A.O. Caldeira and A.J. Leggett, \emph{Physica A,}
\textbf{121,} 587 (1983).

\bibitem {HPZ-92}B.L. Hu, J.P. Paz and Y. Zhang, \emph{Phys. Rev. D,}
\textbf{45,} 2843 (1992).

\bibitem {PZ-86}S.V. Panyukov, A.D. Zaikin, \emph{JETP,} \textbf{91,} 1677 (1986).

\bibitem {Ho-97}W. Brock, C. Hommes, \emph{Econometrica} \textbf{65,} 1059
(1997); C. Hommes, \emph{Quantit. Finance}, \textbf{1,} 149 (2001).

\bibitem {GiBo-03}I. Giardina, J.P. Bouchaud, \emph{Eur. J. of Phys. B}
\textbf{31,} 421 (2003).

\bibitem {LP}E.M. Lifshitz and L.P. Pitaevsky, Physical Kinetics, (Pergamon
Press, 1981).

\bibitem {Ph-E-58}A. Phillips, \emph{Economica}, \textbf{25,} (1958).

\bibitem {DaHaSh-96}S.J. Davis, J. Haltiwanger and S. Schuh, \emph{Job
Creation and Destruction}, MIT Press: Cambridge, Mass.
\end{thebibliography}
\end{document}